\documentclass[a4paper,11pt]{article}

\usepackage{jheppub}

\usepackage[T1]{fontenc} 
\usepackage{dsfont}
\usepackage{framed}

\usepackage{caption}
\usepackage{subcaption}
\usepackage{dsfont}
\usepackage{tikz}
\usepackage{array}
\usepackage{makecell}

\allowdisplaybreaks[1]

\def\be#1\ee{\begin{align}#1\end{align}}

\def\({\left(}
\def\){\right)}
\def\[{\left[}
\def\]{\right]}
\def\<{\langle}
\def\>{\rangle}

\def\nn{\nonumber\\}

\def\rm{\mathrm}
\def\bb{\mathbb}
\def\cal{\mathcal}

\def\pa{\partial}
\def\->{\rightarrow}

\def\xp{x_{\parallel}}
\def\xn{x_{\perp}}
\def\yp{y_{\parallel}}
\def\yn{y_{\perp}}

\def\Ph[#1,#2,#3]{\Phi^{(#2,#3)}_{#1}}
\def\tPh[#1,#2,#3]{\tilde{\Phi}^{(#2,#3)}_{#1}}

\def\Dh[#1]{\hat{\D}_{#1}}
\def\gh[#1]{\hat{\g}_{#1}}
\def\dD{\hat{\D}_{-}^{(p)}}

\def\du{d_\text u}
\def\f{\text f}

\def\a{\alpha}
\def\b{\beta}
\def\g{\gamma}
\def\G{\Gamma}
\def\d{\delta}
\def\D{\Delta}
\def\e{\epsilon}

\def\th{\theta}

\def\k{\kappa}
\def\l{\lambda}

\def\m{\mu}
\def\n{\nu}
\def\x{\xi}

\def\p{\pi}

\def\s{\sigma}

\def\lra{\leftrightarrow}

\begin{document}

\title{Boundary anomalous dimensions from BCFT: 
O($N$)-symmetric \boldmath{$\phi^{2n}$} theories with a boundary 
and higher-derivative generalizations
}
	
\author[a,b]{Yongwei Guo}
\author[a]{and Wenliang Li}
\emailAdd{liwliang3@mail.sysu.edu.cn}
\affiliation[a]{School of Physics, Sun Yat-sen University, Guangzhou 510275, China}
\affiliation[b]{Shing-Tung Yau Center and School of Physics, Southeast University, Nanjing, 210096, China}
	
\abstract{We investigate the $\phi^{2n}$ deformations of the O($N$)-symmetric (generalized) free theories with a flat boundary, where $n\geq 2$ is an integer.
The generalized free theories refer to the $\Box^k$ free scalar theories  
with a higher derivative kinetic term, 
which is related to the multicritical generalizations of the Lifshitz type. 
We assume that the (generalized) free theories and the deformed theories have boundary conformal symmetry and O($N$) global symmetry.
The leading anomalous dimensions of some boundary operators are derived from the bulk multiplet recombination and analyticity constraints.
We find that the $\e^{1/2}$ expansion in the $\phi^6$-tricritical version of the special transition extends to other multicritical cases with larger odd integer $n$, 
and most of the higher derivative cases involve a noninteger power expansion in $\e$. 
Using the analytic bootstrap, we further verify that the multiplet-recombination results are consistent with boundary crossing symmetry. 
}
	
\maketitle 

\section{Introduction}

The O($N$) spin model on a semi-infinite hypercubic lattice consists of the bulk and the boundary.
At high temperatures, the system is in a completely disordered phase.
If the boundary coupling is weak, 
the bulk and boundary of the system can order simultaneously as the temperature is lowered. 
It is expected that the transition point is described by boundary conformal field theory (BCFT).
If the boundary coupling is strong, a surface transition may occur while the bulk remains disordered. 
Accordingly, we may encounter the {\it extraordinary} transition at the bulk critical temperature, 
in contrast to the {\it ordinary} transition at weak boundary coupling.  
The two transition lines may be separated by the so-called {\it special} transition. 
We refer to \cite{Diehl:1996kd} for a review of boundary critical phenomena.\footnote{
In $d=3$ dimensions, the phase diagram of the O($N$) spin model is more subtle at small $N$ 
(see e.g. \cite{Padayasi:2021sik,Hu:2025yrs} for recent conformal bootstrap studies and references therein).}
Besides the application to phase transitions, BCFTs have been proposed as holographic duals of gravity theories \cite{Takayanagi:2011zk,Fujita:2011fp}, 
extending the more standard AdS/CFT correspondence \cite{Maldacena:1997re,Gubser:1998bc,Witten:1998qj}.
See \cite{Andrei:2018die} for a recent review of various aspects of BCFT. 

The boundary mentioned above is a hyperplane.
The bulk conformal symmetry SO($d+1,1$) of $\bb R^d$ is partly broken by the boundary.
We will focus on the cases with boundary conformal symmetry SO($d,1$), 
i.e., the conformal symmetry of $\bb R^{d-1}$. 
In the bulk, the set of CFT data includes the scaling dimensions $\D_i$ of bulk operators and their operator product expansion (OPE) coefficients $\l_{ijk}$.
As the translation symmetry in the normal direction is broken by the presence of a boundary, 
bulk scalar operators can acquire nonzero scale invariant one-point functions.\footnote{The spinning bulk one-point functions vanish due to boundary conformal symmetry.}
We use $a_{i}$ to denote the one-point function coefficients.
On the boundary side, the scaling dimensions $\hat{\D}_{i}$ and OPE coefficients $f_{ijk}$ of boundary operators form a set of boundary data.
Furthermore, we can expand a bulk operator in terms of boundary operators.  
We use $b_{ij}$ to denote the corresponding boundary operator expansion (BOE) coefficients.
\footnote{The bulk one-point functions are related to the boundary identity operator in the BOEs.}
We assume that
the set of bulk data $\{\D_i,\l_{ijk}\}$ coincides with that in the absence of a boundary. 
\footnote{In the region far from the boundary,  
correlation functions can be well approximated by those without a boundary 
because the symmetry breaking effects are negligible. 
The bulk scaling dimensions and OPE coefficients are expected to be the same as those without a boundary.
The boundary effects become more significant as we approach the boundary. 
For example, the effects of nonvanishing bulk one-point functions become more prominent. }
Our main goal is to deduce the leading anomalous dimensions of the boundary scalar operators in the $\e$ expansion.
We also obtain some operator expansion coefficients, 
such as some BOE coefficients
associated with the special transitions. 

Let us start with the basic example of the canonical ($k=1$) free theory. 
We need to clarify the meaning of conformal boundary conditions.
The standard derivation of the Euler-Lagrange equations is based on the fixed boundary conditions 
in which the variations at the boundary are set to zero.
In the context of BCFTs, we also consider other types of boundary conditions.   
In fact, the boundary behaviors are closely related to the choice of boundary terms. 
To illustrate this, consider the O($N$)-symmetric free theory with a boundary:
\be\label{free action k=1}
S_\pm\propto\int_{\bb R_+^d} \rm d^d x \; \phi_a\Box\phi_a + \int_{\bb R^{d-1}} \rm d^{d-1} x \; (\pm)\Ph[a,0,1]\Ph[a,1,1]\,,
\ee
where summation over repeated indices is implied, and $\phi_a$ is an $N$-dimensional O($N$) vector $a=1,2,\ldots N$.
The theory is defined on the Euclidean upper half space
\be
\bb R^d_+=\{(\xp^\m,\xn):\xp^\m\in\bb R^{d-1},\,\xn\geq 0\}
\,,
\ee
where $\xp$ and $\xn$ are the coordinates parallel and perpendicular to the boundary. 
The boundary term in \eqref{free action k=1} involves the boundary fundamental primaries\footnote{A boundary operator is primary if it is annihilated by $K_\m$, where $K_\m$ are the generators of special conformal transformations along the directions parallel to the boundary.} 
\be\label{boundary primary k=1 introduction}
\Ph[a,0,1]=\lim_{\xn\->0}\phi_a \,,\quad
\Ph[a,1,1]=\lim_{\xn\->0}\pa_\perp\phi_a
\,.
\ee
The first superscript $q=0,1$ indicates the order of derivatives.  
The second superscript $1$ implies that they are boundary fundamental operators.
We will also consider the boundary composite operators $\Ph[,q,2p]$, $\Ph[a,q,2p+1]$,  
which are composed of $2p$ and $2p+1$ boundary fundamental operators of the $q$-derivative type.
The boundary conditions can be derived from the requirement that the action is stationary under arbitrary variations of $\phi_a$.
Assuming that the bulk equation of motion is satisfied, we have
\be\label{delta S}
\d S_+\propto\int_{\bb R^{d-1}} \rm d^{d-1} x \; \Ph[a,1,1]\d\Ph[a,0,1]\,,\quad
\d S_-\propto\int_{\bb R^{d-1}} \rm d^{d-1} x \; \Ph[a,0,1]\d\Ph[a,1,1]
\,,
\ee
where $\d\phi_a$ is assumed to vanish at infinity. 
In this way, the boundary conditions are a consequence of $\d S_\pm=0$.
\footnote{In variational problems, the boundary conditions of this type are sometimes referred to as natural boundary conditions \cite{gelfand2000calculus}.}
The $+$ case leads to the Neumann boundary condition $\Ph[a,1,1]=0$, 
while the $-$ case corresponds to the Dirichlet boundary condition $\Ph[a,0,1]=0$. 
In each boundary condition, one operator in the pair $(\Ph[a,0,1],\Ph[a,1,1])$ is set to zero, and the other operator plays the role of a physical fundamental operator.

In $d=4$ dimensions, the Dirichlet and Neumann boundary conditions correspond to the ordinary and  special transitions,\footnote{In the interacting case, one may need to be more careful about the identification if the boundary condition is scale dependent \cite{Diehl:2020rfx}.} 
which belong to different boundary universality classes.
In $d=4-\e$ dimensions, a free BCFT can be deformed into an interacting BCFT by a quartic bulk interaction $\phi^{4}$.
\footnote{This interacting theory describes the lattice spin model mentioned at the beginning of this section.}
The free data will receive corrections in terms of a formal power series in $\e$.
See \cite{Diehl1986} for an early review. 
In this work, we are interested in the multicritical generalization 
associated with a $\bb Z_2$-even bulk interaction $\phi^{2n}$, where $n>2$ is an integer. 

The $\phi^{2n}$ deformations of the Dirichlet and Neumann cases 
describe the ordinary and special transitions associated with an interacting multicritical bulk. 
For instance, the mixture of helium-3 and helium-4 with a boundary may provide an experimental realization of the $\phi^6$-tricritical bulk \cite{blume1971ising,de1977callan} with a boundary \cite{speth1983tricritical, diehl1987walks, eisenriegler1988surface}. 
The $N\rightarrow 0$ limit of an O($N$)-symmetric BCFT is also interesting 
due to the connection to adsorption of polymer chains.
For $n=2$, the $N\rightarrow 0$ limit is related to polymers in good solvents, i.e., self-avoiding walks, near a boundary \cite{de1972exponents,de1976scaling,eisenriegler1982adsorption}. For $n=3$, the $N\rightarrow 0$ limit corresponds to polymers in theta solutions, i.e., random walks, near an attracting wall \cite{FvanDieren1987}.
With a slight abuse of terminology, 
we will use Dirichlet (D) and Neumann (N) to specify the boundary conditions of the deformed BCFTs.
We should emphasize that the interacting boundary conditions are not necessarily the same as the free boundary conditions.  

In the Neumann case, we find that 
a pure bulk $\phi^{2n}$ deformation of the BCFT is inconsistent for odd integer $n$. 
Since $n+1$ is an even integer, 
the classically marginal boundary interaction $h\phi^{n+1}\big|_{\xn=0}$ is also $\bb Z_2$ even 
and can mix with the bulk interaction $\phi^{2n}$.
\footnote{For even $n$ or in the Dirichlet case, there is no marginal $\bb Z_2$-even boundary interaction.}
Accordingly, the critical coupling constant for this boundary interaction is of order $\e^{1/2}$, 
and the $\e$ expansion admits half-integer power terms. 
The $\e^{1/2}$ expansion of the tricritical case,\footnote{The $\b$ function for $h$ is nonzero for $h=0$ or $h=O(\e)$, but it can vanish for $h=O(\e^{1/2})$,
which is associated with an interacting IR fixed point. 
See equation (8) in \cite{diehl1987walks} for more details on the $\b$ function.} 
i.e., $n=3$, was discussed long ago already in \cite{diehl1987walks,eisenriegler1988surface},
which is related to the mixing of the bulk interaction $g\phi^{6}$ and the boundary interaction $h\phi^{4}\big|_{\xn=0}$ in $3-\e$ dimensions. 
\footnote{
In \cite{Herzog:2022jlx}, the $\e^{1/2}$ expansion also appeared in the study of a fermionic boundary conformal field theory in $3-\e$ dimensions.
In fact, the existence of the $\e^{1/2}$ expansion was noticed earlier in the context of the random Ising models in $4-\e$ dimensions \cite{khmelnitskii1975second,shalaev1977phase,aharony1976new}.
} 
We notice that the anomalous dimension of the boundary fundamental operator may require some clarification, 
as the existing results in the literature appear to differ by a factor $C$:
\be
\gh[0,1]=-C\frac{(N+2)(N+4)}{16(3N+22)}\e+O(\e^{3/2})
\,.
\ee
The first subscript $0$ indicates the Neumann case and 
the second subscript $1$ denotes the boundary fundamental operator.
The values for $C$ in the literature are $C=1$ \cite{diehl1987walks,eisenriegler1988surface} and $C=2$ \cite{speth1983tricritical,Dey:2020jlc}.\footnote{It appears to us that the result in the early reference \cite{speth1983tricritical} has a missing factor of $1/2$.
At $d=3$, a missing factor of $1/2$ was also noticed in \cite{eisenriegler1988surface}.
The more recent reference \cite{Dey:2020jlc} derived the same result because their bulk critical coupling constant also has a missing factor of 1/2 (see their 
equation (3.29) with $n=3$).}
Our results are consistent with the $C=1$ case obtained by Diehl and Eisenriegler.\footnote{References \cite{speth1983tricritical,Dey:2020jlc} also considered the Dirichlet case at $n=3$.
Our analyses suggest that their results have a missing factor of $1/2$ as well.}
More details on our multiplet recombination and 
crossing analyses can be found around \eqref{gamma Phi} and \eqref{crossing Neumann}.

Apart from the multicritical generalizations associated with $\phi^{2n}$ interactions, we also consider the higher derivative generalizations, 
which are related to Lifshitz points. 
A Lifshitz point is a multicritical point where a disordered phase, a spatially homogeneous ordered phase, and a spatially modulated ordered phase meet.
They can be realized in diverse physical systems (see e.g. \cite{Diehl:2002ri} and references therein).
The Lifshitz Lagrangians contain higher derivative bilinear terms,  
and the $d$ directions in $\bb R^d$ may be divided into two sets of directions $X$ and $Y$. 
For instance, the Lifshitz theory with a $\phi^4$ interaction reads
\be
S\propto\int\rm d^d x\(\phi_a\,\Box_X\phi_a+g_0\phi_a\,\Box_Y\phi_a+g'_0\phi_a\,\Box_Y^2\phi_a+g_2\phi^2+g_4\phi^4\)
\,,
\ee
where $\Box_X$ and $\Box_Y$ are the Laplacians associated with the $X$ and $Y$ directions.
The case $g_0=g_2=0$ is a Lifshitz point.
This multicriticality is reached by tuning $g_0$ to zero, 
as opposed to the $\phi^6$-tricritical point associated with $g_4=0$. 
For $N=1$, it describes the axial-next-nearest-neighbor Ising (ANNNI) model at criticality, where the modulated phase arises from the competition between different interactions.
If the set $X$ is empty, 
then the only bilinear term is given by the four-derivative term $\phi_a\,\Box^2_Y\phi_a$.  
In this case, the critical point is known as an isotropic Lifshitz point, 
which can be viewed as the $\phi^4$ deformation of the $\Box^2$ higher derivative free theory.
Besides the condensed matter realizations, 
higher derivative theories also appear naturally in the contexts of  elasticity and cosmology. 
(See \cite{Chalabi:2022qit,Herzog:2024zxm} and references therein.) 
In this paper, the multicritical points mainly refer to the $\phi^{2n}$ generalizations, 
while the isotropic Lifshitz type is referred to as higher derivative generalizations.

Along these lines, 
we want to study the $\phi^{2n}$ deformations of the higher derivative free theories with a boundary:\footnote{See also \cite{Brust:2016gjy} for the generalized free theories without a boundary.}
\be\label{gff free action}
S_{\pm,\ldots,\pm}\propto\int_{\bb R_+^d} \rm d^d x \; \phi_a\,\Box^k\phi_a+S_{\text{bdy},\pm,\ldots,\pm}\,,
\ee
where the bulk action has a higher derivative kinetic term and $k$ is a positive integer. 
As in the canonical case \eqref{free action k=1}, the boundary conditions are specified by the boundary term\footnote{
The product of two operators in each summand has a total number of $2k-1$ derivatives.
Since the bulk kinetic term has $2k$ derivatives, the product $\Ph[a,j,1]\Ph[a,2k-1-j,1]$ has dimension $d-1$, where $d$ is the dimension of the bulk kinetic term.
The dimension of the boundary is also $d-1$, so these boundary terms do not introduce 
additional scales.}
\be\label{boundary terms-k}
S_{\text{bdy},\pm,\ldots,\pm}=\sum_{j=0}^{k-1}\int_{\bb R^{d-1}}\rm d^{d-1}x\;(\pm )\Ph[a,j,1]\Ph[a,2k-1-j,1]
\,,
\ee
where the boundary fundamental primaries 
\be\label{boundary primary schematic}
\Ph[a,q,1]=\lim_{\xn\->0}\(\pa_\perp^{\,q}+\ldots\)\phi_a
\ee
are generalizations of the $q=0,1$ cases in \eqref{boundary primary k=1 introduction}. 
The ellipsis in \eqref{boundary primary schematic} represents the derivative terms constructed using both the normal derivative $\pa_\perp$ and the boundary Laplacian $\Box_\parallel\equiv\Box-\pa_\perp^2$. 
\footnote{Their coefficients are fixed by the requirement that $\Ph[a,q,1]$ is a boundary primary.} 
In a $\Box^k$ higher derivative theory, the order of derivatives is at most $2k-1$ for the boundary fundamental operators, 
so $q\in\{0,1,\ldots,2k-1\}$. 
According to the signs of the product terms in \eqref{boundary terms-k}, 
the variational principle leads to different boundary conditions.
As a generalization of the pair formulation at $k=1$, the boundary fundamental primaries are organized into $k$ pairs. 
In each pair, only one operator is physical and the other one should vanish, 
i.e., be associated with a boundary null state. 
The set of vanishing boundary fundamental operators 
is in one-to-one correspondence with   
the conformal boundary conditions for the generalized free theory.
\footnote{This is different from Cardy's conformal boundary condition, where the ``normal-$\m$'' components of the bulk stress tensor vanish at the boundary \cite{Cardy:1984bb}.
Here the ``normal-$\m$'' components are the off-diagonal components associated with the normal direction.
See \cite{Chalabi:2022qit} for a discussion of the relation between Cardy's conformal boundary condition and the conformal boundary conditions here.} 
See \cite{Chalabi:2022qit} and section \ref{Generalized free theory with a boundary} for further discussion.

After a brief overview of the higher derivative free theory with a boundary, 
we move to the deformation induced by a bulk interaction $\phi^{2n}$. 
The $\e$ expansion is formulated slightly below the upper critical dimension
\be
\du=\frac{2nk}{n-1}
\,, 
\ee 
which is a higher derivative generalization of the $k=1$ case \eqref{upper-d}. 
As in \cite{Guo:2023qtt}, we assume that $k$ and $n-1$ have no common divisor to avoid the possibility of derivative interactions in the bulk. 
\footnote{In a very recent work \cite{Herzog:2024zxm}, 
a deformation of the $\Box^2$ BCFT by a two-derivative interaction was considered in $d=6-\e$ dimensions.}
The noninteger power expansion in the canonical $k=1$ case extends to most of the higher derivative theories.
For $k>1$, the noninteger powers in $\e$ are introduced to ensure the consistency of the matching conditions and the analyticity of three-point functions discussed later. 
In some cases, the noninteger power $\e$ expansion may be understood as a consequence of mixing, i.e.,  
the bulk interaction $\phi^{2n}$ mixes with the boundary interaction $\pa^{k-1}\phi^{n+1}|_{\xn=0}$, which is classically marginal and $\bb Z_2$ even 
for odd $k$ and odd $n$.

\begin{figure}
\center
\begin{tikzpicture}[scale = 1.2]
\draw[black, thick] (-4,2) -- (-3,1);
\draw[black, thick] (-2,2) -- (-3,1);
\draw[black, thick] (-3,1) -- (-3,0);
\draw[black, thick] (-4.5,0) -- (-1.5,0);

\draw[black, thick] (1.75,2) -- (1.75,0);
\draw[black, thick] (3.25,2) -- (3.25,0);
\draw[black, thick] (1,0) -- (4,0);

\draw (-5.5,1.5) node [anchor=north west]   {\Large $\sum\limits_i$};
\draw (-0.75,1.5) node [anchor=north west]   {\Large $=\hspace{1em}\sum\limits_j$};

\draw (-4.5,2.5) node [anchor=north west]   {$\phi_a$};
\draw (-2,2.5) node [anchor=north west]   {$\phi_b$};

\draw (1.2,2.5) node [anchor=north west]   {$\phi_a$};
\draw (3.2,2.5) node [anchor=north west]   {$\phi_b$};

\draw (-3,0.8) node [anchor=north west]   {$i$};
\draw (2.3,0.6) node [anchor=north west]   {$j$};

\end{tikzpicture}
\caption{The bootstrap equation for the bulk-bulk two-point function $\<\phi_a\phi_b\>$.
In the bulk channel (left), 
the bulk OPE of $\phi_a\phi_b$ leads to a sum of bulk one-point functions.
In the boundary channel (right), 
we apply the BOE to each $\phi_a$ and obtain a sum of boundary two-point functions.}
\label{bootstrap equation}
\end{figure}
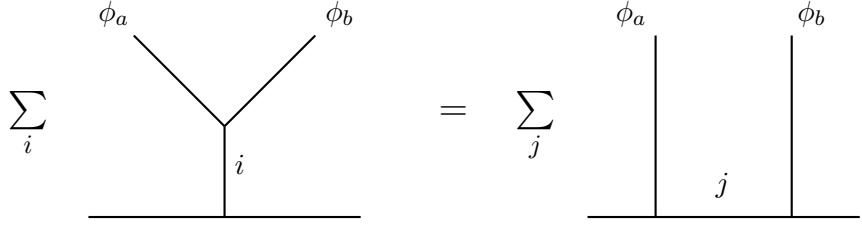

Our procedure for deriving the boundary anomalous dimensions is described below.
Following \cite{Nishioka:2022odm}, we assume that the bulk multiplet recombination \cite{Rychkov:2015naa} is unaffected by the boundary.\footnote{In principle, the leading bulk data can be derived by considering the matching conditions for correlators infinitely far away from the boundary.
The infinite distance limit is equivalent to the case without a boundary. 
See also \cite{Rychkov:2015naa,Basu:2015gpa,Yamaguchi:2016pbj,Ghosh:2015opa,Raju:2015fza,Gliozzi:2016ysv,Roumpedakis:2016qcg,Soderberg:2017oaa,Behan:2017emf,Gliozzi:2017hni,Gliozzi:2017gzh,Nishioka:2022odm,Nishioka:2022qmj,Antunes:2022vtb,Guo:2023qtt,Guo:2024bll,Antunes:2024mfb} for the multiplet recombination method in various CFTs, and \cite{Safari:2017irw,Nii:2016lpa,Hasegawa:2016piv,Hasegawa:2018yqg,Skvortsov:2015pea,Giombi:2016hkj,Giombi:2017rhm,Giombi:2016zwa,Codello:2017qek,Codello:2018nbe,Antipin:2019vdg,Vacca:2019rsh,Giombi:2020rmc,Giombi:2020xah,Dey:2020jlc,Giombi:2021cnr,Safari:2021ocb,Zhou:2022pah,Bissi:2022bgu,Herzog:2022jlx,Giombi:2022vnz,SoderbergRousu:2023pbe,Herzog:2024zxm} for the closely related studies based on the equations of motion.}
The bulk multiplet recombination means that the free multiplets $\{\phi_\text{free}\}$ and $\{\phi^{2n-1}_\text{free}\}$ recombine into a long multiplet $\{\phi\}$ in the deformed theory.
In other words, the interacting counterpart of the free primary $\phi^{2n-1}_\text{free}$ is a descendant $\Box^k\phi$.
This gives rise to a set of matching conditions associated with the smooth Gaussian limit of the deformed correlators $\<\Box^k\phi\ldots\>$, 
and fix some of the leading bulk data.
Since the boundary does not affect the bulk multiplet recombination, 
we can directly make use of the bulk data obtained in our previous work \cite{Guo:2023qtt}. 
We will use the matching conditions for the bulk-boundary two-point functions $\<\Box^k\phi_a(x) \Ph[b,q,1](y)\>$ 
to determine the anomalous dimensions of boundary fundamental primaries $\Ph[a,q,1]$.
For a consistency check, we carry out the analytic bootstrap study 
for the bulk-bulk two-point function $\<\phi_a(x)\phi_b(y)\>$. 
This correlator can be evaluated using the bulk OPE or the BOE, 
which should lead to the same result \cite{Liendo:2012hy}.
We verify that the multiplet-recombination results are consistent with the boundary crossing equation 
(see figure \ref{bootstrap equation}). 
We refer to \cite{Liendo:2012hy,Bissi:2018mcq,Dey:2020jlc,Dey:2020lwp} 
for some previous analytic bootstrap studies of the canonical BCFTs with $k=1$. 

The matching conditions above involve the boundary fundamental operators, 
whose leading anomalous dimensions are always of order $\e^1$. 
However, this does not mean that the leading anomalous dimensions of the boundary composite operators 
are necessarily of integer power in $\e$. 
In fact, we have mentioned that noninteger powers of $\e$ do appear in some cases 
due to the mixing of the bulk $\phi^{2n}$ interaction with a boundary interaction. 
The necessity of the noninteger power $\e$ expansion can be noticed from 
the matching conditions involving certain composite operators. 
At $k=1$, we find that 
the matching condition for the bulk-boundary two-point function $\<\Box\phi_a(x)\Ph[b,0,n](y)\>$ is inconsistent with the integer power $\e$ expansion in the Neumann case with odd $n$. 
For $k>1$, there are more scenarios in which the existence of composite operators implies a noninteger power series in $\e$, which are related to the analyticity requirement below. 

Furthermore, we want to determine the anomalous dimensions of composite operators 
because they are also part of the boundary data. 
As in \cite{Nishioka:2022odm}, we consider the composite operators $\Ph[,q,2p]$ and $\Ph[a,q,2p+1]$, which are O($N$) singlet and vector operators. 
Since these operators do not appear in the free-theory BOE of $\phi_a$, 
the matching conditions for the corresponding bulk-boundary two-point functions lead to constraints on their BOE coefficients, 
which do not directly determine their anomalous dimensions. 
The essential constraints from the bulk multiplet recombination are associated with 
the bulk-boundary two-point functions.
It is not particularly useful to further study the matching conditions for more complicated correlators, 
such as the cases of the bulk-bulk two-point and bulk-boundary-boundary three-point functions. 
According to BOE, the matching conditions for these correlators hold automatically if the bulk-boundary matching conditions are satisfied, 
and thus they do not yield independent constraints. 

To determine the anomalous dimensions of the composite operators, 
we will make use of additional consistency constraints 
from the analyticity of Euclidean quantum field theories  \cite{Lauria:2020emq,Behan:2020nsf,Behan:2021tcn,Nishioka:2022odm,Bartlett-Tisdall:2023ghh}.
Following \cite{Nishioka:2022odm}, we examine the bulk-boundary-boundary three-point functions $\<\phi_a \Ph[,q,2p] \Ph[b,q,2p+1]\>$ and $\<\phi_a \Ph[b,q,2p-1] \Ph[,q,2p]\>$.
They play a similar role as the bulk three-point functions 
$\<\phi_a \phi^{2p}\phi_b^{2p+1}\>$ and $\<\phi_a \phi_b^{2p-1}\phi^{2p}\>$ 
in the study of the bulk-composite anomalous dimensions \cite{Rychkov:2015naa}.
For generic boundary scaling dimensions,
these bulk-boundary-boundary correlators have unphysical singularities 
when the bulk operator and a boundary operator lie along a line perpendicular to the boundary. 
This contradicts the axiom of Euclidean quantum field theories 
that singularities should be associated with the coincidence limit, 
so the coefficients of these nonanalytic terms should vanish.  
In this way, we are able to determine the anomalous dimensions of the boundary composite operators.  
In some sense, we generalize the BCFT results of the canonical $\phi^4$ case in \cite{Nishioka:2022odm} to 
the $\phi^{2n}$-multicritical and higher derivative theories. 
Our main results are summarized in table \ref{summary}.
\begin{table}
	\centering
	\begin{tabular}{|c|c|c|}
		\hline
		Type of the kinetic term & $\gh[q,1]$ & $\gh[q,2p+1]$, $\gh[q,2p]$ \\
		\hline 
		$k=1$ & \eqref{gamma Phi} & 
		\makecell{ $\e$ expansion: \eqref{gamma odd}, \eqref{gamma even} \\ $\e^{1/2}$ expansion: \eqref{epsilon 1/2 odd}, \eqref{epsilon 1/2 even} }\\
		\hline
		$k>1$ & \eqref{gamma fundamental} & \makecell{$(k,n)=(2,2)$, NN: \eqref{k=2 composite vector}, \eqref{k=2 composite singlet}\\
		$(k,n)=(2,2)$, DD: \eqref{k=2 composite vector DD}, \eqref{k=2 composite singlet DD}  } \\
		\hline
	\end{tabular}
	\caption{Summary of the boundary anomalous dimensions $\gh[q,r]=\hat\g_{\Ph[,q,r]}$
	for the $\phi^{2n}$ deformations of the $\Box^k$ free BCFT.
	The second column for $\gh[q,1]$ corresponds to the anomalous dimensions of the boundary fundamental operators $\Ph[a,q,1]$, 
	while the third column for $\gh[q,2p+1]$ and $\gh[q,2p]$ are associated with the boundary composite operators $\Ph[a,q,2p+1]$ and $\Ph[,q,2p]$.
	See \eqref{boundary primary schematic} and \eqref{composite definition} for the definitions of the boundary fundamental and composite operators.
	In the higher derivative case $k>1$, the anomalous dimensions of  composite operators have more involved dependence on the boundary conditions, so we focus on concrete examples: 
	the $\phi^4$ deformation of the $\Box^2$ free BCFT with the NN or DD boundary condition, i.e., $\Ph[{a,\text{free}},3,1]=\Ph[{a,\text{free}},2,1]=0$ or $\Ph[{a,\text{free}},0,1]=\Ph[{a,\text{free}},1,1]=0$.
	}
	\label{summary}
\end{table}

The rest of this paper is organized as follows.
As the case of the canonical kinetic term is more standard and technically simpler, we first study the $k=1$ case in section \ref{sec2}, 
and then extend the discussion to the higher derivative cases with $k>1$ in section \ref{sec3}. 
In section \ref{sec2}, we study the $\phi^{2n}$ deformations of the canonical free theory.
For both the Neumann and Dirichlet cases, we compute the anomalous dimensions of the boundary fundamental primaries using the bulk multiplet recombination.
Then, we deduce the anomalous dimensions of the boundary composite operators 
from the analyticity constraints.
Using the analytic bootstrap, we further verify that the results for the anomalous dimensions of boundary fundamental operators are consistent with boundary crossing symmetry.
In section \ref{sec3}, we extend the analysis to generalized $\phi^{2n}$ theories, where the kinetic term is of higher order in derivatives.
We obtain the general expression for the anomalous dimensions of the boundary fundamental operators.
Then we focus on the NN case of the $(k,n)=(2,2)$ theory as a concrete  example, 
and derive the explicit anomalous dimensions of the boundary composite operators.
As a direct generalization, we also derive the composite anomalous dimensions in the DD case. 
For $k=2$, we also confirm the consistency of the boundary fundamental anomalous dimensions 
with boundary crossing symmetry.
In appendix \ref{Free BOE coefficients and OPE coefficients}, we derive the products of some BOE and OPE coefficients.
In appendix \ref{A bulk OPE coefficient}, we use the multiplet recombination to compute a bulk OPE coefficient, which is used in the analytic bootstrap analysis.

\section{\boldmath Canonical O($N$)-symmetric $\phi^{2n}$ theories with a boundary}
\label{sec2}

The canonical $\phi^{2n}$ theories with a boundary in $d=\du-\e$ dimensions are associated with the action
\be\label{action k=1}
S_{\pm}\propto\int_{\bb R^d_+} \rm d^d x\(\phi_a\Box\phi_a+g\m^{(n-1)\e}\phi^{2n}\)+S_{\text{bdy},\pm}
\,.
\ee
Here $\m$ has mass dimension 1, so the coupling constant $g$ is dimensionless.
The upper critical dimension is given by
\be\label{upper-d}
\du=\frac{2n}{n-1}
\,.
\ee
The bulk interaction is given by the O($N$) singlet
\be
\phi^{2n}=(\phi_a\phi_a)^n
\,.
\ee
The boundary term in the action \eqref{action k=1} can be written as
\be\label{boundary term k=1}
S_{\text{bdy},\pm}=\int_{\bb R^{d-1}}\rm d^{d-1}x\;(\pm)\Ph[a,0,1]\Ph[a,1,1]+\ldots
\ee
Here and below, we use the ellipsis to denote higher order terms in $\e$. 
The explicit definitions of the boundary primaries $\Ph[a,q,1]$ in terms of $\phi_a$ are given in \eqref{boundary primary k=1 introduction}.
As explained in the Introduction, the signs $\pm$ correspond to the Neumann and Dirichlet boundary conditions in the free theory,
\be
\text{Dirichlet}:\;\Ph[{a,\f},0,1]&=0\,,\\
\text{Neumann}:\;\Ph[{a,\f},1,1]&=0
\,,
\ee
where the subscript $\f$ indicates free theory. 
It is useful to introduce the $\s$ variable to label the boundary condition
\be
\s=\begin{cases}
	\; -1 \quad & \text{(Dirichlet)} \\
	\; +1  & \text{(Neumann)}\,. 
\end{cases}
\ee
In both the Dirichlet and Neumann cases, there is only one physical boundary fundamental operator, 
and the other one should vanish in accordance with the boundary condition.
We will use $\Ph[a,q,1]$ to denote the physical operator, 
where the possible values are $q=\frac{1-\s}{2}=0,1$. 

Below, we will frequently use Wick contractions to calculate the free correlators.
Let us provide the ingredients first.
The free bulk-bulk two-point function reads
\be\label{free bulk-bulk}
\<\phi_a(\xp,\xn)\phi_b(\yp,\yn)\>_\f=\(\;\frac{1}{|x-y|^{2\D_{\phi_\f}}}+\frac{\s}{{\(|\xp-\yp|^2+(\xn+\yn)^2\)^{\D_{\phi_\f}}}}\;\)\d_{ab}
\,,
\ee
where $\<\ldots\>_\f$ is the free correlator.
The bulk one-point function of the bilinear operator $(\phi_a\phi_b)$ can be obtained by taking the coincident point limit of \eqref{free bulk-bulk} and discarding the singular term,\footnote{The one-point function $\<\phi_a(x)\>_\f$ vanishes, indicating that the O$(N)$ symmetry is unbroken.}
\be\label{canonical-phiphi-1pt}
\<(\phi_a\phi_b)(\xp,\xn)\>_\f=\frac{\d_{ab}\,\s}{\(2\xn\)^{2\D_{\phi_\f}}}
\,.
\ee
Taking the normal derivative $\frac{\pa^q}{\pa\yn^q}$ and boundary limit $\yn\->0$ of \eqref{free bulk-bulk}, we obtain the bulk-boundary two-point function 
\be\label{bulk boundary k=1}
\<\phi_a(\xp,\xn)\Ph[b,q,1](\yp,0)\>_\f=\frac{(2+2(d-3)q)\d_{ab}}{\(|\xp-\yp|^2+\xn^2\)^{\Dh[q,1,\f]}\xn^{\smash{\D_{\phi_\f}-\Dh[q,1,\f]}}}
\,,
\ee
where $\Dh[q,1,\f]=\D_{\Ph[\f,q,1]}=\D_{\phi_\f}+q$ is the scaling dimension of $\Ph[{a,\f},q,1]$. 
The normal derivative $\frac{\pa^q}{\pa\xn^q}$ and boundary limit $\xn\->0$ of \eqref{bulk boundary k=1} give the boundary-boundary two-point function
\be
\<\Ph[a,q,1](\xp,0)\Ph[b,q,1](\yp,0)\>_\f=\frac{(2+2(d-3)q)\d_{ab}}{|\xp-\yp|^{2\Dh[q,1,\f]}}
\,.
\ee
The free correlators above can be generalized to the higher derivative theories, 
which is slightly more complicated and will be discussed in section \ref{Generalized free theory with a boundary}.

\subsection{Multiplet recombination}
\label{Multiplet recombination}

In this section, we compute the anomalous dimensions of boundary scalars associated with the $\phi^{2n}$ deformations using the bulk multiplet recombination and analyticity constraints.
We are interested in the boundary fundamental primary $\Ph[a,q,1]$ and the boundary composite operators
\be\label{composite definition}
\Ph[,q,2p]=\(\Ph[b,q,1]\Ph[b,q,1]\)^{p}\,,\quad\Ph[a,q,2p+1]=\Ph[a,q,1]\(\Ph[b,q,1]\Ph[b,q,1]\)^{p}
\,, 
\ee
where the second superscript denotes the number of fundamental operators of the $q$-derivative type. 
To be more precise, the operator definitions in \eqref{composite definition} are for the composite operators in free theory, 
and the interacting cases are associated with the deformed counterparts of the free-theory operators. 
The definitions of other operators also follow this convention. 

Let us briefly summarize the constraint from the bulk multiplet recombination.
The bulk operators are subject to the condition
\be\label{recombination k=1}
\lim_{\e\->0}\a^{-1}\Box\phi_a=\phi^{2n-1}_{a,\f}
\,,
\ee
where the right-hand side is defined as $\phi^{2p+1}_{a,\f}\equiv\phi_{a,\f}(\phi_{b,\f}\phi_{b,\f})^p$. 
The singular change in the normalization is given by \cite{Guo:2023qtt}
\be\label{alpha k=1}
\a=\frac{2^{2-n}\, n\,\e}{
(n)_{n}\,{}_3F_{2}\!\[\begin{matrix}\frac{1-n}{2},-\frac{n}{2},1-n-\frac{N}{2}\\ 1\;,\;\frac{1}{2}-n\end{matrix};1\]
}+O(\e^2)
\,,
\ee
where ${}_3F_{2}$ denotes the generalized hypergeometric function.
As mentioned in the Introduction, we assume that the presence of a boundary does not affect the bulk multiplet recombination. 
Roughly speaking, 
the condition \eqref{recombination k=1} specifies the bulk deformations, 
and then determine the boundary deformations through boundary conformal symmetry and analyticity constraints.  
Below, we first study the matching conditions for the bulk-boundary two-point functions $\<\Box\phi_a\,\Ph[b,q,1]\>$ and determine the leading anomalous dimension of $\Ph[a,q,1]$.
Then, we consider the bulk-boundary-boundary three-point functions  $\<\phi_a\,\Ph[,q,2p]\Ph[b,q,2p+1]\>$ and $\<\phi_a\,\Ph[b,q,2p-1]\Ph[,q,2p]\>$.
The leading anomalous dimensions of $\Ph[,q,2p]$ and $\Ph[a,q,2p+1]$ are fixed by 
analyticity of these correlators.

\subsubsection{Boundary fundamental operators}
\label{Boundary fundamental operators}

The matching condition for the bulk-boundary two-point function reads
\be\label{matching bulk boundary k=1}
\lim_{\e\->0}\(\a^{-1}\<\Box\phi_a(\xp,\xn)\Ph[b,q,1](\yp,0)\>\)=\<\phi^{2n-1}_a(\xp,\xn)\Ph[b,q,1](\yp,0)\>_\f
\,.
\ee
On the left-hand side, the two-point function $\<\phi_a(\xp,\xn)\Ph[b,q,1](\yp,0)\>$ is constrained by boundary conformal symmetry up to the BOE coefficient $b_{\phi\Ph[,q,1]}$:
\be\label{bulk boundary 2pt}
\<\phi_a(\xp,\xn)\Ph[b,q,1](\yp,0)\>=\frac{\d_{ab}\,b_{\phi\Ph[,q,1]}}{\(|\xp-\yp|^2+\xn^2\)^{\Dh[q,1]} \xn^{\D_{\phi}-\Dh[q,1]}}
\,.
\ee
The tensor structure $\d_{ab}$ reflects the unbroken O($N$) symmetry.
The action of the Laplacian reads
\be\label{matching condition explicit}
\lim_{\e\->0}\Bigg[&\(\frac{
(\D_\phi-\Dh[q,1])(\D_\phi-\Dh[q,1]+1)
}{\xn^2}+\frac{
4\Dh[q,1](\D_\phi-\frac{d-2}{2})
}{|\xp-\yp|^2+\xn^2}\)
\frac{\a^{-1}\,\d_{ab}\,b_{\phi\Ph[,q,1]}}{
\(|\xp-\yp|^2+\xn^2\)^{\Dh[q,1]}\xn^{\D_\phi-\Dh[q,1]}
}\Bigg]
\nn
&
=\frac{\d_{ab}\,b_{\phi_\f^{2n-1}\Ph[\f,q,1]}}{
\(|\xp-\yp|^2+\xn^2\)^{\Dh[q,1,\f]} \xn^{(2n-1)\D_{\smash{\phi_\f}}-\Dh[q,1,\f]}
}\,,
\ee
where $\Dh[q,1]$ is the scaling dimension of $\Ph[a,q,1]$.
The scaling dimension of the bulk fundamental field $\phi_a$ does not have an order $\e^1$  correction,
\be\label{gamma phi k=1}
\g_\phi=\D_\phi-\frac{d-2}{2}=O(\e^2)
\,,
\ee
so we omit the second term on the left-hand side of \eqref{matching condition explicit}.
The functional forms in \eqref{matching condition explicit} match 
because $\lim_{\e\->0}\D_{\phi}=(2n-1)\D_{\phi_\f}-2$. 
The free BOE coefficient is computed using Wick contractions
\be\label{free coefficient b k=1}
b_{\phi_\f^{2n-1}\Ph[\f,q,1]}=
2^{n-3}\,b_{\phi_\f\Ph[\f,q,1]}\,\(\frac{N}{2}+1\)_{n-1}\s^{n-1}
\,,
\ee
where $(x)_y=\G(x+y)/\G(y)$ denotes the Pochhammer symbol.
We have left $b_{\smash{\phi_\f\Ph[\f,q,1]}}$ implicit because the same coefficient also appears on the left-hand side of \eqref{matching condition explicit}.
The anomalous dimension of the boundary fundamental operator is defined by
\footnote{In the early literature \cite{lubensky1975critical,bray1977critical,reeve1980critical,diehl1981field,reeve1981renormalisation,reeve1981renormalisation2,diehl1987walks,eisenriegler1988surface}, the related results were often expressed in terms of the surface correlation exponent
\be
\eta_\parallel=2\Dh[q,1]-d+2=2q+2\gh[q,1]
\,.
\ee}
\be
\gh[q,1]=\Dh[q,1]-\(\frac{d-2}{2}+q\)
\,.
\ee
The condition \eqref{matching condition explicit} becomes
\be\label{epsilon expansion 2pt k=1}
(-1)^{-q-1}\lim_{\e\->0}\(\a^{-1}\gh[q,1]\)=\s^{n-1}2^{n-3}\(\frac{N}{2}+1\)_{n-1}
\,.
\ee
We obtain the leading anomalous dimension of the boundary fundamental operator $\Ph[a,q,1]$
\be\label{gamma Phi}
\gh[q,1]=(-1)^{q+1}2^{n-3}\(\frac{N}{2}+1\)_{n-1}\s^{n-1}\a+\begin{cases}
	\; O(\e^2) \quad & \text{Dirichlet or even $n$} \\
	\; O(\e^{3/2})  & \text{Neumann with odd $n$}\,,
\end{cases}
\ee
where $\a$ is given in \eqref{alpha k=1}.
There are two scenarios for the higher order terms, corresponding to two different types of $\e$ expansions.
See the second half of section \ref{Boundary composite operators} for an explanation of the $\e^{1/2}$ expansion.
The leading anomalous dimensions of $\Ph[a,q,1]$ in the special and ordinary transitions are equal or differ by an overall sign.
This relation generally does not hold at subleading order in $\e$, 
even for the scenarios associated with only integer powers of $\e$.
Our calculation above is the $\phi^{2n}$-multicritical generalization of the $\phi^4$ result in \cite{Nishioka:2022odm}.
In the general $n$ case, the result \eqref{gamma Phi} involves the hypergeometric function of the ${}_3F_2$ type due to the bulk multiplet recombination coefficient \eqref{alpha k=1}. 
The general $n$ result \eqref{gamma Phi} agrees with \cite{lubensky1975critical,bray1977critical,reeve1980critical,diehl1981field,reeve1981renormalisation,reeve1981renormalisation2,diehl1987walks,eisenriegler1988surface,Nishioka:2022odm} for the $\phi^4$ and $\phi^6$ theories. 
In addition, the $N\->0$ limit yields
\be\label{polymer}
\gh[q,1]\big|_{N=0}=(-1)^{q+1}\frac{n!\,\s^{n-1}}{2\, (n)_n \,
{}_3F_{2}\!\[\begin{matrix}\frac{1-n}{2},-\frac{n}{2},1-n\\ 1\;,\;\frac{1}{2}-n\end{matrix};1\]}\,\e+\begin{cases}
	\; O(\e^2) \quad & \text{Dirichlet or even $n$} \\
	\; O(\e^{3/2})  & \text{Neumann with odd $n$}\,,
\end{cases}
\ee 
which is related to polymer systems at $n=2,3$ mentioned in the Introduction. 
For the ordinary transition at $n=2$, the exponent associated with the layer susceptibility is\footnote{We have used the scaling laws (2.80) and (2.83) in \cite{Diehl1986}, where the bulk exponents are $\eta=2\g_\phi=O(\e^2)$ and $\n^{-1}=d-\D_{\phi^2}=2-\frac 1 4 \e+O(\e^2)$ for $n=2$ and $N=0$.}
\be\label{critical exponent N=0}
\g_1^{\text{ord}}\big|_{N=0}=\frac 1 2+\frac 1 8 \e+O(\e^2) \qquad (q=1,\,n=2)
\,,
\ee
which is in agreement with equation (3.161) in \cite{Diehl1986}.\footnote{There are also other special limits such as $N\rightarrow -2,-4,\ldots,-2n+2$. 
For instance, the limit $N\rightarrow -2$ is related to the loop-erased random walks. 
See appendix H of \cite{Guo:2024bll} for more details.}
To avoid a notational confusion, 
we emphasize that \eqref{critical exponent N=0} is a critical exponent, rather than an anomalous dimension. 

Let us further examine the boundary conditions of the interacting theories. 
For the Dirichlet case ($q=1$), the bulk-boundary two-point function reads
\be
\<\phi_a(\xp,\xn)\cal O(\yp,0)\>=\frac{b_{\phi\Ph[,1,1]}}{\(|\xp-\yp|^2+\xn^2\)^{\Dh[1,1]} \xn^{\D_\phi-\hat{\D}_{\cal O}}}
\,.
\ee
We assume that the interacting deformation of $\Ph[a,0,1]$ remains null to order $\e^1$.
Then, the lowest boundary scaling dimension is $\hat{\D}_{\cal O}=\Dh[1,1]=\D_\phi+1+O(\e)$, so the two-point function above has a vanishing boundary limit $\xn\->0$ for all boundary primaries.
Here we have assumed $\xp\neq\yp$. 
It appears that the Dirichlet boundary condition persists to order $\e^1$ in the interacting theories:
\be\label{D k=1}
\lim_{\xn\->0}\phi(x)=O(\e^2)
\,,
\ee
which is already known in the $\phi^4$ theory with a boundary \cite{Diehl:1996kd}. 
For the Neumann case ($q=0$) with even $n$, the bulk-boundary two-point function is
\be\label{N k=1}
\<\phi_a(\xp,\xn)\Ph[b,0,1](\yp,0)\>=\frac{b_{\phi\Ph[,0,1]}}{\(|\xp-\yp|^2+\xn^2\)^{\Dh[0,1]} \xn^{\g_\phi-\gh[0,1]}}
\,.
\ee
Due to the factor $1/\xn^{\g_\phi-\gh[0,1]}$, the boundary limit $\xn\->0$ is logarithmically divergent at order $\e^1$. 
\footnote{We take the $\xn\rightarrow 0$ limit after the $\e$ expansion.}
Moreover, the normal derivative $\<\pa_\perp\phi_a(\xp,\xn)\Ph[b,0,1](\yp,0)\>$ also diverges in the boundary limit.
This means that the Neumann boundary condition $\lim_{\xn\->0}\pa_\perp\phi_{a,\f}=0$ is no longer satisfied in the interacting theories.
The Neumann case with odd $n$ involves the $\e^{1/2}$ expansion.
We defer the discussion of the corresponding boundary condition to the end of section \ref{Boundary composite operators}.

\subsubsection{Boundary composite operators}
\label{Boundary composite operators}

There are two types of perturbative expansions for the BCFT data in $\e$.
In the Neumann case ($q=0$) with even $n$ or the Dirichlet case ($q=1$), the expansion in integer powers of $\e$ is consistent with the matching condition associated with $\<\phi_a\,\Ph[b,0,n]\>$.
In the Neumann case with odd $n$, we have to introduce noninteger power terms in $\e$, 
otherwise the integer power $\e$ expansion is inconsistent with the aforementioned matching condition. 
Together with the analyticity of the three-point functions, 
we will show that the leading anomalous dimensions of the boundary composite operators are of order $\e^{1/2}$.

\subsubsection*{Integer power $\e$ expansion}

Consider the following bulk-boundary-boundary three-point functions:
\be\label{3pt functions}
\<\phi_a(\xp,\xn)\Ph[,q,2p](0,0)\Ph[b,q,2p+1](\yp,0)\>\,,\quad
\<\phi_a(\xp,\xn)\Ph[b,q,2p-1](0,0)\Ph[,q,2p](\yp,0)\>
\,.
\ee
We will see below that the conformal block expansion may imply nonanalytic terms when the bulk operator $\phi_a$ and the boundary operator $\Ph[,q,2p]$ or $\Ph[a,q,2p-1]$ are on a line perpendicular to the boundary, i.e., in the limit $\xp\->0$.
In this configuration, all the operators are separated, so the nonanalytic behavior contradicts the analyticity of Euclidean correlators for noncoincident points.
We will impose that the coefficients of the nonanalytic terms vanish, 
which determines the anomalous dimensions of boundary composite operators.

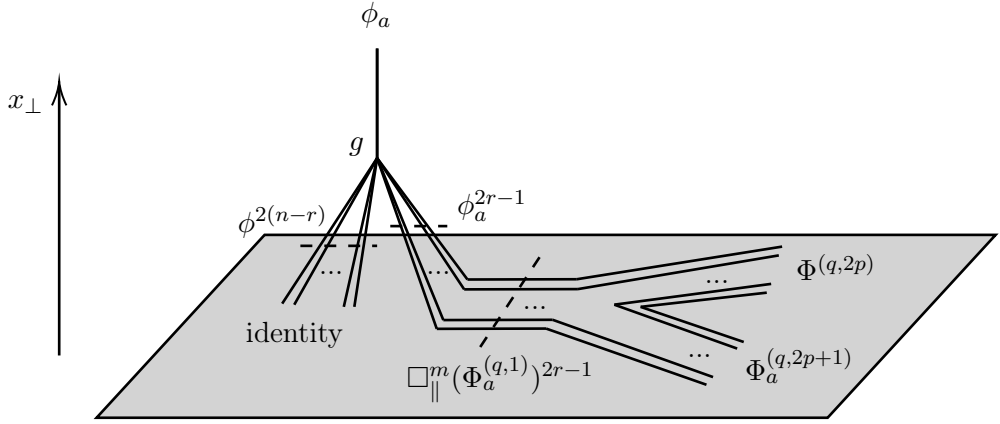
\begin{figure}
	\center
	
\tikzset{every picture/.style={line width=1pt}}       

\begin{tikzpicture}[x=0.75pt,y=0.75pt,yscale=-1.1,xscale=1.1]
	
	\draw  [fill={rgb, 255:red, 211; green, 211; blue, 211 }  ,fill opacity=1 ] (188.68,125) -- (522.08,125) -- (445.45,208.4) -- (112.05,208.4) -- cycle ;
	\draw    (240,40) -- (240,90) ;
	
\draw    (240,40) -- (240,90) ;
\draw    (331.07,145.23) -- (424.7,130.35) ;
\draw    (331.57,149.73) -- (423.2,134.35) ;
\draw    (317.07,167.73) -- (390.2,193.35) ;
\draw    (320.07,163.73) -- (393.2,189.85) ;
\draw    (360.17,157.83) -- (407.2,173.85) ;
\draw    (360.17,157.83) -- (417.7,151.35) ;
\draw    (348.17,156.83) -- (404.2,176.85) ;
\draw    (281.2,145.35) -- (331.07,145.23) ;
\draw    (279.7,149.85) -- (331.57,149.73) ;
\draw    (267.2,167.85) -- (317.07,167.73) ;
\draw    (270.2,163.85) -- (320.07,163.73) ;
\draw    (240,90) -- (281.2,145.35) ;
\draw    (240,90) -- (279.7,149.85) ;
\draw    (240,90) -- (270.2,163.85) ;
\draw    (240,90) -- (267.2,167.85) ;
\draw    (240,90) -- (196.7,156.35) ;
\draw    (240,90) -- (224.8,157.82) ;
\draw    (240,90) -- (229.6,157.82) ;
\draw    (240,90) -- (202.2,156.85) ;
\draw    (348.17,156.83) -- (419.7,147.35) ;

\draw  [dash pattern={on 4.5pt off 4.5pt}]  (314.08,135.18) -- (285.28,178.78) ;

\draw    (95,180) -- (95,117.85) -- (95,57) ;
\draw [shift={(95,55)}, rotate = 90] [color={rgb, 255:red, 0; green, 0; blue, 0 }  ][line width=0.75]    (10.93,-3.29) .. controls (6.95,-1.4) and (3.31,-0.3) .. (0,0) .. controls (3.31,0.3) and (6.95,1.4) .. (10.93,3.29)   ;

\draw  [dash pattern={on 4.5pt off 4.5pt}]  (205,130) -- (240,130) ;
\draw  [dash pattern={on 4.5pt off 4.5pt}]  (246,121) -- (272,121) ;

	\draw (231,17.4) node [anchor=north west][inner sep=0.75pt]    {$\phi_a$};
	\draw (429.67,132) node [anchor=north west][inner sep=0.75pt]    {$\Ph[,q,2p]$};
	\draw (406.3,175.22) node [anchor=north west][inner sep=0.75pt]    {$\Ph[a,q,2p+1] $};
	\draw (212,140) node [anchor=north west][inner sep=0.75pt]    {...};
	\draw (388,144) node [anchor=north west][inner sep=0.75pt]    {...};
	\draw (275,103) node [anchor=north west][inner sep=0.75pt]    {$\phi_a^{2r-1}$};
	\draw (175,110) node [anchor=north west][inner sep=0.75pt]    {$\phi^{2( n-r)}$};
	\draw (70,60) node [anchor=north west][inner sep=0.75pt]    {$\xn$};
	\draw (380,178) node [anchor=north west][inner sep=0.75pt]    {...};
	\draw (261,140) node [anchor=north west][inner sep=0.75pt]    {...};
	\draw (178.67,163.07) node [anchor=north west][inner sep=0.75pt]    {identity};
	\draw (252,178.4) node [anchor=north west][inner sep=0.75pt]    {$\Box_\parallel^m(\Ph[a,q,1])^{2r-1}$};
	\draw (305,156) node [anchor=north west][inner sep=0.75pt]   {...};
	
	\draw (225.73,79) node [anchor=north west][inner sep=0.75pt]    {$g$};
	
\end{tikzpicture}

	\caption{The intermediate states associated with $\Psi^{(q,2r-1,m)}_{a}\sim \Box_\parallel^m(\Ph[a,q,1])^{2r-1}$ in the bulk-boundary-boundary correlator $\<\phi_a\Ph[,q,2p]\Ph[b,q,2p+1]\>$. 
	The number of lines corresponds to the number of fundamental operators. 
	The ellipses indicate potentially more lines that are not drawn explicitly.
	At order $g\sim\e^1$, the BOE of $\phi_a$ contains 
	the operators constructed from $2r-1$ boundary fundamental primaries with $r\leq n$ due to the bulk $g\phi^{2n}$ interaction. 
	They can also involve contracted parallel derivatives, which are denoted schematically by $\Box^m_\parallel$. 
	}
	\label{Feynman 3pt}
\end{figure}

Using the BOE of $\phi_a$ and the boundary OPE $\Ph[,q,2p]\times\Ph[b,q,2p+1]$, the three-point function $\<\phi_a\Ph[,q,2p]\Ph[b,q,2p+1]\>$ becomes a sum of boundary conformal blocks.
Let us find the primaries that contribute at order $\e^1$.
The corresponding products of BOE and boundary OPE coefficients should be of order $\e^1$ or $\e^0$. 
To order $\e^1$, the BOE of $\phi_a$ involves internal operators constructed from $2r-1$ $\phi_a$'s, where the number of $\phi_a$'s is odd due to $\bb Z_2$ symmetry
and the positive integer $r$ satisfies $r\leq n$ due to the $\phi^{2n}$ interaction (see figure \ref{Feynman 3pt}). 
They can also involve contracted derivatives.
The parallel derivatives should be contracted due to boundary Lorentz symmetry, and the orders of the normal derivatives are determined by the boundary OPE.
For $r=1$, the possible contributions are associated with the boundary fundamental primaries $\Ph[a,q,1]$ and $\Ph[a,1-q,1]$. 
Since their parallel derivatives are descendants, 
the derivative contributions are already encoded in the boundary conformal blocks.
In accordance with the free boundary condition, 
the product of the BOE and boundary OPE coefficients associated with $\Ph[a,1-q,1]$ 
is at least of order $\e^2$,
so we only need to consider the physical operator $\Ph[a,q,1]$.
For $2\leq r\leq n$, the BOE coefficient is of order $\e^1$, 
so the boundary OPE coefficients should be of order $\e^0$. 
The corresponding internal operators take the schematic form $\Psi^{(q,2r-1,m)}_{a}\sim \Box_\parallel^m(\Ph[a,q,1])^{2r-1}$, 
where $\Box_\parallel$ indicates that two derivatives associated with the parallel directions have contracted indices
and they can act on the individual fundamental operators in various ways, i.e., not in a total derivative form.
The scaling dimensions of $\Psi^{(q,2r-1,m)}_{a}$ are
\be\label{Delta Psi}
\hat{\D}_{q,2r-1,m}=(2r-1)\(\frac{d-2}{2}+q\)+2m+O(\e)
\,.
\ee
For each $2\leq r\leq n$, there are infinitely many boundary primaries $\Psi^{(q,2r-1,m)}_{a}$ as the number of contracted derivatives can be arbitrarily large.
Note that we do not need to unmix the highly degenerate operators for $m>0$. 

For simplicity, we will focus on the leading term in the $\yp\->\infty$ expansion.
For the first three-point function in \eqref{3pt functions}, the conformal block expansion reads
\be\label{3pt conformal block expansion}
&\<\phi_a(\xp,\xn)\Ph[,q,2p](0,0)\Ph[b,q,2p+1](\yp,0)\>=\frac{\d_{ab}|\yp|^{-2\Dh[q,2p+1]}}{|x|^{\dD}\xn^{\D_\phi}}\Bigg[b_{\phi\tPh[,q,1]}f^{(p)}_{\tPh[,q,1]}G^{(p)}_{\tPh[,q,1]}(v)\nn
&\hspace{4em}+\sum_{r=2}^{n}\sum_{m=0}^{\infty}b_{\phi\tilde\Psi^{(q,2r-1,m)}}f^{(p)}_{\tilde\Psi^{(q,2r-1,m)}}G^{(p)}_{\tilde\Psi^{(q,2r-1,m)}}(v)+O(\e^2)\Bigg]+O(|\yp|^{-2\Dh[q,2p+1]+1})
\,,
\ee
where $f^{(p)}_{\tilde{\cal O}}$ is the OPE coefficient of $\tilde{\cal O}_a\in\Ph[,q,2p]\times\Ph[a,q,2p+1]$.
The cross ratio $v$ and scaling dimension difference $\dD$ are defined as
\be\label{v definition}
v\equiv\frac{\xn^2}{|x|^2}\,, \qquad \dD\equiv \Dh[2]-\Dh[3]
\,.
\ee
Here $\Dh[2]$ and $\Dh[3]$ are the scaling dimensions of the second and third external operators in the three-point function.
Since the three-point function under consideration takes the form $\<\phi_a\Ph[,q,2p]\Ph[b,q,2p+1]\>$, the difference $\dD$ is labeled by the parameter $p$.
We will use these notations again in \eqref{3pt conformal block expansion 2}, which concerns a different three-point function $\<\phi_a\Ph[b,q,2p-1]\Ph[,q,2p]\>$.
These notations will also be used later in section \ref{sec3}.
In \eqref{3pt conformal block expansion}, the internal operators in the boundary channel are tilded, indicating that they have unit-normalized two-point functions:
\be\label{boundary internal operator normalization}
\<\tilde{\cal O}(\xp,0)\tilde{\cal O}(\yp,0)\>=\frac{1}{|\xp-\yp|^{2\D_{\cal O}}}
\,.
\ee
The boundary conformal blocks are given by \cite{Nishioka:2022qmj}
\be\label{conformal block}
G^{(p)}_{\tilde{\cal O}}(v)=v^{\frac 1 2 \hat{\D}_\cal O}{}_2F_1\(\frac{\hat{\D}_\cal O+\dD}{2},\frac{\hat{\D}_\cal O-\dD}{2};\hat{\D}_\cal O-\frac{d-3}{2};v\)
\,,
\ee
where ${}_2F_1$ indicates the Gaussian hypergeometric function. 
In accordance with the left-hand side of  \eqref{3pt conformal block expansion}, 
the superscript $(p)$ of the boundary conformal block tracks the number of boundary fundamental primaries in the external operators. 
For $n>2$, the conformal block expansion \eqref{3pt conformal block expansion} involves a sum over $r$, which leads to $(n-1)$ infinite sums.
This generalizes the conformal block expansion with one infinite sum in \cite{Nishioka:2022odm}, which considered the $\phi^4$ theory, i.e., $n=2$. 

Using the boundary OPE $\Ph[,q,2p]\times\Ph[b,q,2p+1]$, the bulk-boundary-boundary three-point function $\<\phi_a\Ph[,q,2p]\Ph[b,q,2p+1]\>$ can be written as a sum of bulk-boundary two-point functions. 
Later, we will discuss the constraints on these two-point functions from the matching condition in \eqref{composite matching}.
The BOE of $\phi_a$ implies that each bulk-boundary two-point function is associated with a boundary-boundary two-point function and its parallel derivatives, 
which are encoded in the conformal block \eqref{conformal block}. 
We want to extract the nonanalytic part of \eqref{3pt conformal block expansion} at order $\e^1$. 
The order $\e^1$ terms can arise from the corrections to the scaling dimensions and to the coefficients.
The correction to the scaling dimension of $\Ph[a,q,1]$ gives a ${}_3F_2$ function.
The correction to the coefficient $b_{\phi\tPh[,q,1]}f^{(p)}_{\tPh[,q,1]}$ does not lead to any nonanalytic terms.
The products $b_{\phi\tilde\Psi^{(q,2r-1,m)}}f^{(p)}_{\tilde\Psi^{(q,2r-1,m)}}$ are of order $\e^1$, and the corresponding infinite sum yields another ${}_3F_2$ function.
The two aforementioned ${}_3F_2$ functions are nonanalytic in $1-v$. 
To derive the constraints on the anomalous dimensions of the boundary composite operators, 
we perform the $1-v$ expansion and require that 
the leading term with a noninteger power in $1-v$ vanishes.
More details are provided below. 

Let us start with the first term in the square bracket in the conformal block expansion \eqref{3pt conformal block expansion}.
The anomalous dimensions of composite operators $\Ph[,q,2p]$ and $\Ph[a,q,2p+1]$ are defined by
\be
\gh[q,2p+1]=\Dh[q,2p+1]-(2p+1)\(\frac{d-2}{2}+q\)\,,\quad
\gh[q,2p]=\Dh[q,2p]-2p\(\frac{d-2}{2}+q\)
\,,
\ee
where $\Dh[q,2p+1]$ and $\Dh[q,2p]$ are the scaling dimensions of $\Ph[,q,2p]$ and $\Ph[a,q,2p+1]$. 
The conformal block can be written as
\be\label{3pt conformal block expansion first term}
G^{(p)}_{\tPh[,q,1]}(v)
=v^{\Dh[q,1]/2}\(
1+\frac{(nq-q+1)(\gh[q,1]+\gh[q,2p]-\gh[q,2p+1])v}{(n-1)(2q+1)}{}_3F_{2}\!\[\begin{matrix}1,1,\frac{n}{n-1}+q\vspace{0.2em}\\ 2\;,\;\frac{3}{2}+q\end{matrix};v\]+O(\e^2)
\)\,,
\ee
where the hypergeometric function ${}_3F_2$ arises from the perturbative expansion of ${}_2F_1$.
The first term $v^{\Dh[q,1]/2}$ is analytic at $v=1$, but the second term is nonanalytic, which can be seen from the noninteger powers in the expansion
\be\label{3F2 expansion}
{}_3F_{2}\!\[\begin{matrix}a_1,a_2,a_3\\ b_1,b_2\end{matrix};v\]&=\frac{\G(b_1)\G(b_2)}{\G(a_1)\G(a_2)\G(a_3)}(1-v)^{\b}\sum_{i=0}^{\infty}\cal F_i(\b)(1-v)^i+\text{analytic terms}\,,
\ee
where $\b$ and $\cal F_i(\b)$  are defined as
\be
\b&=b_1+b_2-a_1-a_2-a_3\qquad (\b\notin\bb Z)\,,
\ee
\be
\cal F_i(\b)&=\frac{(-1)^i\,\G(-i-\b)\G(a_1+i+\b)\G(a_2+i+\b)}{i!\,\G(a_1+\b)\G(a_2+\b)}{}_3F_{2}\!\[\begin{matrix}b_1-a_3,b_2-a_3,-i\\ a_1+\b,a_2+\b\end{matrix};1\]\,.
\ee
For the ${}_3F_2$ in \eqref{3pt conformal block expansion first term}, we have $\b=\frac{1}{2}-\frac{1}{n-1}$, indicating that $(1-v)^\b$ is nonanalytic for $n=2$ or $n>3$.
The case of $n=3$ is associated with the degenerate limit $\b\rightarrow 0$, 
so the nonanalytic terms are of the logarithmic type.
We will return to this special case after the general discussion.
Let us examine the leading term in the $(1-v)$ expansion:
\be\label{leading non analytic from fundamental}
&b_{\phi\tPh[,q,1]}f^{(p)}_{\tPh[,q,1]}G^{(p)}_{\tPh[,q,1]}(v)\nn
=\;&2^{4p+1}\,p!\(\frac{N}{2}+1\)_p\(1-\frac{(n-3) q}{n-1}\)^{2p+1}\,\frac{(nq-q+1)(\gh[q,1]+\gh[q,2p]-\gh[q,2p+1])v^{\frac{1}{2}(\frac{1}{n-1}+q)+1}}{(n-1)(2q+1)}\nn
&\times\frac{\G(\frac{3}{2}+q)\G(\frac{1}{n-1}-\frac{1}{2})}{\G(\frac{n}{n-1}+q)}(1-v)^{\frac{1}{2}-\frac{1}{n-1}}
+O((1-v)^{\frac{3}{2}-\frac{1}{n-1}})+\text{analytic terms}+O(\e^2)
\,,
\ee
where the order $\e^0$ expression of $b_{\phi\tPh[,q,1]}f^{(p)}_{\tPh[,q,1]}$ is computed using Wick contractions.

The second term in the square bracket in \eqref{3pt conformal block expansion} involve the coefficients $b_{\phi\tilde\Psi^{(q,2r-1,m)}}f^{(p)}_{\tilde\Psi^{(q,2r-1,m)}}$. 
They can be derived from the matching condition
\be\label{composite matching}
\lim_{\e\->0}\(\a^{-1}\<\Box\phi_a(\xp,\xn)\tilde\Psi^{(q,2r-1,m)}_{b}(\yp,0)\>\)=\<\phi^{2n-1}_a(\xp,\xn)\tilde\Psi^{(q,2r-1,m)}_{b}(\yp,0)\>_\f
\,.
\ee
The explicit expression of the boundary operator $\Psi^{(q,2r-1,m)}_b$ is not needed. 
We obtain
\be\label{BOE coefficients k=1}
b_{\phi\tilde\Psi^{(q,2r-1,m)}}=\frac{\a \,b_{\phi^{2n-1}_\f \tilde\Psi^{(q,2r-1,m)}_\f}}{\((1-2r)q-2m-\frac{2(r-1)}{n-1}\)_{2}}+O(\e^2)
\,.
\ee
We do not need to unmix the degenerate operators because this relation holds for each primary. To be more precise, we should interpret $b_{\phi\tilde\Psi^{(q,2r-1,m)}}$ 
as the sum of all contributions associated with $(q,2r-1,m)$. 
Using the expression for $b_{\phi^{2n-1}\tilde\Psi^{(q,2r-1,m)}}f^{(p)}_{\tilde\Psi^{(q,2r-1,m)}}$  in \eqref{free relation}, we obtain
\be\label{bf k=1}
b_{\phi\tilde\Psi^{(q,2r-1,m)}}f^{(p)}_{\tilde\Psi^{(q,2r-1,m)}}\!=\!\frac{(-1)^m \((r-1)(\tfrac{1}{n-1}+q)\)_m \(r(\tfrac{1}{n-1}+q)\)_m \a\,W_{2r-1,p}}{\((1-2r)q \!-\! 2m \!-\! \frac{2(r-1)}{n-1}\)_{2}\(m \!+\! (2 r-1)q \!-\! \frac{n-2 r+1}{n-1} \!+\! \frac{1}{2}\)_m m!} \!+\! O(\e^2)
\,,
\ee
where $W_{2r-1,p}$ is given in \eqref{W}.
The product above is of order $\e^1$, 
so $\Psi^{(q,2r-1,m)}_b$ contributes to the conformal block expansion under consideration.
We carry out the infinite sum over $m$ in \eqref{3pt conformal block expansion}, 
and then use \eqref{3F2 expansion} to extract the leading nonanalytic term,
\be\label{infinite sum}
&\sum_{m=0}^{\infty}b_{\phi\tilde\Psi^{(q,2r-1,m)}}f^{(p)}_{\tilde\Psi^{(q,2r-1,m)}}G^{(p)}_{\tilde\Psi^{(q,2r-1,m)}}(v)\nn
=\;&\frac{\a\,W_{2r-1,p}\,v^{\frac{2r-1}{2}(\frac{1}{n-1}+q)}}{\((1-2r)q-\frac{2(r-1)}{n-1}\)_{2}}\,{}_3F_{2}\!\[\begin{matrix}1,(r-1)(\frac{1}{n-1}+q),r(\frac{1}{n-1}+q)\vspace{0.3em}\\ \frac{r-1}{n-1}+\frac{(2r-1)q+1}{2},\frac{r-1}{n-1}+\frac{(2r-1)q+2}{2}\end{matrix};v\]+O(\e^2)
\nn
=\;&\frac{\a\,W_{2r-1,p}}{\((1-2r)q-\frac{2(r-1)}{n-1}\)_{2}}\times \frac{\G(\frac{r-1}{n-1}+\frac{(2r-1)q+1}{2})\G(\frac{r-1}{n-1}+\frac{(2r-1)q+2}{2})\G(\frac{1}{n-1}-\frac{1}{2})}{\G((r-1)(\frac{1}{n-1}+q))\G(r(\frac{1}{n-1}+q))}(1-v)^{\frac{1}{2}-\frac{1}{n-1}}\nn&+O((1-v)^{\frac{3}{2}-\frac{1}{n-1}})+\text{analytic terms}+O(\e^2)
\,.
\ee
To derive the ${}_3F_2$ hypergeometric function in \eqref{infinite sum}, we make use of the formula
\be
&\sum_{m=0}^{\infty}\frac{(-1)^m \left(a_1\right)_m \left(a_2\right)_m \, v^m}{m! \left(1-2b-2 m\right)_2
	 \left(m+2b-\frac 3 2\right)_m}\, {}_2F_1\left(m+a_1,m+a_2;2 m+2b-1/2;v\right) \nn
 =\;&\frac{1}{ \left(1-2b\right)_2 }
\,{}_3F_{2}\!\[\begin{matrix}1,a_1,a_2\\ b,b+\frac 1 2\end{matrix};v\]
\,. 
\ee

Finally, we require that the three-point function is analytic at $v=1$, so the nonanalytic terms \eqref{leading non analytic from fundamental} and \eqref{infinite sum} should cancel out, yielding a constraint on the anomalous dimensions,
\be\label{recursion 1}
&2^{4p+1}\,p!\(\frac{N}{2}+1\)_p\(1-\frac{(n-3) q}{n-1}\)^{2p+1}\frac{(nq-q+1)(\gh[q,1]+\gh[q,2p]-\gh[q,2p+1])}{(n-1)(2q+1)}\frac{\G(\frac{3}{2}+q)}{\G(\frac{n}{n-1}+q)}\nn
&+\a\sum_{r=2}^{n}\Bigg[\;\frac{\G(\frac{r-1}{n-1}+\frac{(2r-1)q+1}{2})\G(\frac{r-1}{n-1}+\frac{(2r-1)q+2}{2})W_{2r-1,p}}{\G((r-1)(\frac{1}{n-1}+q))\G(r(\frac{1}{n-1}+q))\((1-2r)q-\frac{2(r-1)}{n-1}\)_{2}}\;\Bigg]=O(\e^2)
\,.
\ee
The constraint above gives a relation between the anomalous dimensions of the O($N$) singlet and vector operators.
To solve for the anomalous dimensions, we also need to study the conformal block expansion of the second three-point function in \eqref{3pt functions}:
\be\label{3pt conformal block expansion 2}
&\<\phi_a(\xp,\xn)\Ph[b,q,2p-1](0,0)\Ph[,q,2p](\yp,0)\>
=\frac{\d_{ab}|\yp|^{-2\Dh[q,2p]}}{|x|^{\dD}\xn^{\D_\phi}}\Bigg[b_{\phi\tPh[,q,1]}f^{\prime(p)}_{\tPh[,q,1]}G^{\prime(p)}_{\tPh[,q,1]}(v)\nn
&\hspace{5em}+\sum_{r=2}^{n}\sum_{m=0}^{\infty}b_{\phi\tilde\Psi^{(q,2r-1,m)}}f^{\prime(p)}_{\tilde\Psi^{(q,2r-1,m)}}G^{\prime(p)}_{\tilde\Psi^{(q,2r-1,m)}}(v)+O(\e^2)\Bigg]+O(|\yp|^{-2\Dh[q,2p]+1})
\,.
\ee
Here the boundary OPE coefficients $f'$ and the conformal blocks $G'(v)$ are primed, implying that the external boundary operators are different from those in \eqref{3pt conformal block expansion}.
The analyticity requirement leads to another constraint,
\be\label{recursion 2}
&2^{4p-1}\,p!\(\frac{N}{2}+1\)_{p-1}\(1-\frac{(n-3) q}{n-1}\)^{2p}\frac{(nq-q+1)(\gh[q,1]+\gh[q,2p-1]-\gh[q,2p])}{(n-1)(2q+1)}\frac{\G(\frac{3}{2}+q)}{\G(\frac{n}{n-1}+q)}\nn
&+\a\sum_{r=2}^{n}\Bigg[\;\frac{\G(\frac{r-1}{n-1}+\frac{(2r-1)q+1}{2})\G(\frac{r-1}{n-1}+\frac{(2r-1)q+2}{2})W'_{2r-1,p}}{\G((r-1)(\frac{1}{n-1}+q))\G(r(\frac{1}{n-1}+q))\((1-2r)q-\frac{2(r-1)}{n-1}\)_{2}}\;\Bigg]=O(\e^2)
\,,
\ee
which has almost the same form as \eqref{recursion 1}, 
except that some coefficients are associated with the second three-point function in \eqref{3pt functions}.
Combining the two constraints \eqref{recursion 1} and \eqref{recursion 2}, we derive the anomalous dimensions of boundary composite operators,
\be\label{gamma odd}
\gh[q,2p+1]&=(2p+1)\gh[q,1]+\a\frac{(n-1)(2q+1)\G(\frac{n}{n-1}+q)}{(nq-q+1)\G(\frac{3}{2}+q)}\sum_{s=1}^{p}\Bigg[\;\tfrac{1}{2^{4s+1}\,s!\(\frac{N}{2}+1\)_s\(1-\frac{(n-3) q}{n-1}\)^{2s+1}}\nn
&\hspace{8em}\times\sum_{r=2}^{n}\frac{\G(\frac{r-1}{n-1}+\frac{(2r-1)q+1}{2})\G(\frac{r-1}{n-1}+\frac{(2r-1)q+2}{2})}{\G((r-1)(\frac{1}{n-1}+q))\G(r(\frac{1}{n-1}+q))\((1-2r)q-\frac{2(r-1)}{n-1}\)_{2}}\nn
&\hspace{11em}\times\(W_{2r-1,s}+2(N+2s)(1-\tfrac{(n-3) q}{n-1})W'_{2r-1,s}\)\Bigg]+O(\e^2)\,,\\
\label{gamma even}
\gh[q,2p]&=\gh[q,2p+1]-\gh[q,1]-\frac{(n-1)(2q+1)\G(\frac{n}{n-1}+q)\a}{2^{4p+1}\,p!\(\frac{N}{2}+1\)_p\(1-\frac{(n-3) q}{n-1}\)^{2p+1}(nq-q+1)\G(\frac{3}{2}+q)}\nn
&\hspace{3em}\times\sum_{r=2}^{n}\Bigg[\;\frac{\G(\frac{r-1}{n-1}+\frac{(2r-1)q+1}{2})\G(\frac{r-1}{n-1}+\frac{(2r-1)q+2}{2})W_{2r-1,p}}{\G((r-1)(\frac{1}{n-1}+q))\G(r(\frac{1}{n-1}+q))\((1-2r)q-\frac{2(r-1)}{n-1}\)_{2}}\;\Bigg]+O(\e^2)
\,,
\ee
where $\a$, $\gh[q,1]$, $W_{r,s}$ and $W'_{r,s}$ are given in \eqref{alpha k=1}, \eqref{gamma Phi}, \eqref{W} and \eqref{Wp} with $k=1$.
The results \eqref{gamma odd} and \eqref{gamma even} also lead to the cancellations of the nonanalytic terms at higher orders in $1-v$, so the three-point functions are indeed analytic around $v=1$.
For $n=2$, our results are in agreement with the $\phi^4$-theory result in \cite{Nishioka:2022odm}.
Moreover, we find the crossover exponent for the $(q,n)=(1,2)$ theory:
\be
\text{crossover exponent}=\frac{d-1-\Dh[0,2]}{d-\D_{\phi^2}}=\frac{1}{2}-\frac{N+2}{4(N+8)}\e+O(\e^2)
\,,
\ee
where we have used the bulk data $\g_{\phi^2}=\frac{N+2}{N+8}\e+O(\e^2)$ \cite{Guo:2023qtt}.
The result agrees with the crossover exponent obtained in \cite{diehl1981field2}. 

In the above computation, we have assumed that $n\neq 3$.
For the Dirichlet case with $n=3$, the nonanalytic term from $\Ph[a,1,1]$ is logarithmic:
\be\label{3pt conformal block expansion first term 2}
G^{(p)}_{\tPh[,q,1]}(v)
=v^{\Dh[q,1]/2}\(
1-\frac{\gh[q,1]+\gh[q,2p]-\gh[q,2p+1]}{2}\log(1-v)+O(\e^2)
\)\,.
\ee
The nonanalytic terms from the infinite sums are given by logarithms as well,
\be
&\sum_{m=0}^{\infty}b_{\phi\tilde\Psi^{(1,2r-1,m)}}f^{(p)}_{\tilde\Psi^{(1,2r-1,m)}}G^{(p)}_{\tilde\Psi^{(1,2r-1,m)}}(v)\nn
=\;&-\frac{\a\, W_{2r-1,p}}{6r-4}\log(1-v)(1+O(1-v))+\text{analytic terms}+O(\e^2)
\,.
\ee
The results for the anomalous dimensions $\gh[1,2p+1]$ and $\gh[1,2p]$ are covered by the general expressions \eqref{gamma odd} and \eqref{gamma even} as the special case of $(q,n)=(1,3)$,
\be
\gh[1,2p+1]&=\frac{21 N^2+6 N \left(22 p^2-32 p+21\right)+8 \left(160 p^3-174 p^2-16 p+21\right)}{336 (3
   N+22)}\e+O(\e^2)\,,\\
\gh[1,2p]&=\frac{p(p-1)(33 N+320 p-508)}{84 (3 N+22)}\e+O(\e^2)
\,.
\ee
Curiously, the anomalous dimension of $\Ph[,1,2]$ vanishes to order $\e$.

\subsubsection*{$\e^{1/2}$ expansion}

As mentioned in the Introduction, the bulk interaction $\phi^{2n}$ and the boundary interaction $\phi^{n+1}|_{\xn=0}$ mix in the Neumann case with odd $n$, 
and the resulting $\e$ expansion contains noninteger power terms.
Below, we will show that it is inconsistent to assume the integer power $\e$ expansion 
from the multiplet recombination perspective.
Then we will use the analyticity requirement to derive the leading anomalous dimensions of the composite operators $\Ph[a,0,2p+1]$ and $\Ph[,0,2p]$, which are of order $\e^{1/2}$.

To see the inconsistency, consider the matching condition associated with the boundary composite operator $\Ph[b,0,n]$ 
\be\label{matching special}
\lim_{\e\->0}\(\a^{-1}\<\Box\phi_a(\xp,\xn)\Ph[b,0,n](\yp,0)\>\)=\<\phi^{2n-1}_a(\xp,\xn)\Ph[b,0,n](\yp,0)\>_\f
\,.
\ee
The explicit constraint shares some similarities with the case of the boundary fundamental operator in \eqref{matching condition explicit}:
\be\label{matching special explicit}
&\lim_{\e\->0}\Bigg[\(\frac{
(\D_\phi-\Dh[0,n])(\D_\phi-\Dh[0,n]+1)
}{\xn^2}+\frac{
4\Dh[0,n](\D_\phi-\frac{d-2}{2})
}{|\xp-\yp|^2+\xn^2}\)\frac{\a^{-1}\d_{ab}\,b_{\phi\Ph[,0,n]}}{
\(|\xp-\yp|^2+\xn^2\)^{\Dh[0,n]}\xn^{\D_\phi-\Dh[0,n]}
}\Bigg]\nn
&=\frac{\d_{ab}\,b_{\phi_\f^{2n-1}\Ph[\f,0,n]}}{
\(|\xp-\yp|^2+\xn^2\)^{\Dh[0,n,\f]}\; \xn^{(2n-1)\D_{\phi_\f}-\Dh[0,n,\f]}}
\,. 
\ee
We again omit the second term on the left-hand side, as $\D_\phi-\frac{d-2}{2}=O(\e^2)$ is of order $\e^2$.
The anomalous dimension of $\Ph[a,0,n]$ is defined as
\be
\gh[0,n]=\Dh[0,n]-n\frac{d-2}{2}
\,.
\ee
The crucial point here is that there are two vanishing factors on the left-hand side of \eqref{matching special explicit}:
\begin{itemize}
\item
The first vanishing factor comes from the action of the Laplacian
\be\label{epsilon 1/2}
\D_\phi-\Dh[0,n]+1=\frac{1}{2}(n-1)\e-\gh[0,n]
\,,
\ee
which is related to the classical degeneracy of $\phi^{2n}$ and $\phi^{n+1}|_{\xn=0}$
\be
\D_{\phi_\text{f}^{2n}}=\D_{\phi_\text{f}}+2=\D_{\phi_\text{f}^{n+1}}+1\,. 
\ee
\item
The second vanishing factor is associated with the coefficient of the bulk-boundary two-point function
\be
\lim_{\e\->0}b_{\phi\Ph[,0,n]}=0
\,.
\ee
\end{itemize}
Let us examine the Gaussian limit of the combination 
$\a^{-1}(\frac{1}{2}(n-1)\e-\gh[0,n])b_{\phi\Ph[,0,n]}$. 
If both $\gh[0,n]$ and $b_{\phi\Ph[,0,n]}$ are at least of order $\e^1$, 
then this combination does not have a finite Gaussian limit,
and the left-hand side of \eqref{matching special explicit} vanishes. 
However,  if we compute the free BOE coefficient $b_{\phi^{2n-1}_\f\Ph[\f,0,n]}$ using Wick contractions, 
we find that the right-hand side of \eqref{matching special explicit} is finite for odd $n$.
Therefore, the matching condition cannot hold for odd $n$ if 
both $\gh[0,n]$ and $b_{\phi\Ph[,0,n]}$ are of order $\e^1$ or higher.
To resolve this issue, $\gh[0,n]$ and $b_{\phi\Ph[,0,n]}$ should be of order $\e^{\th}$ and $\e^{1-\th}$, where perturbation theory requires $0<\th<1$.
The matching condition \eqref{matching special} then yields
\be\label{alpha gamma b}
\lim_{\e\->0}\(\a^{-1}\gh[0,n]\,b_{\phi\Ph[,0,n]}\)=b_{\phi^{2n-1}_\f\Ph[\f,0,n]}
\,.
\ee
The solution for $b_{\phi\Ph[,0,n]}$ is
\be\label{BOE coefficient special}
b_{\phi\Ph[,0,n]}=\;& 2^{\frac{5 n-9}{2}}N (2 n+N-2) 
\Big(\frac{n+1}{2}\Big)_{\frac{n-1}{2}} \,
\Big[\Big(\frac N 2+1\Big)_{\frac{n-3}{2}}\Big]^2
{}_2F_1\Big(\frac{1-n}{2},\frac{1-n}{2};\frac{N}{2};1\Big)\frac{\a}{\gh[0,n]}
\nn&+\ldots
\,,
\ee
where higher order terms in $\e$ are denoted by the ellipsis.
The explicit expression of $\gh[0,n]$ can be found in \eqref{gh0n}.

In the derivation of the anomalous dimensions of boundary composite operators, 
we again make use of the analyticity constraints of three-point functions \eqref{3pt functions}.
The main difference from the integer power $\e$ expansion is the following.
For the leading corrections, 
the conformal block expansion here involves fewer terms. 
More specifically, only two terms are of lower order than $\e^1$. 
According to \eqref{BOE coefficients k=1}, 
the BOE coefficients $b_{\phi\Psi^{(0,2r-1,m)}}$ in \eqref{3pt conformal block expansion} are of order $\e^1$, 
so they do not contribute to the leading corrections.
However, the case $(r,m)=(\frac{n+1}{2},0)$ is more subtle because the denominator in \eqref{BOE coefficients k=1} vanishes. 
In fact, $\Ph[a,0,n]$ can have a BOE coefficient of lower order than $\e^1$ and 
be associated with a leading correction term.  
The corresponding conformal block involves a ${}_2F_1$ function, which is nonanalytic in $1-v$.
The other nonanalytic term is similar to \eqref{3pt conformal block expansion first term}, except that here we do not need to include the anomalous dimension $\gh[0,1]$, which is of order $\e^1$.

The conformal block expansion takes the form
\be\label{conformal block expansion special}
&\<\phi_a(\xp,\xn)\Ph[,0,2p](0,0)\Ph[b,0,2p+1](\yp,0)\>
=\frac{\d_{ab}|\yp|^{-2\Dh[0,2p+1]}}{|x|^{\dD}\xn^{\D_\phi}}\nn
&\times\Big[b_{\phi\tPh[,0,1]}f^{(p)}_{\tPh[,0,1]}G^{(p)}_{\tPh[,0,1]}(v)+b_{\phi\tPh[,0,n]}f^{(p)}_{\tPh[,0,n]}G^{(p)}_{\tPh[,0,n]}(v)+\ldots\Big]+O(|\yp|^{-2\Dh[0,2p+1]+1})
\,,
\ee
where the contribution from $\Ph[a,0,1]$ is given by
\be\label{block fundamental special}
G^{(p)}_{\tPh[,0,1]}(v)=v^{\Dh[0,1]/2}\(
1+\frac{(\gh[0,2p]-\gh[0,2p+1])v}{n-1}{}_3F_{2}\!\[\begin{matrix}1,1,\frac{n}{n-1}\vspace{0.2em}\\ \frac{3}{2},2\end{matrix};v\]+\ldots
\)\,,
\ee
with the coefficient
\be
b_{\phi\tPh[,0,1 ]}f^{(p)}_{\tPh[,0,1]}=2^{4 p+1}p!\(\frac{N}{2}+1\)_p+\ldots
\ee
We have omitted the term proportional to $\gh[0,1]$ because \eqref{gamma Phi} implies that $\gh[0,1]$ is of order $\e^1$.
As in the Dirichlet case, let us first assume $n\neq 3$.
According to \eqref{3F2 expansion}, the ${}_3F_2$ function in \eqref{block fundamental special} is nonanalytic at $v=1$.

For the second term in the bracket in \eqref{conformal block expansion special}, we use \eqref{alpha gamma b} and obtain
\be\label{contribution composite special}
b_{\phi\tPh[,0,n]}f^{(p)}_{\tPh[,0,n]}G^{(p)}_{\tPh[,0,n]}(v)=\frac{\a\,b_{\phi^{2n-1}_\f \tPh[\f,0,n]}f^{(p)}_{\tPh[\f,0,n]}}{\gh[0,n]}v^{\frac{n}{2(n-1)}}\,{}_2F_1\!\(\frac{1}{2},\frac{n+1}{2(n-1)};\frac{3}{2};v\)+\ldots
\,,
\ee
where the nonanalytic part of ${}_2F_1$ is given by
\be\label{2F1 expansion}
{}_2F_1(a,b;c;v)=\;&\frac{\G(c)\G(a+b-c)}{\G(a)\G(b)}(1-v)^{c-a-b}{}_2F_1(c-a,c-b;c-a-b+1;1-v)\nn
&+\text{analytic terms} \quad (c-a-b\notin\bb Z)
\,.
\ee
To cancel out the nonanalytic terms, the combination $\gh[0,2p]-\gh[0,2p+1]$ in \eqref{block fundamental special} and $\a/\gh[0,n]$ in \eqref{contribution composite special} should be of the same order in $\e$.
We write
\be\label{recursion}
2^{4 p+1}p!\(\frac{N}{2}+1\)_p\frac{n(\gh[0,2p]-\gh[0,2p+1])}{n-1}\frac{1}{\G(\frac{n}{n-1})}+\frac{\a\,b_{\phi^{2n-1}_\f \tPh[\f,0,n]}f^{(p)}_{\tPh[\f,0,n]}}{\gh[0,n]}\frac{1}{\G(\frac{1}{2})\G(\frac{n+1}{2(n-1)})}=0+\ldots
\ee
For the second three-point function in \eqref{3pt functions}, we have
\be
&\<\phi_a(\xp,\xn)\Ph[b,0,2p-1](0,0)\Ph[,0,2p](\yp,0)\>
=\frac{\d_{ab}|\yp|^{-2\Dh[0,2p]}}{|x|^{\D^-}\xn^{\D_\phi}}\nn
&\times\Big[b_{\phi\tPh[,0,1]}f^{\prime(p)}_{\tPh[,0,1]}G^{\prime(p)}_{\tPh[,0,1]}(v)+b_{\phi\tPh[,0,n]}f^{\prime(p)}_{\tPh[,0,n]}G^{\prime(p)}_{\tPh[,0,n]}(v)+O(\e)\Big]+O(|\yp|^{-2\Dh[0,2p]+1})
\,,
\ee
where the boundary OPE coefficients $f'$ and conformal blocks $G'(v)$ are primed since the external boundary operators are different from those in \eqref{conformal block expansion special}.
The analyticity of this correlator around $v=1$ implies the constraint
\be\label{recursion 2}
2^{4p-1}\,p!\(\frac{N}{2}+1\)_{p-1}\frac{n(\gh[0,2p-1]-\gh[0,2p])}{n-1}\frac{1}{\G(\frac{n}{n-1})}+\frac{\a\,b_{\phi^{2n-1}_\f \tPh[\f,0,n]}f^{\prime(p)}_{\tPh[\f,0,n]}}{\gh[0,n]}\frac{1}{\G(\frac{1}{2})\G(\frac{n+1}{2(n-1)})}=0
\,.
\ee

Solving the two equations \eqref{recursion} and \eqref{recursion 2}, we obtain the anomalous dimension
\be\label{epsilon 1/2 odd}
\gh[0,2p+1]=\frac{\G(\frac{n}{n-1})(n-1)\,\a}{\G(\frac{1}{2})\G(\frac{n+1}{2 (n-1)})\,\gh[0,n]}\sum_{s=1}^{p}\frac{1}{2^{4s+1}s!\(\frac{N}{2}+1\)_s}\(W_{n,s}+2(N+2s)W'_{n,s}\)+\ldots
\ee
Here we have used the condition that $\gh[0,1]$ is of higher order in $\e$.
The expressions of $\a$, $W_{n,p}$, and $W'_{n,p}$ are in \eqref{alpha k=1}, \eqref{W} and \eqref{Wp}.
For $p=\frac{n-1}{2}$, we have a quadratic equation for $\gh[0,n]$,  
whose solution reads
\be\label{gh0n}
\gh[0,n]=\pm\a^{1/2}\[\frac{\G(\frac{n}{n-1})(n-1)}{\G(\frac{1}{2})\G(\frac{n+1}{2 (n-1)})}\sum_{s=1}^{\frac{n-1}{2}}\frac{1}{2^{4s+1}s!\(\frac{N}{2}+1\)_s}\(W_{n,s}+2(N+2s)W'_{n,s}\)\]^{1/2}\!\!+O(\e)
\,,
\ee
which is of order $\e^{1/2}$.
Furthermore, we also obtain
\be\label{epsilon 1/2 even}
\gh[0,2p]=\gh[0,2p+1]-\frac{\a\,W_{n,p}}{\gh[0,n]}\frac{\G(\frac{n}{n-1})(n-1)}{\G(\frac{1}{2})\G(\frac{n+1}{2(n-1)})2^{4 p+1}p!\(\frac{N}{2}+1\)_p}+O(\e)
\,.
\ee

Similar to the Dirichlet case, the nonanalytic terms for $n=3$ are also given by logarithms:
\be
G^{(p)}_{\tPh[,0,3]}(v)&=v^{\Dh[0,1]/2}\(
1-\frac{1}{2}(\gh[0,2p]-\gh[0,2p+1])\log(1-v)+O(\e)\)\,,\\
b_{\phi\tPh[,0,3]}f^{(p)}_{\tPh[,0,3]}G^{(p)}_{\tPh[,0,3]}(v)&=\frac{\a\,b_{\phi^{5}_\f \tPh[\f,0,3]}f^{(p)}_{\tPh[\f,0,3]}}{\gh[0,3]}v^{1/4}\log(1-v)+\text{analytic terms}+O(\e)
\,.
\ee
The anomalous dimensions $\gh[0,2p+1]$ and $\gh[0,2p]$ are
\be
\gh[0,2p+1]&=\pm p (N+6 p+2)\(\frac{(N+4)}{2(3 N+22)(N+8)}\)^{1/2}\e^{1/2}+O(\e)\,,\\
\gh[0,2p]&=\pm p (N+6 p-4)\(\frac{(N+4)}{2(3N+22)(N+8)}\)^{1/2}\e^{1/2}+O(\e)
\label{n=3 k=1 even}
\,,
\ee
which agree with the $(q,n)=(0,3)$ case of \eqref{epsilon 1/2 odd} and \eqref{epsilon 1/2 even}. 
According to \cite{eisenriegler1988surface}, there are two fixed points if the bulk is at $\phi^6$-tricritical point in $d=3-\e$ dimensions.
One fixed point is stable, but the other one is unstable in most directions.
By examining the crossover exponent below, we find that the case of the overall plus signs corresponds to the stable fixed point.
Using \eqref{n=3 k=1 even} in the $+$ case and $\D_{\phi^2}=1+O(\e)$, we obtain
\be
\text{crossover exponent}=\frac{d-1-\Dh[0,2]}{d-\D_{\phi^2}}=\frac{1}{2}-\frac{N+2}{4}\(\frac{2(N+4)}{(3N+22)(N+8)}\)^{1/2}\e^{1/2}+O(\e)
\,.
\ee
This is in agreement with the results at the stable fixed point in \cite{diehl1987walks,eisenriegler1988surface}, which were derived from the standard diagrammatic computation.

In addition, let us examine the boundary condition in the case of $\e^{1/2}$ expansion.
The correction to the Neumann boundary condition is related to the boundary interaction $h\phi^{n+1}|_{\xn=0}=h\Ph[,0,n+1]$, where the coupling constant $h$ is of order $\e^{1/2}$.
Let us use the variational principle to establish this connection.
Consider the action for the $\phi^{2n}$-multicritical theory with odd $n$
\be
S_{+}\propto\int_{\bb R^d_+} \rm d^d x\(\phi_a\Box\phi_a+g\m^{(n-1)\e}\phi^{2n}\)+\int_{\bb R^{d-1}}\rm d^{d-1}x\;\(\Ph[a,0,1]\Ph[a,1,1]+h\Ph[,0,n+1]+O(\e)\)
\,,
\ee
where subscript $+$ corresponds the Neumann boundary condition in the Gaussian limit.
If the bulk equation of motion is satisfied, then the variation of the action is
\be
\d S_+\propto\int_{\bb R^{d-1}} \rm d^{d-1} x \; \(2\pa_\perp\phi_a|_{\xn=0}+(n+1)h\Ph[a,0,n]\)\d\Ph[a,0,1]
\,.
\ee
We require that $\d S_+=0$ for arbitrary $\d\Ph[a,0,1]$, which implies 
\be
\pa_\perp\phi_a|_{\xn=0}=-\frac 1 2 (n+1)h\Ph[a,0,n] +O(\e)\,.\label{partial x perpendicular phi}
\ee 
To find the coefficient, we examine the correlator on the left-hand side of \eqref{matching special}, since it is of order $\e^{1/2}$.
Taking the normal derivative $\frac{\pa}{\pa\xn}$ and the boundary limit $\xn\->0$, we obtain
\be
\frac{\pa}{\pa\xn}\<\phi_a(\xp,\xn)\Ph[b,0,n](\yp,0)\>\Big|_{\xn=0}=\frac{b_{\phi\Ph[,0,n]}}{|\xp-\yp|^{2\Dh[0,n]}}+O(\e)
\,.
\ee 
In accordance with \eqref{partial x perpendicular phi}, we interpret the right-hand side as
$\frac{b_{\phi\Ph[,0,n]}}{b_{\phi^n\Ph[,0,n]}}\<\Ph[,0,n](\xp,0)\Ph[b,0,n](\yp,0)\>+O(\e)$. 
This implies the relation
\be 
\frac{b_{\phi\Ph[,0,n]}}{b_{\phi^n\Ph[,0,n]}}=-\frac 1 2 (n+1)h+O(\e)
\,,\label{b over b}
\ee 
where the expression for $b_{\phi\Ph[,0,n]}$ is given in \eqref{BOE coefficient special}.
The other BOE coefficient $b_{\phi^n\Ph[,0,n]}$ is of order $\e^0$ and can be computed using Wick contractions:
\be 
b_{\phi^n\Ph[,0,n]}=2^{2n-1}\(\frac{n-1}{2}\)!\(1+\frac N 2\)_{\frac{n-1}{2}}+O(\e^{1/2})
\,.
\ee 
As a concrete example, the $n=3$ case at the stable fixed point reads
\be\label{boundary coupling}
\frac{b_{\phi\Ph[,0,3]}}{b_{\phi^3\Ph[,0,3]}}=\(\frac{N+4}{8(3N+22)(N+8)}\)^{1/2}\e^{1/2}+O(\e)
\,,
\ee
which agrees with the boundary coupling constant obtained in \cite{diehl1987walks,eisenriegler1988surface} after changing the normalization.\footnote{The different conventions are related by $\phi_{a,\text{ours}}=2\sqrt \p \phi_{a,\text{theirs}}$.
Their modified Neumann boundary condition is
$\pa_\perp\phi_\text{theirs}|_{\xn=0}=\frac 1 3 \p u \Ph[{a,\text{theirs}},0,3]+O(\e)$, where $u=-288h$ is their boundary coupling constant.
Comparing with \eqref{b over b}, we find that
$u=144\frac{b_{\phi\Ph[,0,3]}}{b_{\phi^3\Ph[,0,3]}}+O(\e)=(\frac{18(N+4)}{(3N+22)(N+8)})^{1/2}\e^{1/2}+O(\e)$, which is precisely the leading order of equation (9) in \cite{diehl1987walks}.
}

\subsection{Analytic bootstrap}
\label{Analytic bootstrap}

In this section, we verify that the multiplet-recombination results are consistent with boundary crossing symmetry.
We consider the bootstrap equation for the bulk-bulk two-point function $\<\phi_a(x)\phi_b(y)\>$.
To order $\e^1$, there are finitely many primaries contributing to the conformal block expansion in both the bulk and boundary channels.
We will solve the bootstrap equation and deduce the anomalous dimensions of the boundary fundamental operators.
The results agree with those in \eqref{gamma Phi} from the multiplet recombination.
We would like to emphasize that the analytic bootstrap approach is independent of the multiplet recombination.
In other words, we do not use any interacting boundary data derived from the multiplet recombination,
i.e., we only make use of the bulk data and some input from the free BCFT.

The bulk-bulk two-point function takes the form
\be
\<\phi_a(x)\phi_b(y)\>=\frac{\d_{ab}F(\x)}{\(4\xn\yn\)^{\D_\phi}}
\,,
\ee
where the conformally invariant variable $\x$ is defined as
\be\label{xi definition}
\x=\frac{|x-y|^2}{4\xn\yn}
\,.
\ee
Let us expand the function $\x^{\D_\phi}F(\x)$ in terms of conformal blocks.
In the bulk channel, the two-point function $\<\phi_a(x)\phi_b(y)\>$ is viewed as a sum of bulk one-point functions, which are associated with operators in the $\phi_a\times\phi_b$ OPE.
The one-point functions of spinning operators vanish due to conformal symmetry \cite{Liendo:2012hy}.
Moreover, although the bulk OPE $\phi_a\times\phi_b$ contains spin-0 symmetric traceless O($N$) tensors such as $\phi_a\phi_b-\frac{\d_{ab}}{N}\phi^2$, the one-point functions of these operators are zero 
because a rank-2 traceless O($N$)-invariant tensor does not exist.
Therefore, we can restrict the internal operators to singlet scalars in the bulk channel.
The bulk channel scalar blocks read \cite{McAvity:1995zd}
\be
f_\text{bulk}(\D,\x)=\x^{\D/2}\,{}_2F_1\!\(\frac{\D}{2},\frac{\D}{2};\D-\frac{d}{2}+1;-\x\)
\,,
\ee
where $\D$ is the scaling dimension of the internal operator.
In the boundary channel, $\<\phi_a(x)\phi_b(y)\>$ is thought of as a sum of boundary two-point functions associated with the operators in the BOEs of $\phi_a$ and $\phi_b$.
The spinning operators do not appear in these BOEs thanks to boundary Lorentz invariance, so we also only need to consider scalar internal operators in the boundary channel.
The boundary channel scalar blocks are \cite{McAvity:1995zd}
\be
f_\text{bdy}(\Dh[],\x)=\x^{-\Dh[]}\,{}_2F_1\!\(\Dh[],\Dh[]-\frac{d}{2}+1;2\Dh[]-d+2;-\frac{1}{\x}\)
\,,
\ee
where $\Dh[]$ denotes the scaling dimension of the internal operator in the boundary channel.

In the free theory, the bootstrap equation takes the form
\be\label{free crossing k=1}
1+\l_{\tilde\phi^2_\f}a_{\tilde\phi^2_\f}f_\text{bulk}(\D_{\phi^2_\f},\x)=\x^{\D_{\phi_\f}}\m_{\tPh[\f,q,1]}^2 f_\text{bdy}(\Dh[q,1,\f],\x)\,,
\ee
where $\l_{\tilde{\cal O}}$ is the bulk OPE coefficient of $\tilde{\cal O}_{ab}\in\phi_a\times\phi_b$, and $\m_{\tilde{\cal O}}$ is related to the BOE coefficient before by
\be\label{mu-b}
\m_{\tilde{\cal O}}=2^{\D_\phi-\Dh[\cal O]}b_{\phi\tilde{\cal O}}
\,.
\ee
The tilded internal operators in the boundary channel have unit-normalized two-point functions as in \eqref{boundary internal operator normalization}.
In the bulk channel, the internal operators are also tilded, 
as they have unit-normalized two-point functions in the limit $\xn,\,\yn\->\infty$:
\be
\<\tilde{\cal O}(\xp,\xn)\tilde{\cal O}(\yp,\yn)\>\sim\frac{1}{|x-y|^{2\D_{\cal O}}}\quad (\xn,\,\yn\->\infty)
\,.
\ee
The bulk one-point function coefficient $a_{\tilde{\cal O}}$ is defined by
\be\label{one-point coefficient definition}
\<\tilde{\cal O}(\xp,\xn)\>=\frac{a_{\tilde{\cal O}}}{\(2\xn\)^{\D_{\cal O}}}
\,.
\ee
There are two sets of free-theory solutions.
The first one is
\be\label{canonical-free-crossing-N}
\l_{\tilde\phi^2_\f}a_{\tilde\phi^2_\f}=1\,,\quad
\Dh[0,1,\f]=\frac{d-2}{2}\,,\quad
\m_{\tPh[\f,0,1]}^2=2
\,.
\ee
As the scaling dimension of $\Ph[{a,\f},0,1]$ in \eqref{canonical-free-crossing-N} 
is the same as that of $\phi_\f$, 
this solution is identified with the Neumann boundary condition.
The second set of solutions reads
\be
\l_{\tilde\phi^2_\f}a_{\tilde\phi^2_\f}=-1\,,\quad
\Dh[1,1,\f]=\frac{d-2}{2}+1\,,\quad
\m_{\tPh[\f,1,1]}^2=\frac{d-2}{2}
\,,
\ee
which corresponds to the Dirichlet boundary condition.
Below, we study the leading corrections in the interacting theories.

\begin{figure}
	\center

\begin{subfigure}{.5\textwidth}
	\center

\tikzset{every picture/.style={line width=1pt}}      

\begin{tikzpicture}[x=0.75pt,y=0.75pt,yscale=-1,xscale=1]
	
	\draw    (210,60) -- (270,120) ;
	\draw    (270,120) -- (330,60) ;
	\draw    (250,200) -- (270,120) ;
	\draw    (240,200) -- (270,120) ;
	\draw    (300,200) -- (270,120) ;
	\draw    (290,200) -- (270,120) ;
	\draw    (180,200) -- (360,200) ;
	\draw  [dash pattern={on 4.5pt off 4.5pt}]  (240,170) -- (303,170) ;
	
	\draw (200,35) node [anchor=north west][inner sep=0.75pt]    {$\phi_a$};
	\draw (326,35) node [anchor=north west][inner sep=0.75pt]    {$\phi_b$};
	\draw (262.5,153) node [anchor=north west][inner sep=0.75pt]    {...};
	\draw (306,152) node [anchor=north west][inner sep=0.75pt]    {$\phi^{2n-2}$};
	\draw (280,115) node [anchor=north west][inner sep=0.75pt]    {$g$};
	
	\draw (245,205) node [anchor=north west][inner sep=0.75pt]    {identity};

\end{tikzpicture}
	\caption{The contribution from $\phi^{2n-2}$.}
\end{subfigure}%
\begin{subfigure}{.5\textwidth}
\center

\tikzset{every picture/.style={line width=1pt}}     

\begin{tikzpicture}[x=0.75pt,y=0.75pt,yscale=-1,xscale=1]
	\draw    (210,60) -- (260,110) ;
	\draw    (270.4,110) -- (320.4,60) ;
	\draw    (280,200) -- (270.4,110) ;
	\draw    (320,200) -- (270,110) ;
	\draw    (190,200) -- (350,200) ;
	\draw    (259.6,110) -- (260,200) ;
	\draw    (270,110) -- (270.4,200) ;
	
	\draw  [dash pattern={on 4.5pt off 4.5pt}]  (250,170) -- (320,170) ;
	\draw (200,35) node [anchor=north west][inner sep=0.75pt]    {$\phi_a$};
	\draw (316.4,35) node [anchor=north west][inner sep=0.75pt]    {$\phi_b$};
	\draw (278,155) node [anchor=north west][inner sep=0.75pt]    {...};
	\draw (318,152) node [anchor=north west][inner sep=0.75pt]    {$\phi^{2n}$};
	\draw (280,105) node [anchor=north west][inner sep=0.75pt]    {$g$};
	\draw (265,205) node [anchor=north west][inner sep=0.75pt]    {identity};	
\end{tikzpicture}

\caption{The contribution from $\phi^{2n}$.}
\end{subfigure}      

	\caption{The new primaries in the bulk-channel expansion of the bulk-bulk correlator 
	$\<\phi_a\phi_b\>$ at order $\e^1$.  We assume that $n>2$.
	Through the $g\phi^{2n}$ vertex,  the operators $\phi^{2n-2}$ and $\phi^{2n}$ contribute at order $g\sim\e^1$. 
	The lines connecting to the boundary represent the BOEs.
	Only the identity operator on the boundary contributes in these BOEs, since the one-point functions of other boundary operators vanish.
	}
	\label{Feynman 2pt}
\end{figure}
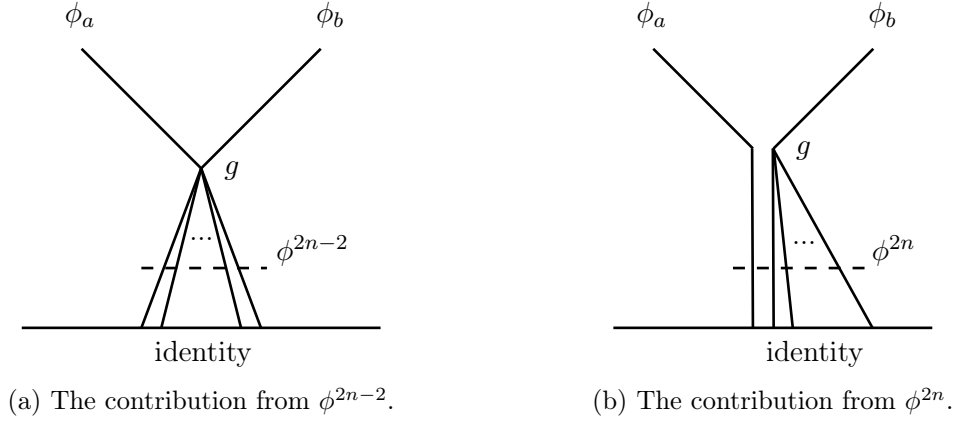

Let us first study the case of integer power $\e$ expansion.
At order $\e^1$, we still only need to consider the contribution from $\Ph[a,q,1]$ in the boundary channel.
This is because the BOE coefficients will be squared in the bootstrap equation, and new boundary operators can only contribute at order $\e^2$.
On the other hand, new operators may contribute to the bulk-channel expansion.
We do not need to consider the primaries constructed from an odd number of $\phi$'s due to $\bb Z_2$ symmetry.
The only physical primary scalars with order $\e^0$ bulk OPE coefficients $\l_{\tilde{\cal O}}$ are the identity and $\phi^2$, which contribute already in the free theory.
So the new operators should have bulk OPE coefficients $\l_{\tilde{\cal O}}$ of order $\e^1$.
They should also have nonvanishing free one-point function coefficients $a_{\tilde{\cal O}}$, so the product $\l_{\tilde{\cal O}}\,a_{\tilde{\cal O}}$ is of order $\e^1$. 
There are two scenarios: $n=2$ and $n>2$.
For $n=2$, the primaries with more than four $\phi$'s do not have order $\e^1$ bulk OPE coefficients, as a single $g\phi^4$ vertex with $g\sim\e^1$ cannot contract all the legs in the Feynman diagrams.
It turns out that the primaries of the schematic forms $\Box^m\phi^4$ with $m>0$ do not contribute at order $\e^1$, so the only new operator in the bulk channel is $\phi^4$. 
\footnote{For $m>0$, the physical bilinear operators of the schematic forms $\phi_a\Box^m\phi_a$ are descendants.}
For $n>2$, two new bulk operators $\phi^{2n-2}$ and $\phi^{2n}$ contribute at order $\e^1$ (see figure \ref{Feynman 2pt}).
Similar to the $n=2$ case, the primaries $\Box^m\phi^{2n-2}$ and $\Box^m\phi^{2n}$ with $m>0$ do not contribute at order $\e^1$. 
\footnote{At order $\e^1$, the bulk-channel contributions can be deduced from the boundary channel contributions to the crossing equation, in which only the boundary fundamental primaries appear.}
For the $\e^{1/2}$ expansion, the bulk channel remains the same as the case of integer power expansion, while a new boundary primary $\Ph[a,0,n]$ contributes at order $\e^1$, since its BOE coefficient is of order $\e^{1/2}$. 

Below, we first solve the bootstrap equation in the $\e$ expansion.
We consider both the $n=2$ and $n>2$ scenarios, as mentioned above.
Then, we turn to the case of the $\e^{1/2}$ expansion, which corresponds to the Neumann case with odd $n$. 

Let us consider the scenario with $n=2$.
The bootstrap equation takes the form
\be\label{crossing n=2}
1+\l_{\tilde\phi^2}a_{\tilde\phi^2}f_\text{bulk}(\D_{\phi^2},\x)+\l_{\tilde\phi^4}a_{\tilde\phi^4}f_\text{bulk}(\D_{\phi^4},\x)=\x^{\D_\phi}\m_{\tPh[,q,1]}^2 f_\text{bdy}(\Dh[q,1],\x)+O(\e^2)
\,.
\ee
We solve this by expanding both sides around $\x=0$, and matching the coefficients order by order.
For the Neumann case, the solution reads
\be\label{crossing solution Neumann}
\gh[0,1]&=-\frac{\g_{\phi^2}}{2}+O(\e^2)\,,\hspace{4em}
\m_{\tPh[,0,1]}^2=2+O(\e^2)\,,\nn
\l_{\tilde\phi^2}a_{\tilde\phi^2}&=1+\frac{\g_{\phi^2}}{2}+O(\e^2)\,,\hspace{3em}
\l_{\tilde\phi^4}a_{\tilde\phi^4}=\frac{\g_{\phi^2}}{4}+O(\e^2)
\,,
\ee
and the solution in the Dirichlet case is
\be\label{crossing solution Dirichlet}
\gh[1,1]&=-\frac{\g_{\phi^2}}{2}+O(\e^2)\,,\hspace{4.8em}
\m_{\tPh[,1,1]}^2=\frac{2-\e}{2}+\frac{\g_{\phi^2}}{2}+O(\e^2)\,,\nn
\l_{\tilde\phi^2}a_{\tilde\phi^2}&=-1+\frac{\g_{\phi^2}}{2}+O(\e^2)\,,\hspace{3em}
\l_{\tilde\phi^4}a_{\tilde\phi^4}=\frac{\g_{\phi^2}}{4}+O(\e^2)
\,.
\ee
We use the input from the bulk theory
\be\label{gamma phi2}
\g_{\phi^2}=\frac{N+2}{N+8}\e+O(\e^2)
\,,
\ee
whose derivation can be found in \cite{Guo:2023qtt}. 
In both the Neumann and Dirichlet cases, the anomalous dimension of $\Ph[a,q,1]$ is
\be
\gh[q,1]=\frac{N+2}{2(N+8)}\e+O(\e^2)
\,,
\ee
which agrees with the multiplet-recombination result \eqref{gamma Phi} at $n=2$.
Apart from verifying the results, the solutions \eqref{crossing solution Neumann} and \eqref{crossing solution Dirichlet} also provide new data of the interacting theory.
For example, we obtain the BOE coefficients
\be
\m_{\tPh[,q,1]}=\begin{cases}
	\; \sqrt{2}+O(\e^2) \quad & \text{Neumann }(q=0) \\
	\; 1-\frac{3\,\e}{2(N+8)}+O(\e^2)  & \text{Dirichlet }\,\;(q=1)\,.
\end{cases}
\ee
Interestingly, the solutions \eqref{crossing solution Neumann} and \eqref{crossing solution Dirichlet} impose a constraint on the interacting bulk data and the free one-point function coefficient.
The bootstrap equation holds only if $\l_{\tilde\phi^4}a_{\tilde\phi^4}=\frac{\g_{\phi^2}}{4}+O(\e^2)$ is satisfied.
Here the one-point function coefficient $a_{\tilde\phi^4}$  at order $\e^0$ can be computed using Wick contractions:
\be
a_{\tilde\phi^4}=\frac{N(N+2)}{2}+O(\e)
\,.
\ee
The bulk OPE coefficient $\l_{\tilde\phi^4}$ is derived in appendix \ref{A bulk OPE coefficient}.
For $(k,n)=(1,2)$, we have
\be
\l_{\tilde\phi^{4}}a_{\tilde\phi^{4}}=\frac{N+2}{8}\a+O(\e^2)
\,.
\ee
Using \eqref{alpha k=1} and \eqref{gamma phi2}, we confirm that the condition $\l_{\tilde\phi^4}a_{\tilde\phi^4}=\frac{\g_{\phi^2}}{4}+O(\e^2)$ is indeed satisfied.

For $n>2$, the bootstrap equation is
\be\label{crossing n>2}
&1+\l_{\tilde\phi^2}a_{\tilde\phi^2}f_\text{bulk}(\D_{\phi^2},\x)+\l_{\tilde\phi^{2n-2}}a_{\tilde\phi^{2n-2}}f_\text{bulk}(\D_{\phi^{2n-2}},\x)+\l_{\tilde\phi^{2n}}a_{\tilde\phi^{2n}}f_\text{bulk}(\D_{\phi^{2n}},\x)\nn
=\;&\x^{\D_\phi}\m_{\tPh[,q,1]}^2 f_\text{bdy}(\Dh[q,1],\x)+O(\e^2)
\,.
\ee
To order $\e^1$, the solution in the Neumann case is given by
\be\label{crossing Neumann}
\gh[0,1]&=\frac{n}{1-n}\l_{\tilde\phi^{2n}}a_{\tilde{\phi}^{2n}}+O(\e^2)\,,\hspace{3em}
\l_{\tilde\phi^{2n-2}}a_{\tilde\phi^{2n-2}}=\frac{n}{n-2}\l_{\tilde\phi^{2n}}a_{\tilde{\phi}^{2n}}+O(\e^2)\,,\nn
\l_{\tilde\phi^2}a_{\tilde\phi^2}&=1+n\l_{\tilde\phi^{2n}}a_{\tilde{\phi}^{2n}}\[1+\frac{1}{n-1}\(H_{\frac{1}{1-n}}-H_{\frac{2-n}{n-1}}\)\]+O(\e^2)\,,\nn
\m_{\tPh[,q,1]}^2&=2+2n\l_{\tilde\phi^{2n}}a_{\tilde{\phi}^{2n}}\(1-\frac{1}{n-1}H_\frac{1}{n-1}\)+O(\e^2)
\,,
\ee
where $H_n$ are the harmonic numbers.
As opposed to the $n=2$ case, here the bulk anomalous dimension $\g_{\phi^2}$ vanishes to order $\e^1$.
The solutions are written in terms of the product $\l_{\tilde\phi^{2n}}a_{\tilde\phi^{2n}}$.
For the Dirichlet case, we find
\be\label{crossing Dirichlet}
\gh[1,1]&=\frac{n}{1-n}\l_{\tilde\phi^{2n}}a_{\tilde{\phi}^{2n}}+O(\e^2)\,,\hspace{3em}
\l_{\tilde\phi^{2n-2}}a_{\tilde\phi^{2n-2}}=-\frac{n}{n-2}\l_{\tilde\phi^{2n}}a_{\tilde{\phi}^{2n}}+O(\e^2)\,,\nn
\l_{\tilde\phi^2}a_{\tilde\phi^2}&=-1+n\l_{\tilde\phi^{2n}}a_{\tilde{\phi}^{2n}}\[1-\frac{1}{n-1}\(H_{\frac{1}{1-n}}-H_{\frac{2-n}{n-1}}\)\]+O(\e^2)\,,\nn
\m_{\tPh[,q,1]}^2&=\frac{d-2}{2}+\frac{n}{(n-1)^2}\l_{\tilde\phi^{2n}}a_{\tilde{\phi}^{2n}}\(2-H_\frac{1}{n-1}\)+O(\e^2)
\,.
\ee
Again, all the variables are expressed in terms of the product $\l_{\tilde\phi^{2n}}a_{\tilde\phi^{2n}}$.
According to \eqref{bulk OPE} and Wick contractions, we have
\be\label{lambda a 2n}
\l_{\tilde\phi^{2n}}a_{\tilde\phi^{2n}}=\frac{2^{n-3}\,(n-1) \(\frac{N}{2}+1\)_{n-1}\,\s^{n}}{n}\a+O(\e^2)
\,.
\ee
In both cases above, the results for the anomalous dimensions $\gh[q,1]$ agree with the result \eqref{gamma Phi} from the multiplet recombination. 
As in the $n=2$ case, the bootstrap equation for $n>2$ imposes a constraint on the bulk data and the free BCFT as well.
The bulk OPE coefficients and one-point function coefficients need to satisfy the relation between
$\l_{\tilde\phi^{2n-2}}a_{\tilde\phi^{2n-2}}$ and $\l_{\tilde\phi^{2n}}a_{\tilde\phi^{2n}}$ in \eqref{crossing Neumann} or \eqref{crossing Dirichlet}.
The product $\l_{\tilde\phi^{2n-2}}a_{\tilde\phi^{2n-2}}$ takes the form
\be
\l_{\tilde\phi^{2n-2}}a_{\tilde\phi^{2n-2}}=\frac{2^{n-3}(n-1)
   \left(\frac{N}{2}+1\right)_{n-1}\,\s^{n-1}}{n-2}\a+O(\e^2)
\,,
\ee
where we have used the bulk OPE coefficient $\l_{\tilde\phi^{2n-2}}$ obtained in \cite{Guo:2023qtt}, and the one-point function coefficient $a_{\tilde\phi^{2n-2}}$ is computed using Wick contractions.
We have verified that these constraints are indeed satisfied.

Let us consider the Neumann case with odd $n$, where we encounter the $\e^{1/2}$ expansion.
In section \ref{Boundary composite operators}, 
the leading corrections in the conformal block expansion \eqref{conformal block expansion special} 
are of order $\e^{1/2}$, so they take a simpler form than the $\e$ expansion case \eqref{3pt conformal block expansion}. 
However, 
for the bulk-bulk correlator, 
the leading corrections remain of order $\e^1$ in the $\e^{1/2}$ expansion. 
In fact, there is one more primary $\Ph[,0,n]$ in the boundary channel expansion.
The bootstrap equation reads
\be
&1+\l_{\tilde\phi^2}a_{\tilde\phi^2}f_\text{bulk}(\D_{\phi^2},\x)+\l_{\tilde\phi^{2n-2}}a_{\tilde\phi^{2n-2}}f_\text{bulk}(\D_{\phi^{2n-2}},\x)+\l_{\tilde\phi^{2n}}a_{\tilde\phi^{2n}}f_\text{bulk}(\D_{\phi^{2n}},\x)\nn
=\;&\x^{\D_\phi}\[\m_{\tPh[,0,1]}^2 f_\text{bdy}(\Dh[0,1],\x)+\m_{\tPh[,0,n]}^2 f_\text{bdy}(\Dh[0,n],\x)+O(\e^2)\]
\,,
\ee
where the BOE coefficient $\m_{\tPh[,0,n]}$ is of order $\e^{1/2}$. 
The solution to the bootstrap equation above is
\be
\gh[0,1]&=\frac{n}{1-n}\l_{\tilde\phi^{2n}}a_{\tilde{\phi^{2n}}}+O(\e^{3/2})\,,\hspace{3em}
\l_{\tilde\phi^{2n-2}}a_{\tilde\phi^{2n-2}}=\frac{n}{n-2}\l_{\tilde\phi^{2n}}a_{\tilde{\phi^{2n}}}+O(\e^{3/2})\,,\nn
\l_{\tilde\phi^2}a_{\tilde\phi^2}&=1+n\l_{\tilde\phi^{2n}}a_{\tilde{\phi}^{2n}}\[1+\frac{1}{n-1}\(H_{\frac{1}{1-n}}-H_{\frac{2-n}{n-1}}\)\]-2(n-1)\m_{\tPh[,0,n]}^2+O(\e^{3/2})\,,\nn
\m_{\Ph[,0,1]}^2&=2+2n\l_{\tilde\phi^{2n}}a_{\tilde{\phi^{2n}}}\(1-\frac{1}{n-1}H_{\frac{1}{n-1}}\)-2(n-1)\m_{\tPh[,0,n]}^2+O(\e^{3/2})
\,.
\ee
The solution is written in terms of $\l_{\tilde\phi^{2n}}a_{\tilde{\phi^{2n}}}$, which is given by \eqref{lambda a 2n}. 
The solution for $\gh[0,1]$ is again in agreement with that from multiplet recombination \eqref{gamma Phi}. 
The squared BOE coefficient $\m_{\Ph[,0,n]}^2$ associated with the new operator $\Ph[a,0,n]$ appears in the solutions for the coefficients $\l_{\tilde\phi^{2}}a_{\tilde\phi^{2}}$ and $\m_{\Ph[,0,1]}^2$, 
but not in the solution for the anomalous dimension $\gh[0,1]$. 
In section \ref{Boundary composite operators}, the BOE coefficient of $\Ph[a,0,n]$ has been determined 
by the multiplet recombination and analyticity constraints, which is given in \eqref{BOE coefficient special}. 
It is straightforward to further determine $\l_{\tilde\phi^2}a_{\tilde\phi^2}$ and $\m_{\Ph[,0,1]}^2$ to order $\e^1$. 
Note that the different conventions $b_{\phi\Ph[,0,n]}$ and $\m_{\Ph[,0,n]}$ are related by \eqref{mu-b}.

\section{\boldmath Higher-derivative O($N$)-symmetric $\phi^{2n}$ theories with a boundary}
\label{sec3}

In this section, we study the higher derivative O($N$)-symmetric $\phi^{2n}$ theories with a boundary.
At $d=\du-\e$, the actions for these generalized theories take the form
\be\label{action}
S_{\pm,\ldots,\pm}\propto\int_{\bb R^d_+} \rm d^d x\(\phi_a\Box^k\phi_a+g\m^{(n-1)\e}\phi^{2n}\)+S_{\text{bdy},\pm,\ldots,\pm}
\,,
\ee
which is a higher derivative generalization of the canonical case in \eqref{action k=1}.
The upper critical dimension is given by
\be\label{upper critical dim}
\du=\frac{2nk}{n-1}
\,.
\ee
The boundary term is a generalization of \eqref{boundary term k=1}:
\be
S_{\text{bdy},\pm,\ldots,\pm}=\sum_{j=0}^{k-1}\int_{\bb R^{d-1}} \rm d^{d-1}x\;(\pm)\Ph[a,j,1]\Ph[a,2k-1-j,1]+\ldots
\,,
\ee
where the higher order terms in $\e$ are not written explicitly.
The general expression of the free boundary fundamental primaries can be found in \eqref{boundary primary}.

\subsection{Higher-derivative free theories with a boundary}
\label{Generalized free theory with a boundary}

To generalize the $k=1$ results in section \ref{sec2} to 
the $k>1$ case with higher derivatives, 
it is crucial to have a good understanding of the higher derivative free theories with a boundary. 
Below we give a brief summary of the $\Box^k$ free scalar BCFTs, 
where the subscript f is omitted for simplicity. 
We refer to \cite{Chalabi:2022qit} for more details. 
We will present a new result for the bulk-bulk two point function $\<\phi_a(x)\phi_b(y)\>$, 
i.e.,  the general solution for all $k$ and all the possible conformal boundary conditions. 

An important distinction from the canonical case is that 
the number of possible boundary fundamental primaries in a $\Box^k$ BCFT is $2k$.
After imposing the conformal boundary conditions, 
the number of physical boundary fundamental primaries with finite two-point functions reduces to $k$, 
and they will be labeled by the subscript $i=0,1,\ldots, k-1$. 
Note that the fundamental boundary operator $\Ph[a,q_i,1]$ is of the $q_i$-derivative type, 
whose schematic form is given in \eqref{boundary primary schematic}. 
The boundary fundamental primaries form $k$ pairs
\be\label{pair}
(\Ph[a,0,1],\Ph[a,2k-1,1])\,,\;
(\Ph[a,1,1],\Ph[a,2k-2,1])\,,\;
\ldots\,,\;
(\Ph[a,k-1,1],\Ph[a,k,1])
\,,
\ee 
where only one operator in each pair is physical. 
In other words, the derivative order of the $i$th operator can be either $q_i=i$ or $ q_i=2k-1-i$.  
The choice of $\{q_0, q_1,\dots, q_{k-1}\}$ for the derivative orders of the physical operators 
specifies the conformal boundary conditions. 

Let us consider the $\Box^2$ theory as a concrete example. 
There are four fundamental primaries, 
but only two of them are physical, 
so the nonvanishing fundamental primaries are labeled by $i=0,1$. 
If we choose $q_0=0$ and $q_1=2k-1-1=2$, 
then $\{\Ph[a,0,1], \Ph[a,2,1]\}$ are the physical boundary fundamental primaries.  
Accordingly, the conformal boundary conditions are given by the null-state conditions 
$\Ph[a,3,1]=\Ph[a,1,1]=0$. 
From the perspective of the variational principle, 
the variation of the higher derivative free action \eqref{gff free action} with $k=2$ reads
\be\label{deltaS k=2}
\d S_{\pm,\pm}\propto\int_{\bb R^{d-1}} \rm d^{d-1} x \sum_{j=0}^3 (-1)^{j+1} \Ph[a,j,1]\d\Ph[a,3-j,1]+\d S_{\text{bdy},\pm,\pm}
\,,
\ee
where we have used the bulk equation of motion $\Box^2\phi_a=0$.
The explicit definitions of the boundary fundamental primaries are 
\be
\Ph[a,0,1]=\lim_{\xn\rightarrow 0}\phi_a, 
\quad
\Ph[a,1,1]=\lim_{\xn\rightarrow 0}\pa_\perp\phi_a, 
\ee
\be\label{boundary primary 32 k=2}
\Ph[a,3,1]=\lim_{\xn\rightarrow 0}(\pa_\perp^3+3\pa_\perp\Box_\parallel)\phi_a\,,
\quad
\Ph[a,2,1]=\lim_{\xn\rightarrow 0}(\pa_\perp^2-\Box_\parallel)\phi_a\,. 
\ee
If we choose the $\{+,+\}$ boundary terms
\be
S_{\text{bdy},+,+}=\int_{\bb R^{d-1}} \rm d^{d-1} x \; (\Ph[a,0,1]\Ph[a,3,1]+\Ph[a,1,1]\Ph[a,2,1])\,,
\ee 
then the variation $\d S_{\text{bdy},+,+}$ cancels the $j=0,2$ terms in \eqref{deltaS k=2}.
We require that $\d S_{+,+}=0$ for arbitrary variations and 
derive the conformal boundary conditions $\Ph[a,3,1]=\Ph[a,1,1]=0$.

As in the canonical case $k=1$, our focus is on the boundary scalar primaries, which include the boundary fundamental primary operators $\{\Ph[a,q_i,1]\}$ and the corresponding composite operators $\{\Ph[,q_i,2p]$, $\Ph[a,q_i,2p+1]\}$.
Since $\{\Ph[a,q_i,1]\}$ can be constructed from the bulk fundamental operator $\phi_a$, we will first derive the bulk-bulk two-point function $\<\phi_a(x)\phi_b(y)\>$ 
from the equation of motion. 
Then we use $\<\phi_a(x)\phi_b(y)\>$ to deduce the bulk-boundary and boundary-boundary correlators $\<\phi_a(x)\Ph[b,q_i,1](y)\>$ and $\<\Ph[a,q_i,1](x)\Ph[b,q_j,1](y)\>$.

\subsubsection*{Equation of motion}

The actions for the generalized free theories are given in \eqref{gff free action}.
The invariance of the generating functional $Z[J]=\int D\phi \,\exp[{-S+\int_{\xn>0} \rm d^d x \, J(x)\phi(x)}]$ under an infinitesimal change $\phi(x)\->\phi(x)+\d\phi(x)$ implies 
the quantum equation of motion
\be
\<\d S/\d\phi(x)\>=\<J(x)\>,\quad \xn>0
\,,
\ee
where $\frac{\d}{\d\phi(x)}$ denotes the functional derivative with respect to $\phi(x)$, and $J$ is the source for the bulk field $\phi$. 
We assume that $\d\phi|_{\xn=0}=0$, so the boundary term does not contribute to the equation of motion 
$\d S_\text{bdy}/\d\phi(x)=0$ 
and we can omit the label $\pm,\ldots,\pm$.
Taking the functional derivative $\frac{\d}{\d J(y)}$ with $\yn>0$ and setting $J=0$, 
we obtain the bulk equation of motion for the two-point function
\be\label{equation of motion}
\Box_x^k\<\phi_a(x)\phi_b(y)\>=\k\,\d^{(d)}(x-y) \quad (\xn,\yn>0)
\,,
\ee
where $\d^{(d)}$ is the $d$-dimensional Dirac delta function.
The subscript $x$ of the Laplacians implies that they act on the $x$ coordinates.
The normalization factor $\k$ is not specified by \eqref{gff free action}.
We will choose the normalization convention 
such that $\<\phi_a(x)\phi_b(y)\>$ is unit normalized in the far-from-boundary limit $\xn,\yn\->\infty$.
On the other hand, 
the complete functional form of the two-point function is fixed by boundary conformal symmetry
\be\label{2pt k}
\<\phi_a(x)\phi_b(y)\>=\frac{\d_{ab}F(\x)}{\(4\xn\yn\)^{\D_\phi}}
\,.
\ee
Note that $\xi$ is the conformally invariant variable defined in \eqref{xi definition}.
For $x\neq y$, we have $\x\neq 0$ and the equation of motion \eqref{equation of motion} becomes an ordinary differential equation.
For example, the case of $k=2$ corresponds to
\be
&\frac{1}{16}d(d-4)(d^2-4) F(\x)+\frac{1}{4} d(d^2-4)(2\x+1)F'(\x)+\frac{1}{4} d (d+2) (6 \xi  (\xi +1)+1) F''(\x)\nn
&+(d+2)\x(2\x+1)(\x+1)F^{(3)}(\x)+\xi ^2 (\xi +1)^2 F^{(4)}(\x)=0 \qquad(\x\neq 0)
\,.
\ee
More generally, for noncoincident points, the equation of motion is a $2k$th order homogeneous ordinary differential equation.
The general solution contains $2k$ free parameters:
\be\label{solution homogeneous}
F(\x)=\sum_{j=0}^{k-1}\(\frac{C_j}{\x^{\frac{d}{2}-k+j}}+\frac{B_j}{(\x+1)^{\frac{d}{2}-k+j}}\)
\,.
\ee
The coefficients $C_j$ are related to the singular behavior in the coincidence limit $y\->x$. 
The $1/\x^{\frac{d}{2}-k+j}$ terms with $j>0$ are inconsistent with the delta function behavior.  
Therefore, the contact term in \eqref{equation of motion} implies that only $C_0$ is nonzero, 
i.e., $C_j=0$ for $j>0$. 
We choose the normalization factor $\k$ such that
\be
C_0=1
\,, 
\ee
which corresponds to $\k={(-1)^k 2^{2k}\p^{d/2}\G(k)}/{\G(\frac{d-2k}{2})}$. 

The remaining coefficients $B_j$ are determined by the conformal boundary conditions. 
The general expression of a boundary fundamental primary is \cite{Chalabi:2022qit}
\be\label{boundary primary}
\Ph[a,q_i,1](\xp,0)=\lim_{\xn\->0}\sum_{j=0}^{\lfloor\frac{q_i}{2}\rfloor}\frac{(-1)^j}{2^j}\frac{q_i!}{j!(q_i-2j)!}\frac{(2k-2q_i-1)!!}{(2k+2j-2q_i-1)!!}\Box_\parallel^{2j}\pa_\perp^{q_i-2j}\phi_a(\xp,\xn)
\,.
\ee
As mentioned around \eqref{pair}, these boundary fundamental primaries form $k$ pairs of operators.
The conformal boundary conditions are specified by 
the requirement that only one operator is nonvanishing in each pair.
In \cite{Chalabi:2022qit}, the coefficients $B_j$ were solved explicitly for the concrete case of $k=2$, 
which has four distinct conformal boundary conditions. 
\footnote{The very recent work \cite{Herzog:2024zxm} was based on these $k=2$ solutions. }

We remind the reader that
one of our goals is to derive the general-$k$ results for the $\phi^{2n}$ deformations, 
such as the leading anomalous dimension of the boundary fundamental primaries,  
which will involve the general-$k$ solutions for the coefficients $B_j$. 
At first sight, this may seem to be a formidable task,   
as the number of possible boundary conditions, i.e., $2^k$, grows rapidly with $k$. 
Nevertheless, we manage to achieve this task.
In \eqref{free coefficients}, we will present our new result on the general-$k$ solutions for the coefficient $B_j$.

As in the $k=1$ case, we also introduce the variable $\s_i$ to label the choice of the $(i+1)$-th boundary condition:
\be
\s_i=\begin{cases}
	\; +1 \quad & \text{(N)} \\
	\; -1  & \text{(D)}\,.
\end{cases}
\ee
In terms of $\s_i$, the two possible choices for the $(i+1)$th pair become more symmetric. 
In accordance with \cite{Chalabi:2022qit},  
N means that the boundary primary of derivative order $(2k-1-i)$ vanishes,
while D implies that the boundary primary of derivative order $i$ is set to zero.
These labels generalize the Neumann and Dirichlet boundary conditions in the canonical case with $k=1$.
The variables $\s_i$ are in one-to-one correspondence with the nonvanishing boundary fundamental primaries $\Ph[a,q_i,1]$. 
Accordingly, the derivative order of the $(i+1)$th physical operator is given by 
\be
q_i=i\frac{1+\s_i}{2}+(2k-1-i)\frac{1-\s_i}{2}
\,.
\ee
For example, the NN boundary condition at $k=2$ is associated with $(\s_0,\s_1)=(1,1)$.
Accordingly, the nonvanishing operators are $\{\Ph[a,0,1], \Ph[a,1,1]\}$, 
and the conformal boundary conditions are given by $\Ph[a,3,1]=\Ph[a,2,1]=0$.

In terms of the variables $\{\s_0,\s_1,\ldots,\s_{k-1}\}$, 
it turns out that 
the solutions for $B_j$ in \eqref{solution homogeneous}   
have a nice general expression
\be\label{free coefficients}
B_j=\;\(\frac{d-2k}{2}\)_j\,\sum _{i=0}^{j}\frac{(2i-2k+1)(1-k)_j}{i!(j-i)!(i-2k+1)_{j+1}}\,\s_i
\,,
\ee
which only depends on the first $j+1$ boundary conditions. 
\footnote{The general solution \eqref{free coefficients} exhibits some interesting structures.
In \eqref{solution homogeneous}, if we set all $C_j$ including $C_0$ to zero, the coefficients $B_j$ can be chosen such that $2k-2$ sets of boundary conditions are satisfied simultaneously.
The remaining ones correspond to the $(i+1)$th pair of conjugate boundary conditions.
All $B_j$ can be fixed in this way up to normalization.
Furthermore, together with a $C_0=1$ term, 
we can determine the normalization of $B_j$ by requiring that the $(i+1)$th boundary condition is satisfied, i.e., $\Ph[a,2k-1-i,1]=0$ for N or $\Ph[a,i,1]=0$ for D.
It turns out that the corresponding normalization factors differ by a sign, 
so they can be encoded in $\s_i$. 
In this way, we can deduce the same coefficients of $\s_i$ as in \eqref{free coefficients}. 
The final result \eqref{free coefficients} can be viewed as a superposition of these solutions.
}
Based on this result, we will derive the general expression of free correlation functions, 
which are the basic ingredients of the free-theory input.

As an aside, we briefly comment on the NDND$\ldots$ and DNDN$\ldots$ boundary conditions.
In these two cases, the bulk-bulk two-point function $\<\phi_a(x)\phi_b(y)\>$ simplifies to
\be
F(\x)&=\frac{1}{\x^{\frac{d}{2}-k}}+\frac{1}{(\x+1)^{\frac{d}{2}-k}}\quad (\text{NDND}\ldots)\,,\\
F(\x)&=\frac{1}{\x^{\frac{d}{2}-k}}-\frac{1}{(\x+1)^{\frac{d}{2}-k}}\quad (\text{DNDN}\ldots)
\,,
\ee
where $B_j=0$ for $j\geq 1$. 
They resemble the cases of Neumann and Dirichlet boundary conditions 
\eqref{free bulk-bulk} in the canonical case.
Besides $C_0$, there is only one nonzero coefficient $B_0$ in the two-point function \eqref{2pt k}.
In \cite{Chalabi:2022qit}, the two sets of boundary conditions $\text{NDND}\ldots$ and $\text{DNDN}\ldots$ are called the generalized Neumann and generalized Dirichlet boundary conditions.

\subsubsection*{Correlation functions}

To deduce the bulk one-point function of $(\phi^2)_{ab}$, 
we take the coincident point limit $y\->x$, i.e., $\x\->0$, of the bulk-bulk two-point function \eqref{2pt k} with \eqref{solution homogeneous}. 
Discarding the singular term associated with $C_0$, we obtain
\be\label{1pt}
\<(\phi_a\phi_b)(\xp,\xn)\>=\frac{\d_{ab}\sum_{j=0}^{k-1}B_j}{\(2\xn\)^{2\D_{\phi}}}
\,.
\ee
The summation of the coefficients $B_j$ in \eqref{1pt} 
generalizes the $k=1$ case in \eqref{canonical-phiphi-1pt} 
with only one coefficient $\s_0=B_0$. 
According to \eqref{2pt k} and \eqref{boundary primary}, the bulk-boundary two-point function is
\be\label{bulk boundary free}
\<\phi_a(\xp,\xn)\Ph[b,q_i,1](\yp,0)\>&=\frac{2(2k-q_i)_{q_i}\(\frac{d-2k}{2}\)_{q_i}}{\(k-q_i+\frac{1}{2}\)_{q_i}}\frac{\d_{ab}}{\(|\xp-\yp|^2+\xn^2\)^{\Dh[q_i,1]}\xn^{\D_\phi-\Dh[q_i,1]}}
\,,
\ee
where we have taken the boundary limit $\yn\->0$.
According to \eqref{bulk boundary free}, 
the BOE of $\phi_a$ contains the boundary primaries $\Ph[a,q_0,1],\,\Ph[a,q_1,1],\,\ldots,\,\Ph[a,q_{k-1},1]$.
We can also read off the BOE coefficients $b_{\phi\Ph[,q_i,1]}$ from the correlator \eqref{bulk boundary free}:
\be 
b_{\phi\Ph[,q_i,1]}=\frac{2(2k-q_i)_{q_i}\(\frac{d-2k}{2}\)_{q_i}}{\(k-q_i+\frac{1}{2}\)_{q_i}}
\,.
\ee
According to the definition of $\Ph[a,q_i,1]$ in \eqref{boundary primary}, 
we can derive the boundary-boundary two-point function 
from  the bulk-boundary correlator \eqref{bulk boundary free} 
by taking parallel and perpendicular derivatives and the boundary limit $\xn\->0$: 
\be\label{boundary boundary free}
\<\Ph[a,q_j,1](\xp,0)\Ph[b,q_i,1](\yp,0)\>&=\frac{2q_i!(2k-q_i)_{q_i}\(\frac{d-2k}{2}\)_{q_i}}{\(k-q_i+\frac{1}{2}\)_{q_i}}\frac{\d_{ij}\d_{ab}}{|\xp-\yp|^{2\Dh[q_i,1]}}
\,,
\ee
where $\d_{ij}$ implies that the boundary-boundary two-point functions of different boundary fundamental primaries vanish.
Since the exponent $\D_\phi-\Dh[q_i,1]$ in \eqref{bulk boundary free} is a negative integer, we need to take at least a $q_i$th order derivative with respect to $\xn$, i.e., $q_j \geq q_i$, otherwise the $\xn\->0$ limit is vanishing. 
On the other hand, the correlator $\<\Ph[a,q_j,1](\xp,0)\Ph[b,q_i,1](\yp,0)\>$ can also be derived from $\<\Ph[a,q_j,1](\xp,0)\phi_b(\yp,0)\>$, and then we have $q_i \geq q_j$.
Therefore, the boundary-boundary two-point functions are nonzero only when $q_i=q_j$.
The factorial $q_i!$ in \eqref{boundary boundary free} follows from applying the normal derivative $q_i$ times. 

Using the boundary fundamental primaries, we can construct the boundary composite operators.
As a generalization of the canonical case in section \ref{Boundary composite operators}, we will also study the O($N$) singlet $\Ph[,q_i,2p]$ and vector $\Ph[a,q_i,2p+1]$ in the deformed theory in section \ref{Boundary composite operators k>1}.
We will investigate the leading corrections to the free bulk-boundary-boundary three-point function
\be\label{conformal block expansion free}
&\<\phi_a(\xp,\xn)\Ph[,q_i,2p](0,0)\Ph[b,q_i,2p+1](\yp,0)\>\nn=\;&\frac{\d_{ab}|\yp|^{-2\Dh[q_i,2p+1]}}{|x|^{\dD}\xn^{\D_\phi}}\Big[b_{\phi\tPh[,q_i,1]}f^{(p)}_{\tPh[,q_i,1]}G^{(p)}_{\tPh[,q_i,1]}(v)\Big]+O(|\yp|^{-2\Dh[q_i,2p+1]+1})
\,.
\ee
In the free-theory BOE of $\phi_a$, only the boundary fundamental operators associated with the same $q_i$ as the external operators appear in the boundary OPE of $\Ph[,q_i,2p]\times\Ph[a,q_i,2p+1]$.
Therefore, there is only one term in the free-theory conformal block expansion above.
Another useful three-point function is
\be
&\<\phi_a(\xp,\xn)\Ph[b,q_i,2p-1](0,0)\Ph[,q_i,2p](\yp,0)\>\nn=\;&\frac{\d_{ab}|\yp|^{-2\Dh[q_i,2p]}}{|x|^{\dD}\xn^{\D_\phi}}\bigg[b_{\phi\tPh[,q_i,1]}f^{\prime(p)}_{\tPh[,q_i,1]}G^{\prime(p)}_{\tPh[,q_i,1]}(v)\bigg]+O(|\yp|^{-2\Dh[q_i,2p]+1})
\,.
\ee
The primes indicate that the boundary OPE coefficient $f'$ and conformal block $G'(v)$ are different from those in \eqref{conformal block expansion free}, since the external operators are different.

\subsection{Multiplet recombination}
\label{Multiplet recombination k>1}

Now that we have the free data, we want to generalize the analysis in section \ref{Multiplet recombination} to higher derivative theories.
The constraint from the bulk multiplet recombination reads
\be\label{recombination}
\lim_{\e\->0}\a^{-1}\Box^k\phi_a=\phi^{2n-1}_{a,\f}
\,,
\ee
which is the $k\geq 1$ generalization of \eqref{recombination k=1}.
Here $\a$ is given by \cite{Guo:2023qtt}
\be\label{alpha}
\a=\frac{(-1)^{k+1}2^{2k-n}(n-1)(k-1)!\(\frac{k}{n-1}+1\)_{k}}{(n)_{n}\,_{3}F_{2}\[\begin{matrix}\frac{1}{2}-\frac{n}{2},-\frac{n}{2},1-n-\frac{N}{2}\\ 1\;,\;\frac{1}{2}-n\end{matrix};1\]}\,\e+O(\e^2)
\,.
\ee
Again, we assume that the bulk multiplet recombination remains unaffected by the boundary.

\subsubsection{Boundary fundamental operators}

The generalization of \eqref{matching bulk boundary k=1} reads
\be\label{matching fundamental k>1}
\lim_{\e\->0}\(\a^{-1}\<\Box^k\phi_a(\xp,\xn)\Ph[b,q_i,1](\yp,0)\>\)=\<\phi^{2n-1}_a(\xp,\xn)\Ph[b,q_i,1](\yp,0)\>_\f
\,.
\ee
On the left-hand side, $\<\phi_a(\xp,\xn)\Ph[b,q_i,1](\yp,0)\>$ takes the same form as \eqref{bulk boundary 2pt}.
To derive the action of $\Box^k$ on this correlator, 
we interpret the $k=1$ result in \eqref{matching condition explicit} as a sum of two bulk-boundary two-point functions.
Applying the action of $\Box$ once, the bulk and boundary effective scaling dimensions changes as
\be 
\D&\->\D+2,\quad  \Dh[]\->\hat{\D}\,,
\label{20} \\
\D&\->\D+1,\quad  \Dh[]\->\hat{\D}+1
\,.
\label{11}
\ee 
The initial condition is given by $\D=\D_\phi$ and $\hat{\D}=\D_{q_i,1}$. 
By repeatedly using the $k=1$ formula, we derive the explicit constraint for general $k$
\be\label{matching explicit k}
&\lim_{\e\->0}\Bigg[\a^{-1}\sum_{l=0}^k\Bigg(\;\frac{(-1)^l 4^{l}k!}{(k-l)!\,l!}\frac{(\D_\phi-\Dh[q_i,1])_{2k-2l}\,(\Dh[q_i,1])_l\,(-\g_\phi)_l\,b_{\phi\Ph[,q_i,1]}\d_{ab}}{\(|\xp-\yp|^2+\xn^2\)^{\Dh[q_i,1]+l}\xn^{\D_\phi-\Dh[q_i,1]+2k-2l}}\,\Bigg)\Bigg]\nn
=\;&\frac{b_{\phi^{2n-1}_\f\Ph[\f,q_i,1]}\d_{ab}}{\(|\xp-\yp|^2+\xn^2\)^{\Dh[q_i,1,\f]}\xn^{(2n-1)\D_{\phi_\f}-\Dh[q_i,1,\f]}}
\,,
\ee
where $\g_\phi=\D_\phi-\frac{d-2k}{2}$ is the anomalous dimension of $\phi_a$.
On the left-hand side, the summand corresponds to using \eqref{11} $l$ times and \eqref{20} $k-l$ times.
The sum over all $\frac{k!}{(k-l)!\,l!}$ possible orderings leads to the factorization as in \eqref{matching explicit k}.
Unexpectedly, they organize into a factorized form with the factor $(-\g_\phi)_l$,
which is not manifest in the individual terms.
For example, the coefficient of the $l=1$ term at $k=2$ is given by
\be 
&(\D_\phi-\Dh[q_i,1])(\D_\phi-\Dh[q_i,1]+1)\times 4\Dh[q_i,1]\Big(\D_\phi+2-\frac{d-2}{2}\Big)\nn
&+4\Dh[q_i,1]\Big(\D_\phi-\frac{d-2}{2}\Big)\times(\D_\phi-\Dh[q_i,1])(\D_\phi-\Dh[q_i,1]+1)\nn
=\;&8\Dh[q_i,1]\Big(\D_\phi-\frac{d-4}{2}\Big)(\D_\phi-\Dh[q_i,1])(\D_\phi-\Dh[q_i,1]+1)
\,,
\ee
where $\D_\phi-\frac{d-4}{2}=\g_\phi$ for $k=2$ on the last line.

For general $k$, The factor $(-\g_\phi)_l$ with $l>0$ is proportional to the bulk anomalous dimension $\g_\phi=O(\e^2)$.
As a generalization of \eqref{gamma phi k=1}, the anomalous dimension $\g_\phi$ starts at order $\e^2$ for $k>1$ as well.
Since $\a$ is of order $\e^1$, the terms with $l>0$ do not contribute to the left-hand side of \eqref{matching explicit k}.
So we only need to consider the $l=0$ term.
Then, the functional forms in \eqref{matching explicit k} match as $\lim_{\e\->0}(\D_{\phi}+2k)=(2n-1)\D_{\phi_\f}$. 
The free coefficient $b_{\phi^{2n-1}_\f\Ph[\f,q_i,1]}$ on the right-hand side is derived from Wick contractions:
\be
b_{\phi^{2n-1}_\f\Ph[\f,q_i,1]}=2^{n-2k-1}\,b_{\phi_\f\Ph[\f,q_i,1]}\(\frac{N}{2}+1\)_{n-1}\(\sum_{j=0}^{k-1}B_j\)^{n-1}
\,. 
\ee
As a generalization of \eqref{free coefficient b k=1}, 
the boundary-condition-dependent factor $\s$ is replaced by $\sum_{j=0}^{k-1}B_j$. 
The anomalous dimensions of the boundary fundamental operators are defined by
\be
\gh[q_i,1]=\Dh[q_i,1]-\(\frac{d-2k}{2}+q_i\)
\,.
\ee
The matching condition \eqref{matching explicit k} becomes
\be
(-1)^{q_i+1}q_i!(2k-1-q_i)!\lim_{\e\->0}\(\a^{-1}\gh[q_i,1]\)
=2^{n-2k-1}\Big(\frac{N}{2}+1\Big)_{n-1}\Big(\sum_{j=0}^{k-1}B_j\Big)^{n-1}
\,.
\ee
This equation has some symmetry properties.
The factor $(-1)^{q_i+1}q_i!(2k-1-q_i)!$ on the left-hand side changes sign under $q_i\lra 2k-1-q_i$.
The factor $(\sum_{j=0}^{k-1}B_j)^{n-1}$ on the right-hand side yields a factor $(-1)^{n-1}$ under N$\lra$D.
The general solution for the anomalous dimensions of boundary fundamental operators is
\be\label{gamma fundamental}
\gh[q_i,1]=(-1)^{q_i+1}\frac{2^{n-2k-1}\(\frac{N}{2}+1\)_{n-1}\(\sum_{j=0}^{k-1}B_j\)^{n-1}}{q_i!(2k-1-q_i)!}\a+\ldots
\,,
\ee 
where the expressions of $B_j$ and $\a$ can be found in \eqref{free coefficients} and \eqref{alpha}.
The ellipsis denotes higher order terms in $\e$.
During the symmetry properties mentioned above, 
the fundamental anomalous dimension $\gh[q_i,1]$ in one theory is equal to $(-1)^{n}\gh[2k-1-q_i,1]$ in the other theory where N and D are exchanged.
The solution \eqref{gamma fundamental} generalizes the $k=1$ result in \eqref{gamma Phi}. 
In addition, the $N\->0$ limit of \eqref{gamma fundamental} reads
\be\label{polymer k>1}
\gh[q_i,1]\big|_{N=0}=(-1)^{q_i+k}\frac{ k! \, n! \left(\frac{k}{n-1}+1\right)_{k-1} \(\sum _{j=0}^{k-1} B_j\)^{n-1}}{2q_i!(2k-1-q_i)! (n)_n \, {}_3F_{2}\!\[\begin{matrix}\frac{1-n}{2},-\frac{n}{2},1-n\\ 1\;,\;\frac{1}{2}-n\end{matrix};1\]}\e+\ldots
\,,
\ee 
which is the higher derivative generalization of the result \eqref{polymer} for polymer systems.

In addition, let us examine the deformed boundary conditions at $k=2$.
Consider the bulk-boundary two-point function
\be 
\<\phi_a(\xp,\xn)\Ph[b,q_i,1](\yp,0)\>&=\frac{\d_{ab}\,b_{\phi\Ph[,q_i,1]}\xn^{q_i+\gh[q_i,1]-\g_\phi}}{\(|\xp-\yp|^2+\xn^2\)^{\Dh[q_i,1]}}
\,. \label{DD1 k>1}
\ee
The normal derivative with respect to $x$ gives 
\be
&\<\pa_\perp\phi_a(\xp,\xn)\Ph[b,q_i,1](\yp,0)\>\nn
=\;&\d_{ab}\,b_{\phi\Ph[,q_i,1]}\[\frac{(-q_i-\gh[q_i,1]+\g_\phi)\xn^{q_i-1+\gh[q_i,1]-\g_\phi}}{\(|\xp-\yp|^2+\xn^2\)^{\Dh[q_i,1]}}-\frac{2\Dh[q_i,1]\,\xn^{q_i+\gh[q_i,1]-\g_\phi}}{\(|\xp-\yp|^2+\xn^2\)^{\Dh[q_i,1]+1}}\] 
\,. \label{DD2 k>1}
\ee
To order $\e^1$, we can assume that the boundary fundamental operators $\Ph[a,0,1]$ and $\Ph[a,1,1]$ are null in the DD case.
So $q_i$ is at least $2$, implying that \eqref{DD1 k>1} and \eqref{DD2 k>1} vanish in the $\xn\->0$ limit.
For boundary composite operators, the corresponding bulk-boundary two-point functions and their normal derivatives are of higher order in $\xn$ than \eqref{DD1 k>1} and \eqref{DD2 k>1}, so they vanish in the $\xn\->0$ limit as well.
It seems that the DD boundary condition holds to order $\e^1$.
This may be seen as a generalization of \eqref{D k=1} at $k=1$.
For the NN case, we examine the higher order derivatives in accordance with \eqref{boundary primary 32 k=2}
\be
&\<(\pa_\perp^2-\Box_\parallel)\phi_a(\xp,\xn)\Ph[b,q_i,1](\yp,0)\> \nn
=\;&\d_{ab}\,b_{\phi\Ph[,q_i,1]}\Bigg[\frac{(-q_i-\gh[q_i,1]+\g_\phi)_2 \, \xn^{q_i-2+\gh[q_i,1]-\g_\phi}}{\(|\xp-\yp|^2+\xn^2\)^{\Dh[q_i,1]}}\nn
&\hspace{5em}+\frac{2\Dh[q_i,1](2\D_\phi-4\Dh[q_i,1]+d-4) \xn^{q_i+\gh[q_i,1]-\g_\phi}}{\(|\xp-\yp|^2+\xn^2\)^{\Dh[q_i,1]+1}}+O(\xn^{q_i+2+\gh[q_i,1]-\g_\phi})\Bigg]\,, \label{NN1 k>1}\\
&\<(\pa_\perp^3+3\pa_\perp\Box_\parallel)\phi_a(\xp,\xn)\Ph[b,q_i,1](\yp,0)\> \nn
=\;&\d_{ab}\,b_{\phi\Ph[,q_i,1]}\Bigg[\frac{(-q_i-\gh[q_i,1]+\g_\phi)_3\, \xn^{q_i-3+\gh[q_i,1]-\g_\phi}}{\(|\xp-\yp|^2+\xn^2\)^{\Dh[q_i,1]}} \nn
&-\frac{6(-q-\gh[q_i,1]+\g_\phi)\Dh[q_i,1](\D_\phi+\Dh[q_i,1]-d+3) \xn^{q_i-1+\gh[q_i,1]-\g_\phi}}{\(|\xp-\yp|^2+\xn^2\)^{\Dh[q_i,1]+1}}+O(\xn^{q_i+1+\gh[q_i,1]-\g_\phi})\Bigg]
\,. \label{NN2 k>1}
\ee 
For $q_i\in\{0,1\}$, the right-hand sides above are singular at order $\e^1$ in the boundary limit $\xn\->0$. 
We conclude that the interacting theory is not associated with the NN boundary condition. 

For the ND case, the $\xn\->0$ limit of \eqref{NN2 k>1} and \eqref{DD2 k>1} vanishes at order $\e^0$, i.e., $\Ph[a,1,1]$ and $\Ph[a,3,1]$ are null in the free theory.
In the deformed theory, the $\e^\th$ expansion with $0<\th<1$ arises, and the BOE coefficients $b_{\phi\Ph[,1,1]}$ and $b_{\phi\Ph[,3,1]}$ may be nonzero at order $\e^\th$.
(See the second half of section \ref{Boundary composite operators k>1}.)
If $b_{\phi\Ph[,3,1]}$ is nonzero at order $\e^\th$, then the $q_i=3$ case of \eqref{NN2 k>1} leads to a nonzero term in the $\xn\->0$ limit, so the N condition in the ND case does not hold at order $\e^\th$.
If $b_{\phi\Ph[,1,1]}$ is nonzero at order $\e^\th$, then the D condition in the ND case does not hold.
This can be seen by substituting $q_i=1$ into \eqref{DD2 k>1}, which corresponds to a nonvanishing term in the $\xn\->0$ limit at order $\e^\th$.

For the DN case, the null boundary fundamental primaries are $\Ph[a,0,1]$ and $\Ph[a,2,1]$ in the free theory, which corresponds to the vanishing $\xn\->0$ limit of \eqref{DD1 k>1} and \eqref{NN1 k>1} at order $\e^0$.
Similar to the ND case above, here the $\e^\th$ expansion also arises in the deformed theory.
At order $\e^\th$, the BOE coefficients $b_{\phi\Ph[,0,1]}$ and $b_{\phi\Ph[,2,1]}$ may be nonzero.
Supposing that $b_{\phi\Ph[,0,1]}$ is nonzero at order $\e^\th$, the D condition in the DN case no longer holds since the $\xn\->0$ limit of \eqref{DD1 k>1} has a nonvanishing term at order $\e^\th$ for $q_i=0$.
If $b_{\phi\Ph[,2,1]}$ is nonzero at order $\e^\th$, then the N condition associated with \eqref{NN1 k>1} does not hold as the $\xn\->0$ limit leads to a nonzero term for $q_i=2$.

\subsubsection{Boundary composite operators}
\label{Boundary composite operators k>1}

As in section \ref{Boundary composite operators}, we consider the boundary composite operators as well.
We will use the matching conditions and the analyticity requirement to compute their anomalous dimensions.
We also encounter the noninteger power $\e$ expansion, but two scenarios appear for $k>1$.
The first one arises due to the consistency requirement of the matching conditions.
In some cases, this may be understood as a direct generalization of the $\e^{1/2}$ expansion in the $k=1$ case, where  
the bulk interaction $\phi^{2n}$ mixes with the boundary interaction $\pa^{k-1}\phi^{n+1}|_{\xn=0}$. 
In the second scenario, the noninteger power $\e$ expansion is introduced to meet the analyticity requirement for the bulk-boundary-boundary three-point functions, whose physical interpretation is less clear to us.

\subsubsection*{Integer power $\e$ expansion}

To be more concrete, below we mainly consider a specific example, namely the NN case of the $(k,n)=(2,2)$ theory.
In this theory, there are two physical boundary fundamental operators, i.e., $i=0,1$, and their derivative orders are $(q_0,q_1)=(0,1)$. 
We assume that the deformed data is given by integer power $\e$ expansion.
One of the novel features is the following. 
The conformal block expansion of the three-point function $\<\phi_a\Ph[,q_i,2p]\Ph[b,q_i,2p+1]\>$ contains the contribution from the other boundary fundamental operator $\Ph[a,q_{1-i},1]$ at order $\e^1$.
In other words, the product $b_{\phi\tPh[,q_{1-i},1]}f_{\Ph[,q_{1-i},1]}^{(p)}$ is of order $\e^1$.
Since the BOE coefficient $b_{\phi\tPh[,q_{1-i},1]}$ is of order $\e^0$, this means that the boundary OPE coefficient $f_{\Ph[,q_{1-i},1]}^{(p)}$ acquires a correction at order $\e^1$ in the deformed theory. 

The conformal block expansion \eqref{conformal block expansion free} receives corrections as follows:
\be\label{k=2 conformal block expansion}
&\<\phi_a(\xp,\xn)\Ph[,q_i,2p](0,0)\Ph[b,q_i,2p+1](\yp,0)\>=\frac{\d_{ab}|\yp|^{-2\Dh[q_i,2p+1]}}{|x|^{\dD}\xn^{\D_\phi}}\Bigg[b_{\phi\tPh[,q_i,1]}f^{(p)}_{\tPh[,q_i,1]}G^{(p)}_{\tPh[,q_i,1]}(v)\nn
&+b_{\phi\tPh[,q_{1-i},1]}f^{(p)}_{\tPh[,q_{1-i},1]}G^{(p)}_{\tPh[,q_{1-i},1]}(v)+\sum_{m=0}^{\infty}b_{\phi\tilde\Psi^{(q_i,3,m)}}f^{(p)}_{\tilde\Psi^{(q_i,3,m)}}G^{(p)}_{\tilde\Psi^{(q_i,3,m)}}(v)+O(\e^2)\Bigg]\nn
&+O(|\yp|^{-2\Dh[q_i,2p+1]+1})
\,.
\ee
In the Gaussian limit, only the first term in the square bracket is nonzero, and the rest of the terms vanish.
As in \eqref{3pt conformal block expansion}, we need to account for the contribution from infinitely many boundary composite operators $\Psi_a^{(q_i,3,m)}\sim \Box_\parallel^m(\Ph[a,q_i,1])^{3}$ in the interacting theory.
Their scaling dimensions are
\be
\hat{\D}_{q_i,3,m}=3\(\frac{d-4}{2}+q_i\)+2m+O(\e)
\,.
\ee
In addition, the boundary fundamental operator $\Ph[a,q_{1-i},1]$ contributes to the conformal block expansion, as mentioned above. 

Using the explicit expression for the conformal blocks in \eqref{conformal block},
we find that the terms in the square bracket in \eqref{k=2 conformal block expansion} are nonanalytic in $1-v$ at order $\e^1$.
The contribution from $\Ph[a,q_i,1]$ has a correction due to anomalous dimensions, yielding a generalized hypergeometric function of the ${}_3F_2$ type:
\be\label{composite block k>1}
G^{(p)}_{\tPh[,q_i,1]}(v)=v^{\Dh[q_i,1]/2}\(1+\frac{(q_i+2)(\gh[q_i,1]+\gh[q_i,2p]-\gh[q_i,2p+1])v}{2q_i-1}\,{}_3F_2\[\begin{matrix}1,1,3+q_i\\ 2,\frac{1}{2}+q_i\end{matrix};v\]+O(\e^2)\)
\,.
\ee
As a generalization of \eqref{composite matching}, the coefficients $b_{\phi\tilde\Psi^{(q_i,3,m)}}f^{(p)}_{\tilde\Psi^{(q_i,3,m)}}$ can be expressed in terms of $\a$ using the matching condition
\be\label{k=2 composite matching}
\lim_{\e\->0}\(\a^{-1}\<\Box^2\phi_a(\xp,\xn)\tilde\Psi^{(q,3,m)}_{b}(\yp,0)\>\)=\<\phi^3_a(\xp,\xn)\tilde\Psi^{(q,3,m)}_{b}(\yp,0)\>_\f
\,.
\ee
Then, the infinite sum over $m$ leads to 
\be\label{composite block 2 k>1}
\sum_{m=0}^{\infty}b_{\phi\tilde\Psi^{(q_i,3,m)}}f^{(p)}_{\tilde\Psi^{(q_i,3,m)}}G^{(p)}_{\tilde\Psi^{(q_i,3,m)}}(v)=\frac{\a\,W_{3,p}}{24(1+34q_i)}\,v^{3+\frac{3q_i}{2}}\,{}_3F_{2}\!\[\begin{matrix}2,2+q_i,4+2q_i\vspace{0.3em}\\ \frac{5}{2}+\frac{3q_i}{2},3+\frac{3q_i}{2}\end{matrix};v\]+O(\e)
\,,
\ee
which has been simplified using $q_i\in\{0,1\}$. 
In contrast to the canonical case $k=1$, the nonanalytic terms given by the two ${}_3F_2$ functions in \eqref{composite block k>1} and \eqref{composite block 2 k>1} cannot cancel.
In order for the correlator to be analytic around $v=1$, we need to take into account the contribution from $\Ph[a,q_{1-i},1]$, which is of the form 
\be\label{fundamental block k>1}
G^{(p)}_{\tPh[,q_{1-i},1]}(v)=v^{1+\frac{q_{1-i}}{2}}\,{}_2F_1\!\(\frac{q_{1-i}-q_i}{2},\frac{q_{1-i}+q_i+4}{2};q_{1-i}-\frac{1}{2},v\)
\,.
\ee
The nonanalytic terms in \eqref{composite block k>1}, \eqref{composite block 2 k>1} and \eqref{fundamental block k>1} can be found using \eqref{3F2 expansion} and \eqref{2F1 expansion}. 
Then, we require that the terms with noninteger powers of $1-v$ cancel, and obtain the constraints on the anomalous dimensions and the product $b_{\phi\tPh[,q_{1-i},1]}f^{(p)}_{\tPh[,q_{1-i},1]}$:
\be\label{k=2 NN recursion}
\gh[q_i,1]+\gh[q_i,2p]-\gh[q_i,2p+1]&=\frac{5-2q_i}{20}p\,\a +O(\e^2)\,,\\
b_{\phi\tPh[,q_{1-i},1]}f^{(p)}_{\tPh[,q_{1-i},1]}&=3\p\times 2^{4p-9}(4-3q_{1-i})^{2p+2} (35q_{1-i}-3)p^2 (p-1)!\,(\tfrac{N}{2}+1)_p\;\a+O(\e^2)
\,.
\ee
As opposed to the canonical case $k=1$, we obtain two constraints due to the following reason.
For the nonanalytic part, the subleading term in $1-v$ does not cancel automatically once the leading term cancels.
If the leading and subleading terms in $1-v$ vanish, the higher order terms in $1-v$ become zero automatically and do not yield additional constraints.
Likewise, for the correlator $\<\phi_a(\xp,\xn)\Ph[a,2,2p-1](0,0)\Ph[,2,2p](\yp,0)\>$, we obtain
\be
\gh[q_i,1]+\gh[q_i,2p-1]-\gh[q_i,2p]&=\frac{N+6 p-4}{8(3+2q_i)}\a+O(\e^2)\,,\\
b_{\phi\tPh[,q_{1-i},1]}f^{\prime(p)}_{\tPh[,q_{1-i},1]}=\p\times2^{4p-7}&(2-q_{1-i})^{4p-7} (49q_{1-i}-48)(N+6p-4)\,p!(\tfrac{N}{2}+1)_{p-1}\;\a+O(\e^2)
\,.
\ee
Together with \eqref{k=2 NN recursion}, 
we derive the solutions for the anomalous dimensions
\be\label{k=2 composite vector}
\gh[q_i,2p+1]&=(2p+1)\gh[q_i,1]-\frac{5 (N+6 p+2)}{40 (3+2
   q_i)}p\,\a+O(\e^2)\,,\\
\label{k=2 composite singlet}
\gh[q_i,2p]&=\gh[q_i,2p+1]-\gh[q_i,1]+\frac{5-2q_i}{20}p\,\a+O(\e^2)
\,,
\ee
where $\a$ and $\gh[q_i,1]$ are given by \eqref{alpha} and \eqref{gamma fundamental} with $n=2$.

The above derivation of the composite anomalous dimensions extends straightforwardly to the DD case.
We obtain
\be
\gh[q_i,2p+1]&=(2p+1)\gh[q_i,1]+\frac{p}{48} \[22 (N-4) (q-2)+9 (q-3) (N+6p+2)\] \a +O(\e^2)\,, \label{k=2 composite vector DD} \\
\gh[q_i,2p]&=\gh[q_i,2p+1]-\gh[q_i,1]+\frac {13q_i-17}{8} p\,\a+O(\e^2)
\,, \label{k=2 composite singlet DD}
\ee
where $q_i\in\{2,3\}$.
The procedure can be further extended to higher $n$.
The difference is that more infinite sums contribute at order $\e^1$ for $n>2$.
More specifically, there are $n-1$ infinite sums associated with $\Psi^{(q_i,2r-1,m)}$ blocks, where $r=2,3,\ldots,n$.
The increase in $n$ leads to more complicated expressions for the composite anomalous dimensions.  
We do not present the explicit results for higher $n$. 

\subsubsection*{Noninteger-power $\e$ expansion}

An inconsistency arises for \eqref{k=2 composite matching} with odd $k$ and odd $n$ in the integer power $\e$ expansion.
The matching condition is 
\be\label{matching k}
\lim_{\e\->0}\(\a^{-1}\<\Box^k\phi_a(\xp,\xn)\Psi_{b}^{(q_i,2r-1,m)}(\yp,0)\>\)=\<\phi^{2n-1}_a(\xp,\xn)\Psi_b^{(q_i,2r-1,m)}(\yp,0)\>_\f
\,,
\ee
where $\Psi_{b}^{(q_i,2r-1,m)}\sim \Box_\parallel^m(\Ph[b,q_i,1])^{2r-1}$ is a scalar primary with the scaling dimension
\be
\hat{\D}_{q_i,2r-1,m}=(2r-1)\(\frac{d-2k}{2}+q_i\)+2m+\ldots
\ee
This is the $k\geq 1$ generalization of \eqref{Delta Psi}.
Here we assume $r>1$, so the BOE coefficient $b_{\phi\Psi^{(q_i,2r-1,m)}}$ vanishes at order $\e^0$.
For $r=1$, the $m=0$ case has already been considered in \eqref{matching fundamental k>1}, and the $m>0$ case corresponds to descendants. 
For a more complete analysis, we will keep $(k,n)$ and $(q_i,r,m)$ general, 
and find out the potentially problematic cases. \footnote{Note that we have not considered the possibility where the composite operator is constructed from more than one types of fundamental operators.}
As a generalization of \eqref{matching special} in the canonical case, 
we will show that for $2r-1=n$ the matching condition \eqref{matching k} is inconsistent with the integer power $\e$ expansion at odd $k$ and odd $n$. 

Replacing $\Dh[q_i,1]$ by $\Dh[q_i,2r-1,m]$ on the left-hand side of \eqref{matching explicit k}, we obtain the result for the action of the Laplacians in \eqref{matching k}.
Again, the leading-order term corresponds to $l=0$, and has the factor
\be\label{epsilon 1/2 k}
\(\D_\phi-\Dh[q_i,2r-1,m]\)_{2k}=\(-\frac{2k}{n-1}(r-1)-(2r-1)q_i-2m\)_{2k}+\ldots
\,,
\ee
As in \eqref{epsilon 1/2}, this factor can vanish at order $\e^0$. 
Suppose that this is the case. 
Note that $\a$ is of order $\e^1$ and the BOE coefficient associated with $\<\phi_a(x)\Psi_{b}^{(q_i,2r-1,m)}(y)\>$ does not have an order $\e^0$ term. 
If the perturbative corrections are positive integer power terms in $\e$, 
then the left-hand side of \eqref{matching k} does not have a finite Gaussian limit.
For $r\leq n$, this leads to a contradiction as the right-hand side of \eqref{matching k} is finite.
Therefore, the perturbative series should contain noninteger powers in $\e$ 
if \eqref{epsilon 1/2 k} vanishes at order $\e^0$.

In \eqref{epsilon 1/2 k}, the zeroth order term in $\e$ can vanish only when $\frac{2k(r-1)}{n-1}$ is an integer.
Since $k$ and $n-1$ have no common divisor and $r\leq n$, the only possibility is $2(r-1)=n-1$, implying that $2r-1=n$ and $n$ is odd.
\footnote{The case of $r=n$ leads to $(q_i-2m-2q_i r-2k)_{2k}$ in \eqref{epsilon 1/2 k}.
This Pochhammer symbol cannot be zero since $q_i-2m-2q_i r-2k\leq -2k$.}
This also indicates that $k$ should also be odd, 
otherwise $k$ and $n-1$ would have a common divisor, i.e., $2$.
Furthermore, the existence of a vanishing term at order $\e^0$ implies 
\be\label{q-m-condition}
q_i \leq \frac{k-1-2m}{n}\,,
\ee
so $q$ and $m$ should be sufficiently small. 
\footnote{For the operators with fewer derivatives, the corrections to the anomalous dimensions can be of lower order in $\e$.
This is reminiscent of the bulk $\phi^4$ theory.
There the anomalous dimensions of the bilinear operators $\cal J_{ab}\sim\phi_a\Box^{m'}\phi_b$ are of order $\e^2$ for $m'>0$, but the $m'=0$ case is of order $\e^1$.
}
We remind the reader that we assume that $k$ and $n-1$ have no common divisor throughout this paper. 
As in the $k=1$ case, $q_i=m=0$ is always a solution, 
but there are other possible choices.
For $k\geq 3$, the case $(q_i,m)=(0,1)$ also satisfies \eqref{q-m-condition}. 
For $k\geq n+1$, the case $(q_i,m)=(1,0)$ is also a solution. 
For $k=1$, the inconsistency of the integer power $\e$ expansion is noticed from the matching condition \eqref{matching special}, which involves the boundary composite operator $\Psi_{b}^{(0,n,0)}=\Ph[b,0,n]$.
For $k>1$, the inconsistency can be seen from the matching condition for $\<\phi_a\Psi_{b}^{(q_i,2r-1,m)}\>$ with higher $q_i$ and $m$, as indicated by \eqref{q-m-condition}. 
For a given pair $(k,n)$, the noninteger power $\e$ expansion appears if there is a physical boundary fundamental operator associated with
\be\label{epsilon 1/2 condition}
q_i \leq \frac{k-1}{n}
\,,
\ee
which corresponds to \eqref{q-m-condition} in the most dangerous case $m=0$. 

For $k=1$, the noninteger power $\e$ expansion is understood as a consequence of the mixing between the bulk interaction $\phi^{2n}$ and a $\bb Z_2$-even boundary interaction.
However, for $k>1$, some cases of \eqref{epsilon 1/2 condition} cannot be associated with such a mixing, 
as classically marginal boundary interactions cannot be constructed from the physical fundamental operators.
To see this, let us determine the schematic form of the classically marginal boundary interaction. 
The general schematic form for a boundary interaction is
\be
\int \rm d^{d-1}x \; \pa^{t_2}\phi^{t_1}|_{\xn=0}
\,,
\ee
where the integers $t_1$ and $t_2$ are the numbers of $\phi$'s and derivatives in the boundary interaction.
We assume that $t_1$ is even, so the boundary interaction is $\bb Z_2$ even.
The classical marginality implies
\be\label{marginal boundary}
t_1 \frac{d_\text{u}-2k}{2}+t_2=d_\text{u}-1
\,,
\ee
so we have $d_\text{u}=\frac{2(kt_1-t_2-1)}{t_1-2}$. 
Since the upper critical dimension is given by $\du=\frac{2nk}{n-1}$, 
the solutions of \eqref{marginal boundary} are associated with $t_1-2=(n-1)t_3$ and $kt_1-t_2-1=nk t_3$, 
and thus $t_1=2+(n-1)t_3$ and $t_2=2k-1-k t_3$. 
The physical solutions with $t_2\geq 0$ are $t_3=0,1$. Therefore, we have
\be
(t_1,t_2)=(2,2k-1), (n+1,k-1)\,.
\ee
The first case is associated with the bilinear boundary terms 
$S_{\text{bdy},\pm,\ldots,\pm}$ in \eqref{boundary terms-k}.
For $k>1$, we can use two different physical fundamental operators to construct the bilinear marginal operators.
The second case $h\pa^{k-1}\phi^{n+1}|_{\xn=0}$ is a $\bb Z_2$-even boundary interaction if $n$ is odd.
The mixing between this boundary interaction and the bulk interaction $g\phi^{2n}$ can lead to the noninteger power $\e$ expansion, 
generalizing the $k=1$ case associated with $h\phi^{n+1}|_{\xn=0}$.

Let us explain why the noninteger power $\e$ expansion cannot always be associated with the mixing between $\phi^{2n}$ and $\pa^{k-1}\phi^{n+1}|_{\xn=0}$.
The boundary interaction $\pa^{k-1}\phi^{n+1}|_{\xn=0}$ cannot be constructed from physical boundary fundamental operators if all of them have derivative orders $q_i>\frac{k-1}{n+1}$.
In other words, the mixing can occur only if there is a physical boundary fundamental operator with $q_i\leq \frac{k-1}{n+1}$.
This inequality is stronger than \eqref{epsilon 1/2 condition}.
For example, suppose that the theory corresponds to $(k,n)=(7,3)$ and the physical fundamental operator with the smallest $q_i$ is $\Ph[a,2,1]$.
Then, the condition \eqref{epsilon 1/2 condition} can be satisfied for $q_i=2$.
However, the boundary interaction $\pa^{6}\phi^{4}|_{\xn=0}$ cannot be constructed using only physical operators, since the smallest $q_i=2$ is larger than $\frac{k-1}{n+1}=\frac 3 2$. 

In fact, the inequality \eqref{epsilon 1/2 condition} is a sufficient condition for the noninteger power expansion, but not a necessary condition.
The noninteger power terms in $\e$ may appear in other cases, where the integer power $\e$ expansion contradicts the analyticity of the three-point functions.
The conflict arises when we try to repeat the analysis below \eqref{k=2 conformal block expansion} in the more general scenario with arbitrary $k$, $n$ and boundary conditions. 
The problem is that the conformal blocks of the boundary fundamental operators with $q_j\neq q_i$ could be analytic at order $\e^0$.
Since their coefficients are at least of order $\e^1$ in the integer power $\e$ expansion, their contributions at order $\e^1$ are analytic and cannot cancel the nonanalytic terms. 
This problem arises if the free theory has two physical boundary fundamental operators with derivative orders $q_i$ and $q_i-2M$, where $M$ is an integer.
To see this, consider the conformal block expansion of $\<\phi_a \Ph[,q_i,2p]\Ph[b,q_i,2p+1]\>$. 
According to \eqref{conformal block}, 
the conformal block associated with $\Ph[a,q_j,1]$ is
\be
G^{(p)}_{\tPh[,q_i-2M,1]}(v)=v^{\frac 1 2 \Dh[q_i-2M,1]}\,{}_2F_1(-M,\, d/2-k-M;q_i-2M-k+3/2;v)+\ldots
\,,\label{G q-2M}
\ee
where the ${}_2F_1$ series terminates due to the argument $-M$.
So the block is analytic in $1-v$ at order $\e^0$.

We believe that if there are two physical boundary fundamental primaries in the free theory with derivative orders differ by an even number, say $\Ph[,q_i,1]$ and $\Ph[,q_i-2M,1]$, then the bulk-boundary-boundary three-point function $\<\phi_a \Ph[,q_i,2p]\Ph[b,q_i,2p+1]\>$ cannot be analytic around $v=1$ in the integer power $\e$ expansion.
We have verified this conjecture for some concrete cases with low $k$ and $n$, such as $(k,n)=(2,2),(2,4),(2,6),(3,2),(3,5),(3,6)$.
Here, we have avoided the cases where $k$ and $n-1$ have a common divisor and the cases where the \eqref{epsilon 1/2 condition} is satisfied.
For the ND and DN cases at $k=2$, the derivative orders of the two physical fundamental primaries  differ by two in the free theory.
For $k>2$, there always exist two physical operators of the forms $\Ph[,q_i,1]$ and $\Ph[,q_i-2M,1]$ in the free theory, since $(q_0,q_1,\ldots,q_{k-1})$ contains either more than one odd numbers or more than one even numbers. 

Below, let us examine the ND case with $(k,n)=(2,2)$ as a simple example.
The physical fundamental primaries in the free theory correspond to $\Ph[a,0,1]$ and $\Ph[a,2,1]$.
According to the $q_i=2$ and $M=1$ case of \eqref{G q-2M}, the correlator $\<\phi_a \Ph[,2,2p]\Ph[b,2,2p+1]\>$ is problematic in the integer power $\e$ expansion.
So the noninteger power terms in $\e$ should arise in this case.
Below, we will find the possible leading corrections to $\<\phi_a \Ph[,2,2p]\Ph[b,2,2p+1]\>$.
In the conformal block expansion, only O($N$) vectors $\cal O_a$ can contribute.

We first consider the case where $\cal O_a$ is composite. 
The order of the BOE coefficient $b_{\phi\cal O}$ may be constrained by the matching condition
\be
\lim_{\e\->0}\(\a^{-1}\<\Box^2\phi_a(\xp,\xn)\cal O_b(\yp,0)\>\)
=\<\phi^3_a(\xp,\xn)\cal O_b(\yp,0)\>_\f
\,,\label{box2 O}
\ee
which is analogous to the $(k,n)=(2,2)$ case of \eqref{matching k}.
The Laplacian leads to a factor $(\D_{\cal O}-\D_{\phi})_4$ given by the left-hand side of \eqref{epsilon 1/2 k} with $k=2$ and $\Dh[q_i,2r-1,m]$ replaced by $\D_{\cal O}$. 
Due to O($N$) symmetry, the composite operator $\cal O_b$ should contain an odd number of boundary fundamental fields in order for $b_{\phi\cal O}$ to be nonzero, so the lowest number of boundary fundamental fields in $\cal O_b$ is three.
The factor $(\D_{\phi}-\D_{\cal O})_4$ is finite at order $\e^0$, since the lowest composite scaling dimension is $\D_{\cal O}=3\D_{\phi}=6$ at order $\e^0$, where $\D_{\phi}=2$ in the free theory.
This implies that $\<\Box^2\phi_a(\xp,\xn)\cal O_b(\yp,0)\>$ is of the same order as $b_{\phi\cal O}$.
If $\cal O$ contains more than three fundamental fields, then the right-hand side of \eqref{box2 O} is zero, so $b_{\phi\cal O}$ is of higher order than $\a\sim\e^1$.
If $\cal O$ contains three fundamental fields, then the right-hand side of \eqref{box2 O} is finite, and the BOE coefficient $b_{\phi\cal O}$ is of $\e^1$.
For the leading corrections to $\<\phi_a \Ph[,2,2p]\Ph[b,2,2p+1]\>$, more possibilities can be excluded since the boundary OPE coefficient $f_{\cal O}^{(p)}$ may be of higher order in $\e$.
If $\cal O_b$ contains a boundary fundamental field $\Ph[a,q_i,1]$ with $q_i\neq 2$, then the boundary OPE coefficient $f_{\cal O}^{(p)}$ is of higher order than $\e^0$.
In this case, the product $b_{\phi\cal O}f_{\cal O}^{(p)}$ is of higher order than $\e^1$ as $b_{\phi\cal O}$ is at least of order $\e^1$.
We conclude that for boundary composite primaries only $\Psi_a^{(2,3,m)}$ contribute at order $\e^1$, where $\Psi_a^{(2,3,m)}$ is of the schematic form $\Box^m_\parallel(\Ph[a,2,1])^3$.
All other boundary composite primaries contribute at higher orders.

Now, we discuss the boundary fundamental primaries $\Ph[a,0,1]$ and $\Ph[a,2,1]$, which are physical in the Gaussian limit.
The nonanalytic term from $\Ph[a,0,1]$ is of higher order in $\e$ for the following reason.
The $\Ph[a,0,1]$ block is analytic in $1-v$ at order $\e^0$.
This means that the nonanalytic term from $\Ph[a,0,1]$ is proportional to $\gh[q_i,1]+\gh[q_i,2p]-\gh[q_i,2p+1]$, which results from the $\e$ expansion of the $\Ph[a,0,1]$ block, as in \eqref{composite block k>1}.
Since the boundary OPE coefficient $f_{\Ph[,0,1]}^{(p)}$ is of higher order than $\e^0$, the nonanalytic term from $\Ph[a,0,1]$ is of higher order than $\gh[q_i,1]+\gh[q_i,2p]-\gh[q_i,2p+1]$.
On the other hand, the nonanalytic term from $\Ph[a,2,1]$ is of the same order as $\gh[q_i,1]+\gh[q_i,2p]-\gh[q_i,2p+1]$, since $b_{\phi\Ph[,2,1]}f_{\Ph[,2,1]}^{(p)}$ is of order $\e^0$.
Therefore, the nonanalytic term from $\Ph[a,0,1]$ is of higher order than that from $\Ph[a,2,1]$.
Since we want to focus on the leading corrections that are nonanalytic in $1-v$, we can ignore the $\Ph[a,0,1]$ contribution. 

There are also the boundary fundamental primaries $\Ph[a,1,1]$ and $\Ph[a,3,1]$, which are null in the free theory.
According to the discussion above, the possible nonanalytic terms in for the leading corrections are associated with the contributions from $\Ph[a,1,1]$, $\Ph[a,2,1]$, $\Ph[a,3,1]$, and $\Psi_a^{(2,3,m)}$.
As the orders of the BOE and boundary OPE coefficients associated with $\Ph[a,1,1]$ and $\Ph[a,3,1]$ are unknown, there are multiple possibilities for the nonanalytic terms to cancel out. 
A simple scenario is that only $\Ph[a,1,1]$ and $\Ph[a,2,1]$ contribute nonanalytic terms in $1-v$ at lower order than $\e^1$.
For the leading corrections to $\<\phi_a \Ph[,2,2p]\Ph[b,2,2p+1]\>$, the nonanalytic terms are associated with
\be
G^{(p)}_{\Ph[,2,1]}(v)&=\frac {4(\gh[2,2p]-\gh[2,2p+1])} 3 v^3\,{}_3F_2\[\begin{matrix}1,1,5\\ 2,\frac 5 2\end{matrix};v\]+\text{analytic terms}+\ldots\,,\label{G21} \\
G^{(p)}_{\Ph[,1,1]}(v)&=v^{3/2}{}_2F_1\!\(-\frac 1 2,\, \frac 7 2; \frac 1 2;v\)+\ldots
\,,\label{G11}
\ee
which are obtained by substituting $q_i=2$ and $q_{1-i}=1$ into \eqref{composite block k>1} and \eqref{fundamental block k>1}.
Here, the term with $\gh[2,1]$ is omitted because it is of order $\e^1$.
Using \eqref{3F2 expansion} and \eqref{2F1 expansion}, one can check that \eqref{G21} and \eqref{G11} have the same noninteger power terms in $1-v$ up to an overall factor.
So the analyticity requirement implies $\gh[2,2p]-\gh[2,2p+1]\propto b_{\phi\Ph[,1,1]}f_{\Ph[,1,1]}^{(p)}$, 
where the product $b_{\phi\Ph[,1,1]}f_{\Ph[,1,1]}^{(p)}$ is not determined by these constraints. 
If the leading nonanalytic corrections are of order $\e^1$, 
we need to take into account the nonanalytic terms from $\Ph[a,3,1]$ and the infinite sum associated with the composite operators $\Psi_a^{(2,3,m)}$ as in \eqref{k=2 conformal block expansion}. 

In summary, there are two arguments for the occurrence of the noninteger power $\e$ expansion.
The first argument is associated with the consistency of the matching condition \eqref{matching k}.
When both $k$ and $n$ are odd integers, the noninteger power terms in $\e$ arise  if \eqref{epsilon 1/2 condition} is satisfied.
For $k=1$, this leads to the $\e^{1/2}$ expansion in the second half of section \ref{Boundary composite operators}.
The second argument makes use of the analyticity requirement of the bulk-boundary-boundary three-point function $\<\phi_a \Ph[,2,2p]\Ph[b,2,2p+1]\>$.
According to our analyses, this may occur in the ND and DN cases at $k=2$, and all of the $k>2$ theories. 

\subsection{Analytic bootstrap}

In this section, we will consider the analytic bootstrap for $\<\phi_a(x)\phi_b(y)\>$ in the higher derivative theories, i.e., $k>1$.
We will present the bootstrap equations for general $k$, 
and solve them explicitly for $k=2$.
The results for the fundamental anomalous dimensions $\gh[q_i,1]$ agree with the multiplet-recombination result \eqref{gamma fundamental}. 

In a higher derivative free theory,  
more operators contribute to the bootstrap equation in both channels.
In the bulk channel, the bulk bilinear operators can have higher twists, 
so more bulk primaries can contribute to the bulk channel expansion. 
They are the internal operators of the form $\psi^{(2,m)}\sim\phi_a\,\Box^m\phi_a$, where $m=0,1,\ldots k-1$.
\footnote{As opposed to the boundary composite operators $\Psi^{(q_i,2r-1,m)}_a$, the bulk composite operators $\psi^{(2,m)}$ are not labeled by $q_i$, i.e., are not constructed using the normal and parallel derivatives separately. }
On the other hand, there exist $k$ boundary fundamental operators $\Ph[\f,q_i,1]$ in the boundary channel, where $i=0,1,\ldots k-1$.
As a result, the order-$\e^0$ bootstrap equation takes the form
\be\label{crossing free k>1}
1+\sum_{m=0}^{k-1}\l_{\tilde\psi_{\f}^{(2,m)}}a_{\tilde\psi_{\f}^{(2,m)}}f_\text{bulk}(\D_{\tilde\psi_{\f}^{(2,m)}},\x)=\sum_{i=0}^{k-1}\m_{\tPh[\f,q_i,1]}^2 f_\text{bdy}(\Dh[q_i,1,\f],\x)
\,.
\ee
The free-theory solutions can be classified by the scaling dimensions of the boundary fundamental primaries
\be
\Dh[q_i,1,\f]=\frac{d-2k}{2}+q_i
\,.
\ee
Each set of solution for $(q_0,q_1,\ldots,q_{k-1})$ corresponds to a set of physical boundary fundamental primaries.
For example, we obtain four sets of solutions at $k=2$.
The choices for physical fundamental primaries are given by 
$(q_0,q_1)=(0,1), (0,2), (3,1), (3,2)$.
For the $(0,1)$ case, the boundary condition is NN, since the boundary fundamental primaries are associated with the zeroth- and first-order derivatives of $\phi$.
The solutions for the free data read
\be
\l_{\tilde\psi^{(2,0)}}a_{\tilde\psi^{(2,0)}}=\frac{d-2}{2}\,,\quad
\l_{\tilde\psi^{(2,1)}}a_{\tilde\psi^{(2,1)}}=2\frac{d-4}{d-6}\,,\quad
\m_{\tPh[,0,1]}^2=2\,,\quad
\m_{\tPh[,1,1]}^2=\frac{d-4}{2}
\,.
\ee
The case of $(0,2)$ corresponds to the ND boundary condition, and we obtain
\be
\l_{\tilde\psi^{(2,0)}}a_{\tilde\psi^{(2,0)}}=1\,,\quad
\l_{\tilde\psi^{(2,1)}}a_{\tilde\psi^{(2,1)}}=\frac{d-4}{d-6}\,,\quad
\m_{\tPh[,0,1]}^2=2\,,\quad
\m_{\tPh[,2,1]}^2=\frac{(d-4)(d-2)}{8}
\,.
\ee
The $(3,1)$ case is associated with the DN boundary condition, and the set of solutions is given by
\be
\l_{\tilde\psi^{(2,0)}}a_{\tilde\psi^{(2,0)}}=-1\,,\quad
\l_{\tilde\psi^{(2,1)}}a_{\tilde\psi^{(2,1)}}=-\frac{d-4}{d-6}\,,\quad
\m_{\tPh[,1,1]}^2=\frac{d-4}{2}\,,\quad
\m_{\tPh[,3,1]}^2=-\frac{(d-4)(d-2)d}{96}
\,.
\ee
Finally, the DD boundary condition is given by the $(3,2)$ solution.
In this case, we have
\be
\l_{\tilde\psi^{(2,0)}}a_{\tilde\psi^{(2,0)}}&=-\frac{d-2}{2}\,,\qquad\qquad\;\;
\l_{\tilde\psi^{(2,1)}}a_{\tilde\psi^{(2,1)}}=-2\frac{d-4}{d-6}\,,\nn
\m_{\tPh[,2,1]}^2&=\frac{(d-4)(d-2)}{8}\,,\quad\qquad
\m_{\tPh[,3,1]}^2=-\frac{(d-4)(d-2)d}{96}
\,.
\ee
The four sets of solutions are in agreement with table 1 in \cite{Herzog:2024zxm}.\footnote{The free data can also be derived using the free correlators in section \ref{Generalized free theory with a boundary}.
For example, the BOE coefficients $b_{\phi\Ph[,q_i,1]}$ can be read off from \eqref{bulk boundary free}.
After changing the normalization according to \eqref{boundary internal operator normalization} and \eqref{mu-b}, we confirm that the results agree with the analytic bootstrap results.}

In the interacting theories, the bootstrap equation \eqref{crossing free k>1} receives corrections in the $\e$ expansion.
As in section \ref{Analytic bootstrap}, there are two scenarios: $n=2$ and $n>2$. 
In the generalized $\phi^4$ theory, we need to take into account the new operators $\psi^{(4,m)}\sim\Box^m\phi^4$ in the bulk channel.
It turns out that we only need to consider $m=0,1,\ldots k-1$ in the $\Box^k$ theory. 
The bootstrap equation takes the form
\be
&1+\sum_{m=0}^{k-1}\l_{\tilde\psi^{(2,m)}}a_{\tilde\psi^{(2,m)}}f_\text{bulk}(\D_{\tilde\psi^{(2,m)}},\x)+\sum_{m=0}^{k-1}\l_{\tilde\psi^{(4,m)}}a_{\tilde\psi^{(4,m)}}f_\text{bulk}(\D_{\tilde\psi^{(4,m)}},\x)\nn
=\;&\x^{\D_\phi}\sum_{i=0}^{k-1}\m_{\tPh[,q_i,1]}^2 f_\text{bdy}(\Dh[q_i,1],\x)+O(\e^2)
\,,
\ee
which is the higher derivative generalization of \eqref{crossing n=2}. 
Here, we have assumed that the $\e$-expansion series contain only integer powers. 
As the BOE coefficients are squared, we do not need to consider new operators in the boundary channel. 
The solutions for the anomalous dimensions of boundary fundamental operators are written in terms of the anomalous dimension of the bulk operator $\g_{\phi^2}$.
For $k=2$, we obtain
\be
\gh[0,1]&=\frac{3}{2}\g_{\phi^2}+O(\e^2)\,,\quad
\gh[1,1]=-\frac{9}{2}\g_{\phi^2}+O(\e^2)\,,\quad \text{(NN)}\,,\label{crossing k=2 NN solution}\\
\gh[3,1]&=\frac{3}{2}\g_{\phi^2}+O(\e^2)\,,\quad
\gh[2,1]=-\frac{9}{2}\g_{\phi^2}+O(\e^2)\,,\quad \text{(DD)}
\,. \label{crossing k=2 DD solution}
\ee
The noninteger power terms in $\e$ may arise in the ND and DN cases.
We defer the related discussions to the end of this section.
To leading order in $\e$, the anomalous dimension of $\phi^2$ is $k$ independent \cite{Guo:2023qtt}:
\be\label{bulk input phi2}
\g_{\phi^2}=\frac{N+2}{N+8}\e+O(\e^2)
\,.
\ee
Using this input,
we confirm that \eqref{crossing k=2 NN solution} and \eqref{crossing k=2 DD solution} agree with the results from the multiplet recombination \eqref{gamma fundamental}.
Furthermore, all solutions require that the bulk anomalous dimension $\g_{\Box\phi^2}$ vanishes at order $\e^1$.
This is consistent with the result in \cite{Guo:2023qtt}.

For $n>2$, the bootstrap equation is
\be\label{crossing n>2 k>1}
&1+\sum_{m=0}^{k-1}\l_{\tilde\psi^{(2,m)}}a_{\tilde\psi^{(2,m)}}f_\text{bulk}(\D_{\tilde\psi^{(2,m)}},\x)+\sum_{m=0}^{k-1}\l_{\tilde\psi^{(2n-2,m)}}a_{\tilde\psi^{(2n-2,m)}}f_\text{bulk}(\D_{\tilde\psi^{(2n-2,m)}},\x)\nn
&\hspace{1em}+\sum_{m=0}^{k-1}\l_{\tilde\psi^{(2n,m)}}a_{\tilde\psi^{(2n,m)}}f_\text{bulk}(\D_{\tilde\psi^{(2n,m)}},\x)
=\x^{\D_\phi}\sum_{i=0}^{k-1}\m_{\tPh[,q_i,1]}^2 f_\text{bdy}(\Dh[q_i,1],\x)+O(\e^2)
\,,
\ee
which is analogous to \eqref{crossing n>2}.
As above, the internal operators $\psi^{(p,m)}\sim\Box^m\phi^p$ in the bulk channel can have higher twists.
Similar to the generalized $\phi^4$ theory, we find that the primaries $\psi^{(2n-2,m)}$ and $\psi^{(2n,m)}$ with $m>k-1$ do not contribute at order $\e^1$. 
The solutions to the bootstrap equation at order $\e^1$ are written in terms of
\be\label{lambda a k n}
\l_{\tilde{\phi}^{2n}}a_{\tilde\phi^{2n}}=\frac{(-1)^k2^{n-2 k} \(\frac{N}{2}+1\)_{n-1}\(\sum_{j=0}^{k-1}B_j\)^n}{k! \(\frac{(1-2n)k}{n-1}+1\)_k}\a+O(\e^2)
\,,
\ee
where we have used \eqref{bulk OPE} and Wick contractions.
This is the $k>1$ generalization of \eqref{lambda a 2n}. 
For example, the case of $k=2$ corresponds to the solutions
\be\label{k=2 n=4 NN}
\gh[0,1]&=\frac{2 n (1-3n)}{3 \left(n^2-1\right)}\l_{\tilde{\phi}^{2n}}a_{\tilde\phi^{2n}}+O(\e^2)\,,\quad
\gh[1,1]=\frac{2 n (3 n-1)}{n^2-1}\l_{\tilde{\phi}^{2n}}a_{\tilde\phi^{2n}}+O(\e^2)\,, \quad \text{(NN)}\,,\\
\gh[3,1]&=\frac{2 n (1-3n)}{3 \left(n^2-1\right)}\l_{\tilde{\phi}^{2n}}a_{\tilde\phi^{2n}}+O(\e^2)
\,,\quad
\gh[2,1]=\frac{2 n (3 n-1)}{n^2-1}\l_{\tilde{\phi}^{2n}}a_{\tilde\phi^{2n}}+O(\e^2)\,,
 \quad \text{(DD)}
\,.
\ee
These results are in agreement with \eqref{gamma fundamental} from the multiplet recombination.

In the case of noninteger power $\e$ expansion, there can be boundary fundamental operators with BOE coefficients of lower order than $\e^1$. 
For $n=2$, we have
\be\label{crossing noninteger n=2}
&1+\sum_{m=0}^{k-1}\l_{\tilde\psi^{(2,m)}}a_{\tilde\psi^{(2,m)}}f_\text{bulk}(\D_{\tilde\psi^{(2,m)}},\x)+\sum_{m=0}^{k-1}\l_{\tilde\psi^{(4,m)}}a_{\tilde\psi^{(4,m)}}f_\text{bulk}(\D_{\tilde\psi^{(4,m)}},\x)\nn
=\;&\x^{\D_\phi}\[\sum_{i=0}^{k-1}\(\m_{\tPh[,q_i,1]}^2 f_\text{bdy}(\Dh[q_i,1],\x)+\m_{\tPh[,2k-1-q_i,1]}^2 f_\text{bdy}(\Dh[2k-1-q_i,1],\x)\)+\ldots\]
\,.
\ee
Here we have assumed that the BOE coefficients $\m_{\tPh[,2k-1-q_i,1]}$ are of order $\e^{1/2}$ for concreteness.
If $\m_{\tPh[,2k-1-q_i,1]}$ are of higher order than $\e^{1/2}$, then we can ignore the $\Ph[,2k-1-q_i,1]$ contributions on the right-hand side of \eqref{crossing noninteger n=2}.
In the more complicated case, the BOE coefficients $\m_{\tPh[,2k-1-q_i,1]}$ are of lower order than $\e^{1/2}$, and $a_{\tilde\psi^{(2,m)}}$ should admit noninteger power $\e$ expansions in order for the bootstrap equation to be consistent to order $\e^1$.
At $k=2$, all the cases lead to the same solutions for the fundamental anomalous dimensions:
\be
\gh[0,1]&=\frac{1}{2}\g_{\phi^2}+\ldots\,,\quad\;\;\;
\gh[2,1]=\frac{3}{2}\g_{\phi^2}+\ldots\,,\quad\,\text{(ND)}\,, \label{crossing k=2 ND solution}\\
\gh[3,1]&=\frac{1}{2}\g_{\phi^2}+\ldots\,,\quad\;\;\;
\gh[1,1]=\frac{3}{2}\g_{\phi^2}+\ldots\,,\quad\;\text{(DN)}
\,,\label{crossing k=2 DN solution}
\ee
Using \eqref{bulk input phi2}, we verify that \eqref{crossing k=2 ND solution} and \eqref{crossing k=2 DN solution} agree with those from the multiplet-recombination result \eqref{gamma fundamental}.
For $n>2$, the bulk channel expansion is the same as that in \eqref{crossing n>2 k>1}, but more boundary fundamental operators contribute to the boundary channel expansion,
\be
&1+\sum_{m=0}^{k-1}\l_{\tilde\psi^{(2,m)}}a_{\tilde\psi^{(2,m)}}f_\text{bulk}(\D_{\tilde\psi^{(2,m)}},\x)+\sum_{m=0}^{k-1}\l_{\tilde\psi^{(2n-2,m)}}a_{\tilde\psi^{(2n-2,m)}}f_\text{bulk}(\D_{\tilde\psi^{(2n-2,m)}},\x)\nn
&+\sum_{m=0}^{k-1}\l_{\tilde\psi^{(2n,m)}}a_{\tilde\psi^{(2n,m)}}f_\text{bulk}(\D_{\tilde\psi^{(2n,m)}},\x)
\nn
=\;&\x^{\D_\phi}\[\sum_{i=0}^{k-1}\(\m_{\tPh[,q_i,1]}^2 f_\text{bdy}(\Dh[q_i,1],\x)+\m_{\tPh[,2k-1-q_i,1]}^2 f_\text{bdy}(\Dh[2k-1-q_i,1],\x)\)+\ldots\]
\,,
\ee
where $\m_{\tPh[,2k-1-q_i,1]}$ are assumed to be of order $\e^{1/2}$. 
For $k=2$, the solutions for the anomalous dimensions are
\be
\gh[0,1]&=\frac{2 n (1-3n)}{3 (n-1)^2}\l_{\tilde{\phi}^{2n}}a_{\tilde\phi^{2n}}+\ldots\,,\quad
\gh[2,1]=\frac{2 n (1-3n)}{(n-1)^2}\l_{\tilde{\phi}^{2n}}a_{\tilde\phi^{2n}}+\ldots\,, \quad \text{(ND)}\,,\label{crossing k=2 ND solution n>2}\\
\gh[3,1]&=\frac{2 n (1-3n)}{3 (n-1)^2}\l_{\tilde{\phi}^{2n}}a_{\tilde\phi^{2n}}+\ldots
\,,\quad
\gh[1,1]=\frac{2 n (1-3 n)}{(n-1)^2}\l_{\tilde{\phi}^{2n}}a_{\tilde\phi^{2n}}+\ldots\,,
\quad \text{(DN)}
\,. \label{crossing k=2 DN solution n>2}
\ee 
Using \eqref{lambda a k n}, we verify that the solutions above agree with the multiplet-recombination results \eqref{gamma fundamental}. 
Similar to the $n=2$ case, the results \eqref{crossing k=2 ND solution n>2} and \eqref{crossing k=2 DN solution n>2} remain the same if the BOE coefficients $\m_{\tPh[,2k-1-q_i,1]}$ are not of order $\e^{1/2}$.
Again, we emphasize that if $\m_{\tPh[,2k-1-q_i,1]}$ are of lower order than $\e^{1/2}$, then $a_{\tilde\psi^{(2,m)}}$ admit noninteger power expansions in $\e$ and the bootstrap equation is solved to order $\e^1$. 

Note that the solutions for $\g_{q_i,1}$ and $\g_{2k-1-q_i,1}$ are related by the transformation $\text{N}\leftrightarrow \text{D}$. 
This is related to the duality between $\g_{q_i,1}$ and $\g_{2k-1-q_i,1}$
mentioned around \eqref{gamma fundamental}. 
To avoid the case where $n$ and $k-1$ have a common divisor, $n$ should be an even integer for $k=2$, 
so we do not write the factor $(-1)^n$ explicitly. 
Although the crossing solutions in this section correspond to $k=2$, the procedure can be straightforwardly extended to higher $k$ solutions.

\section{Discussion}

In this paper, we have considered the O($N$)-symmetric $\phi^{2n}$ theories with a boundary and their higher derivative generalizations.
We have obtained the anomalous dimension of the boundary fundamental operator using the multiplet recombination, and the result is given in \eqref{gamma Phi}.
We have also considered the O($N$) vector and singlet boundary composite operators.
Their anomalous dimensions are obtained by requiring the analyticity of the bulk-boundary-boundary three-point functions.
The results with the integer power $\e$ expansion are given in \eqref{gamma odd} and \eqref{gamma even}.
We have also obtained the results involving the $\e^{1/2}$ expansion, which are given in \eqref{epsilon 1/2 odd} and \eqref{epsilon 1/2 even}.
For the higher derivative $\phi^{2n}$ theories with a boundary, we have derived the anomalous dimensions \eqref{gamma fundamental} of the boundary fundamental operators.
For the boundary composite operators, we have studied the $\phi^4$ deformation of the free $\Box^2$ theory with the NN or DD boundary condition.
The anomalous dimensions of boundary composite operators are given in \eqref{k=2 composite vector} and \eqref{k=2 composite singlet}.
We have extended the $\e^{1/2}$ expansion in the canonical case to the higher derivative theories around \eqref{epsilon 1/2 condition}. \footnote{Besides \eqref{epsilon 1/2 condition}, the noninteger power $\e$ expansion may appear in other cases for $k>1$.
	We have not found a physical explanation for this.
	It might be useful to consider the analytic bootstrap to higher orders in $\e$.}
In addition, the results for the anomalous dimensions of boundary fundamental operators have been verified for $k=2$ using the analytic bootstrap.

We have not considered the extraordinary transitions, in which the O($N$) symmetry is spontaneously broken. 
They have been studied recently in perturbation theory \cite{Shpot:2019iwk,Dey:2020lwp,Sun:2025ihw}  
for the canonical case with a $\phi^{4}$-bicritical or $\phi^{6}$-tricritical bulk.
It would be interesting to generalize these results to higher derivative cases or $\phi^{2n}$ theories with higher $n$. 
One obstacle for the multiplet recombination approach is that the Gaussian limit $\e\->0$ diverges, as the free theories do not have extraordinary transitions. 
Nevertheless, the analytic bootstrap method may be applicable \cite{Liendo:2012hy, Dey:2020lwp}. 

It would also be interesting to consider anisotropic Lifshitz points and their generalizations 
based on the anisotropic free theory. 
After determining the bulk data, one can further study the case with a boundary, 
which can be either parallel to all the modulation axes or perpendicular to the one of the modulation axes.
These two different orientations of the boundary are expected to lead to distinct universality classes. See for example \cite{Diehl:2004kj,Diehl_2006} and references therein. 

Besides the $\phi^{2n}$ deformations, another class of theories are associated with $\bb Z_2$-odd interactions of the form $\phi^{2n+1}$.
In the case without a boundary, the $\phi^{2n+1}$ theories with a generic positive integer $n$ has been investigated using the multiplet recombination or equation of motion \cite{Codello:2017qek,Guo:2024bll}.
These theories are of physical relevance as well.
For instance, in statistical physics, the $\phi^3$ theory describes the Yang-Lee edge singularity \cite{Fisher:1978pf}.
The $S_{N+1}$-symmetric version of the $\phi^3$ theory describes the Potts model, 
whose $N\rightarrow 0$ and $N\rightarrow -1$ limits correspond to the percolation problem and spanning forest.
It would be interesting to study these systems with a boundary (see e.g. \cite{diehl1989semi}). 
Moreover, the non-Hermitian $\phi^3$ theory with an imaginary coupling constant is $\cal {PT}$-symmetric \cite{Bender:2012ea}.
The spectra of $\cal {PT}$-symmetric theories are real and bounded from below \cite{Bender:1998ke,Bender:2007nj,Bender:2023cem}.
In low dimensions, the $\cal {PT}$-symmetric theories have been studied using bootstrap methods \cite{Li:2022prn,Khan:2022uyz,Li:2023nip,John:2023him,Li:2023ewe,Li:2024rod}.
Higher dimensional bootstrap studies of these theories are more challenging and require further investigations.
The presence of a boundary or defect can have interesting physical effects.

As in nonunitary CFTs, the bootstrap equations for BCFTs do not need to satisfy positivity conditions 
because the conformal block expansion in the bulk channel does not lead to squared expansion coefficients.
In the nonperturbative studies, we may need to use the conformal bootstrap methods that do not rely on positivity conditions \cite{Gliozzi:2015qsa,Padayasi:2021sik}. (See also  \cite{Gliozzi:2013ysa,Gliozzi:2014jsa,El-Showk:2016mxr,Esterlis:2016psv,Li:2017agi,Li:2017ukc,Li:2021uki,Kantor:2021kbx,Kantor:2021jpz,Afkhami-Jeddi:2021iuw,Laio:2022ayq,Li:2023tic} for the case without a boundary.)
It would be interesting and useful to further explore these nonpositive bootstrap methods.

Boundaries are codimension-one defects.
A natural continuation would be to consider the cases of higher codimensional defects.
In the canonical case $k=1$, the multiplet recombination has been used to study the Wilson-Fisher CFT with a localized magnetic field \cite{Nishioka:2022qmj} or with a monodromy defect \cite{Yamaguchi:2016pbj,Soderberg:2017oaa}.
We plan to extend these results to the multicritical generalizations of the $\phi^{2n}$ and higher derivative types.

\section*{Acknowledgments}

We would like to thank Chris Herzog for helpful correspondence.
This work was supported by the Natural Science Foundation of China (Grant No. 12205386) and the Guangzhou Municipal Science and Technology Project (Grant No. 2023A04J0006).

\appendix

\section{\boldmath Free BOE coefficients and boundary OPE coefficients}
\label{Free BOE coefficients and OPE coefficients}

Using Wick contractions, we can write the three-point function as
\be\label{free 3pt}
&\<\phi_a^{2n-1}(\xp,\xn)\Ph[,q,2p](0,0)\Ph[b,q,2p+1](\yp,0)\>\nn
=\;&\frac{\d_{ab}|\yp|^{-2\Dh[2p+1]}}{|x|^{\dD}\xn^{\smash{\D_{\phi^{2n-1}}}}}\sum_{r=1}^{n}W_{2r-1,p}\, v^{\frac{(2r-1)(d-2k+2q)}{4}}+O(|\yp|^{-2\Dh[2p+1]+1})
\,,
\ee
where the cross ratio $v$ is defined in \eqref{v definition} and the expansion coefficient is
\be\label{W}
W_{2r-1,p}=\;&\frac{2^{n+2p-2}p!(n-1)!(\frac{N}{2}+1)_{p}(r+1)_r(p-r+2)_{r-1}(\frac{N}{2})_{n-r}(N+2n-2)}{r!(r-1)!(n-r)!(N+2r-2)}\nn
&\times\(\frac{2q!(1-2k)_{q}\(\frac{d-2k}{2}\)_{q}}{\(\frac{1}{2}-k\)_{q}}\)^{2p-r+1}\(\frac{2(1-2k)_{q}\(\frac{d-2k}{2}\)_{q}}{\(\frac{1}{2}-k\)_{q}}\)^{2r-1}\(\frac{\sum_{j=0}^{k-1}B_j}{2^{d-2k}}\)^{n-r}\nn
&\times{}_2F_1\!\(1-r,r-n;\frac{N}{2};1\) {}_3F_{2}\!\[\begin{matrix}1-r-\frac{N}{2},\frac{1-r}{2},1-\frac{r}{2}\\\frac{1}{2}-r,p-r+2\end{matrix};1\]
\,.
\ee
Here we have used \eqref{1pt}, \eqref{bulk boundary free}, and \eqref{boundary boundary free}.
The Gaussian hypergeometric function satisfies the identity
\be
\sum_{m=0}^{\infty}\frac{(-1)^m(a)_m(b)_m}{(m+c)_m\,m!}v^m\,{}_2F_1(m+a,m+b;2m+1+c;v)=1
\,.
\ee
Setting $a=\frac{(r-1)(2q-2k+d)}{2}$, $b=\frac{r(d-2 k+2 q)}{2}$, and $c=\frac{2(r-1)d+2(2r-1)(q-k)+1}{2}$, we write the terms with $r>1$ in \eqref{free 3pt} as infinite sums over $m$.
Then the sum over $r$ in \eqref{free 3pt} becomes
\be\label{3pt wick}
&W_{1,p}\, v^{\frac{(d-2k+2q)}{4}}+\sum_{r=2}^{n}W_{2r-1,p}\,v^{\frac{(2r-1)(d-2k+2q)}{4}}\sum_{m=0}^{\infty}\frac{(-1)^m \(\frac{(r-1)(d-2 k+2 q)}{2}\)_m \(\frac{r(d-2 k+2 q)}{2}\)_m}{\(m+\frac{2(r-1)d+2(2r-1)(q-k)+1}{2}\)_m m!}\nn
&\hspace{3em}\times v^m\,{}_2F_1\!\(m+\tfrac{(r-1)(2q-2k+d)}{2},m+\tfrac{r(d-2 k+2 q)}{2};2m+1+\tfrac{2(r-1)d+2(2r-1)(q-k)+1}{2};v\)
\,.
\ee
On the other hand, the three-point function has the conformal block expansion
\be
&\<\phi_a^{2n-1}(\xp,\xn)\Ph[,q,2p](0,0)\Ph[b,q,2p+1](\yp,0)\>=\frac{\d_{ab}|\yp|^{-2\Dh[2p+1]}}{|x|^{\dD}\xn^{\D_\phi}}\Bigg[b_{\phi^{2n-1}\tilde\Phi}f_{\tilde\Phi}^{(p)}G^{(p)}_{\tilde\Phi}(v)\nn
&\hspace{7em}+\sum_{r=2}^{n}\sum_{m=0}^{\infty}b_{\phi^{2n-1}\tilde\Psi^{(q,2r-1,m)}}f^{(p)}_{\tilde\Psi^{(q,2r-1,m)}}G^{(p)}_{\tilde\Psi^{(q,2r-1,m)}}(v)\Bigg]+O(|\yp|^{-2\Dh[2p+1]+1})
\,.
\ee
According to \eqref{conformal block}, the terms in the square bracket are
\be\label{3pt conformal block}
&b_{\phi^{2n-1}\tilde\Psi^{(q,2r-1,m)}}f^{(p)}_{\tilde\Psi^{(q,2r-1,m)}}v^{\frac{(d-2k+2q)}{4}}+\sum_{r=2}^{n}v^{\frac{(2r-1)(d-2k+2q)}{4}}\sum_{m=0}^{\infty}b_{\phi^{2n-1}\tilde\Psi^{(q,2r-1,m)}}f^{(p)}_{\tilde\Psi^{(q,2r-1,m)}}\nn
&\hspace{3em}\times v^m\,{}_2F_1\!\(m+\tfrac{(r-1)(2q-2k+d)}{2},m+\tfrac{r(d-2 k+2 q)}{2};2m+1+\tfrac{2(r-1)d+2(2r-1)(q-k)+1}{2};v\)
\,.
\ee
For the free theory at the upper critical dimension \eqref{upper critical dim}, we have
\be
v^{\frac{(2r-1)(d-2k+2q)}{4}}=v^{\frac{2r-1}{2}(\frac{k}{n-1}+q)}
\,.
\ee
Since $k$ and $n-1$ have no common divisor and $q$ is an integer, the exponents of $v$ do not differ by integers for distinct $r\in\{1,2,\ldots,n\}$.
For this reason, the infinite sums in \eqref{3pt wick} or \eqref{3pt conformal block} do not mix with each other.
We can thus match \eqref{3pt wick} and \eqref{3pt conformal block} order by order in $v$, and determine the products of BOE and boundary OPE coefficients
\be
b_{\phi^{2n-1}\tPh[,q,1]}f_{\tPh[,q,1]}&=W_{1,p}\,,\\
\label{free relation}
b_{\phi^{2n-1}\tilde\Psi^{(q,2r-1,m)}}f^{(p)}_{\tilde\Psi^{(q,2r-1,m)}}&=\frac{(-1)^m \((r-1)(\tfrac{k}{n-1}+q)\)_m \(r(\tfrac{k}{n-1}+q)\)_m}{\(m+(2 r-1)q-\frac{k (n-2 r+1)}{n-1}+\frac{1}{2}\)_m m!}\,W_{2r-1,p}
\,.
\ee
Here $d$ is set to the upper critical dimension: $d=\frac{2nk}{n-1}$.
For the other three-point function $\<\phi_a^{2n-1}(\xp,\xn)\Ph[,q,2p-1](0,0)\Ph[,q,2p](\yp,0)\>$, we replace $W_{2r-1,p}$ by
\be\label{Wp}
W'_{2r-1,p}=\;&\frac{2^{n+2p-3}p!(n-1)!(\frac{N}{2}+1)_{p-1}(r+1)_r(p-r+1)_{r-1}(\frac{N}{2})_{n-r}(N+2n-2)}{r!(r-1)!(n-r)!(N+2r-2)}\nn
&\times\(\frac{2q!(1-2k)_{q}\(\frac{d-2k}{2}\)_{q}}{\(\frac{1}{2}-k\)_{q}}\)^{2p-r}\(\frac{2(1-2k)_{q}\(\frac{d-2k}{2}\)_{q}}{\(\frac{1}{2}-k\)_{q}}\)^{2r-1}\(\frac{\sum_{j=0}^{k-1}B_j}{2^{d-2k}}\)^{n-r}\nn
&\times{}_2F_1\!\(1-r,r-n;\frac{N}{2};1\) {}_3F_{2}\!\[\begin{matrix}1-r-\frac{N}{2},\frac{1-r}{2},-\frac{r}{2}\\\frac{1}{2}-r,p-r+1\end{matrix};1\]
\,.
\ee

\section{A bulk OPE coefficient}
\label{A bulk OPE coefficient}

In the analytic bootstrap for $n>2$, the solutions to the bootstrap equation are written in terms of the bulk data $\l_{\tilde{\phi}^{2n}}a_{\tilde\phi^{2n}}$.
In Appendix A of our previous paper \cite{Guo:2023qtt}, we obtained the bulk OPE coefficient $\l_{\tilde{\phi}^{2n-2}}$ in the case without a boundary, but we did not derive the result for $\l_{\tilde{\phi}^{2n}}$.
We derive $\l_{\tilde{\phi}^{2n}}$ in this appendix.
Consider the matching condition in the generalized $\phi^{2n}$ theories without a boundary:
\be
\lim_{\e\->0}\(\a^{-1}\<\Box^k\phi(x_1)\phi(x_2)\phi^{2n}(x_3)\>\)=\<\phi^{2n-1}(x_1)\phi(x_2)\phi^{2n}(x_3)\>_\f
\,.
\ee
This equation yields
\be
\lim\[\a^{-1}4^k\(-k\)_k\(\frac{(1-2n)k}{n-1}+1\)_k
\l_{\phi^{2n}}\]=2^{2n-1}n!\(\frac{N}{2}+1\)_{n-1}
\,,
\ee
which gives the bulk OPE coefficient
\be\label{bulk OPE}
\l_{\phi^{2n}}=\frac{ (-1)^k 2^{2 n-2 k-1}n!
   \left(\frac{N}{2}+1\right
   )_{n-1}}{k! \left(\frac{(1-2n)k
   }{n-1}+1\right)_k}\a+O(\e^2)
   \,.
\ee

\providecommand{\href}[2]{#2}\begingroup\raggedright\endgroup

\end{document}